\documentclass[a4paper]{article}
\usepackage[ruled,vlined]{algorithm2e}
\usepackage{a4wide}
\usepackage{amssymb, amsmath, graphicx,epstopdf, amsfonts}
\usepackage[T1]{fontenc}
\usepackage{lmodern} 
\usepackage{enumitem}
\usepackage{mathpartir}
\usepackage{bm}
\usepackage{url}
\usepackage{cite}
\newcounter{qcounter}

\SetCommentSty{mycommfont}
\usepackage{xcolor}

\newtheorem{ttd}{Definition}

\newtheorem{ttp}{Property}

\newenvironment{proof}{\paragraph{Proof:}}{\hfill$\square$}

\usepackage{hyperref,url}

\definecolor{darkgreen}{RGB}{0,127,173}
\definecolor{darkblue}{RGB}{0,0,200}

\hypersetup{
               colorlinks,
               linkcolor={darkblue},
               citecolor={darkgreen},
               urlcolor={darkblue}
}

\begin{document}

\title{DataProVe: A \textbf{Data} \textbf{Pr}otection P\textbf{o}licy and System Architecture \textbf{Ve}rification Tool} 

\author{V. T. Ta\\
vtta@uclan.ac.uk}

\maketitle

\begin{abstract}
In this paper, we propose a tool, called DataProVe, for specifying high-level data protection policies and system architectures, as well as verifying the conformance between them in a fully  automated way. The syntax of the policies and the architectures is based on semi-formal languages, and the automated verification engine relies on logic and resolution based proofs. The functionality  and operation of the tool are presented using different examples.  
\end{abstract}

\section{Introduction}
\label{sec:int}
Under the General Data Protection Regulation (GDPR) \cite{Gdpr4}, personal data is defined as ``any information relating to an identified or identifiable natural person''\footnote{In the US, personally identifiable information is used  with a similar interpretation \cite{NIST}.}.  The GDPR specifies the rights for living individuals who have their personal data processed, and enforce responsibilities for the data controllers and the data processors who store, process or transmit such data.  

Despite the data protection laws, there were several data breaches incidents in the past (e.g. \cite{Bloomberg-012014, Guardian-062015,  Engaget-072015}) and nowadays, such as the Cambridge Analytica scandal of Facebook \cite{Faceanalytica17}, where personal data of more than 87 millions Facebook users has been collected and used for advertising and election campaign purposes without a clear data usage consent. One of the main problems was the insufficient check by Facebook on the third party applications. Google also faced lawsuit over collecting personal data without permission, and has been reported to illegally gather the personal data of millions of iPhone users in the UK \cite{Google}. 

The GDPR took effect in May 2018, and hence, designing compliant data protection policies and system architectures became even more important for organizations to avoid penalties. Data protection by design, under the Article 25 of the GDPR \cite{Gdpr25}, requires the design of data protection measures  into the development of business processes of service providers.   
The regulation also limits businesses from performing user profiling and demanding appropriate consents before personal data collection (Article 6 of the GDPR \cite{Gdpr6}).

Unfortunately, in textual format, the data protection laws are sometimes ambiguous and can be misinterpreted by the policy and system designers. From the technical perspective, to the best of our knowledge,  only a small number of studies can be found in the literature that investigate the formal or automated method to design and verify  policies and architectures in the context of data protection and privacy. The main advantage of using formal approaches during system design is that data protection properties can be mathematically proved, and design flaws can be detected at an early stage, which can save time and money. 

On the other hand, using formal method for this purpose is also challenging, as  abstraction is required, which is difficult in case of complex laws. In this paper, we address this problem, and model some simple data protection requirements of GDPR with regards to the data collection, usage, storage, deletion, and transfer phases.  Privacy requirements are also considered such as the right to have certain data and link certain data types.  We focus on the policy and architecture levels, and  propose a variant of policy and architecture language, specifically designed for specifying and verifying data protection and privacy requirements. In addition, we propose a fully automated algorithm, for verifying three types of conformance relations between a policy and an architecture specified in our language. Our theoretical methods are implemented in the form of a software tool, called DataProVe, for demonstration purposes.  

The main goals of our policy and architecture languages and software tool include helping a system designer at the higher level specification (compared to the other tools that mainly focus on the protocol level), such as with the policy and architecture design, to spot any potential errors prior the concrete lower level system specification. Besides, our tool can be used for education or research purposes as well. To the best of our knowledge, this is the first work that addresses  the problem of fully automated conformance check between a policy and an architecture in the context of data protection and privacy requirements.  


This paper includes the following contributions: 
\begin{enumerate}
\item We propose a variant of privacy policy language (in Section~\ref{sec:polgen}). 

\item We propose a variant of privacy architecture language (in Section~\ref{sec:arch0}). 

\item We propose the definition of three conformance relations between a policy and architecture (in Section~\ref{conformance}), namely, the privacy, data protection, and functional conformance relations.   

\item We propose a logic based fully automated conformance verification procedure (in Section~\ref{sec:aut}) for the above three conformance relations.  

\item Finally, we propose a (prototype) tool, called DataProVe, based on the theoretical foundations (in Section~\ref{sec:imp}).  
\end{enumerate} 

\begin{figure}[htb!]
    \begin{center}
        \includegraphics[width=0.65\textwidth]{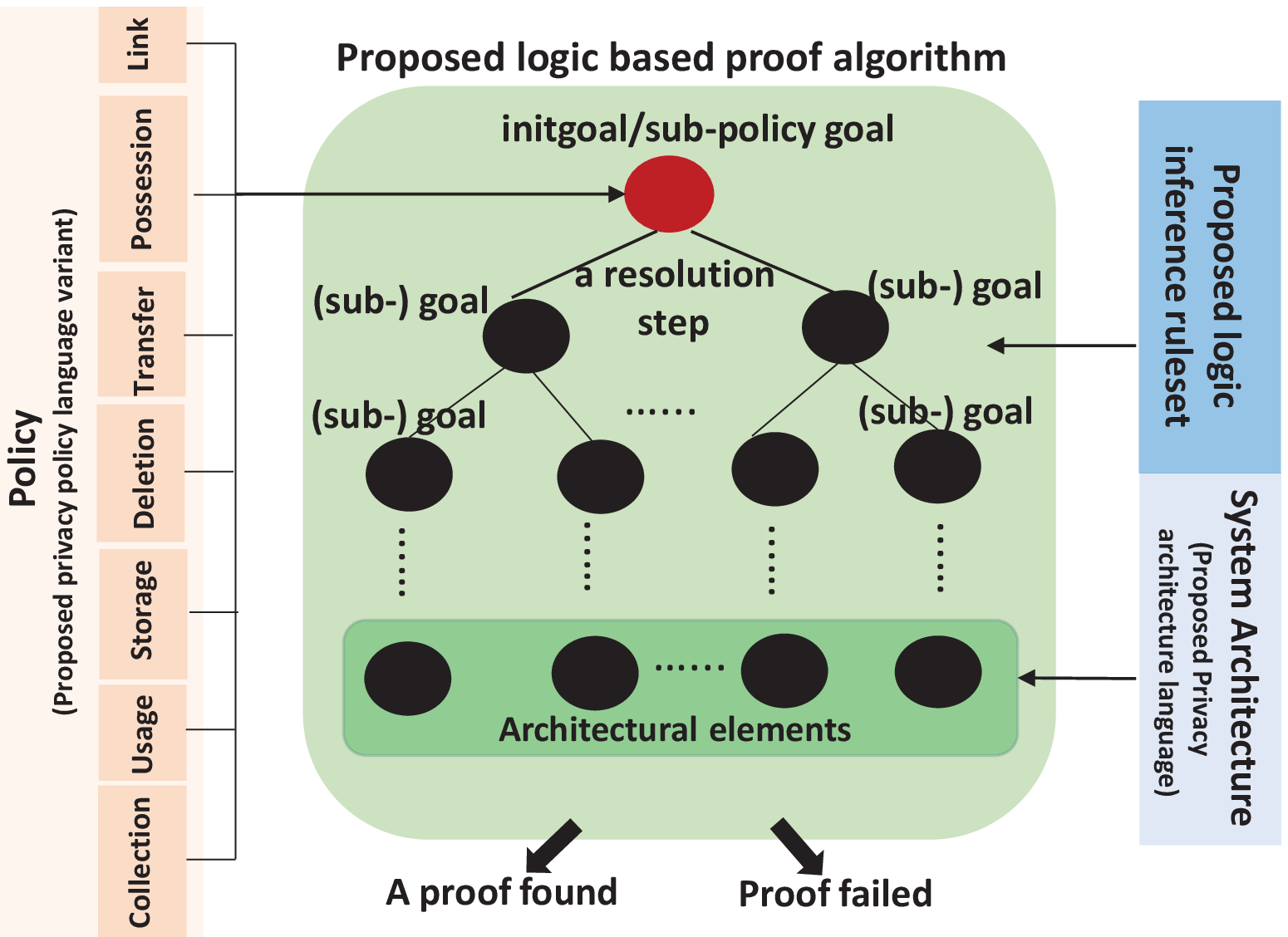}
    \end{center}
    \caption{An overview and the intuition behind the contributions of this paper.}
    \label{fig:contribution}
\end{figure} 
 
In Figure~\ref{fig:contribution}, a policy is specified using our proposed language variant, which covers seven sub-policies (data collection, usage, storage, deletion, transfer, possession and link). Each sub-policy is mapped to a logic goal that reflects the requirement in the policy.  The verification engine attempts to prove each goal based on a set of logic inference rules and architectural elements (specified in our language).  The proposed verification algorithm is based on a series of resolution steps, represented as a derivation tree, where the root is a goal to be proved, and the leaves are the architectural elements used to prove the goal.

The paper is structured as follows: In Section~\ref{sec:related}, we discuss the related policy and architecture languages. In Sections~\ref{sec:polgen}-\ref{sec:arch0}, we present our policy and architecture languages, respectively. The automated conformance verification engine is detailed in Section~\ref{sec:aut}. In Section~\ref{sec:imp} we present the DataProVe tool and its operation using two simple examples. Finally, we discuss the results and conclude the paper in Sections~\ref{sec:discussion} and \ref{sec:conc}. 

\section{Related Works}
\label{sec:related}


\subsection{Policy Languages} 
The Platform for Privacy Preferences (P3P) \cite{P3P} enables web users 
to gain control over their private information on online services. On a website users can express their privacy practices in a standard format that can be retrieved automatically and interpreted by web client applications. Users are notified about certain website's privacy policies and have a 
chance to make decision on that. To match the privacy preferences of the users and web services, the authors proposed the Preference Exchange Language (APPEL) \cite{APPEL} integrated into the web clients, with 
which the user can express their privacy preferences that can be matched against the practices set by the online services. According to the study \cite{XPref}, in APPEL, users can only specify what is unacceptable in a policy. Identifying this, the authors in \cite{XPref} proposed a more expressive preference language called XPref giving more freedom for the users, such as allowing acceptable preferences.

The Customer Profile Exchange (CPExchange) language \cite{CPexchange}, is a XML-based policy language, which was designed 
to facilitate business-to-business communication privacy policies (i.e. the  privacy-enabled global exchange of customer profile 
information). The eXtensible Access Control Markup Language (XACML) \cite{XACML} is a de-facto, XML-based
policy language, specifically designed for access control management in distributed systems. The latest version was approved by the 
OASIS standards organization as an international standard in 2017. The 
Enterprise Privacy Authorisation Language (EPAL) of IBM \cite{EPAL} was designed to regulate an organisation's 
internal privacy policies. EPAL is partly similar to XACML, however, it mainly focuses on privacy policies instead of 
access control policies in XACML.

A-PPL \cite{APPL} is an accountability policy language specifically designed for modelling data accountability (such as data 
retention, logging and notification) in the cloud. A-PPL is an extension of the the PrimeLife Privacy Policy 
Language (PPL) \cite{PPL}, which enables specification of access and usage control rules for the data subjects and the data 
controller. PPL is built upon XACML, and allows users to define the so-called sticky policies on personal data based on obligations. 
Obligation defines whether the policy language can trigger tasks that must be performed by a server and a client, once some event occurs and the related 
condition is fulfilled. This is also referred to as the Event-Action-Condition paradigm. The Policy Description Language (PDL) \cite{PDL}, 
proposed by Bell Labs, is one of the first policy-based management languages, specifically for network administration. It is declarative and 
is based on the Event-Action-Condition paradigm like PPL. 

RBAC (Role-Based Access Control) \cite{RBAC} is one of the most well-known role-based access control policy languages.
It uses roles and permissions in the enforced policies, namely, a subject can be assigned roles, and roles can be assigned certain access control permissions. 
ASL (Authorization Specification Language) \cite{ASL} is another Role-based access control language based on first order logic, and RBAC components. 
Ponder \cite{Ponder} is a declarative and object-oriented policy language, and designed for defining and modelling security policies
using RBAC, and security management policies for distributed systems. The policies are defined on roles or group 
of roles. Rei \cite{Rei} is a policy language based on deontic logic, designed mainly for modelling security and privacy properties of 
pervasive computing environments. Its syntax involves obligation and permission, where policies are defined as constraints over permitted and obligated actions on resources.

\subsection{Architecture Description Languages (ADLs)} 

Research on formal specification of architectures can be categorised into two groups of languages for software and hardware architectures, respectively. Darwin \cite{darwin}, one of the first languages for architectures, defined interaction of components through \textit{bindings}. Bindings associate services required by a component with the services provided by others. Its semantics is based on the  $\pi$-calculus \cite{pi}, a process algebra that makes Darwin capable of modelling dynamic architectures. In Wright \cite{wright}, components are associated via the \textit{connector} elements instead of bindings. Its semantics is defined in another process algebra, CSP \cite{csp}, with the architecture specific \textit{port processes} that specify external behaviour of a component, and \textit{spec process}, the internal behaviour of a component. 

Similar to Darwin, Rapide \cite{rapide} defines connections between the required service  and provided service ``ports" of components. Similar to Wright, Rapide also supports connectors, but in a more limited way (e.g. no first class connector elements), and hence, the user can only specify explicit links between the the required and provided services. Unlike Wright, Rapide also defines the actions \textit{in} and \textit{out} for asynchronous communication. The semantics of Rapide is based on the event pattern language \cite{rapide}, and is defined as a partially ordered set of events.         
Among the more recent ADLs, SOFA \cite{sofa} also defines connectors, which the user can specify based on four types of communication, a procedure call,
messaging, streaming, and  blackboard. The semantics of SOFA is based on Behaviour Protocol \cite{bp}, a simplified version of CSP. 

AADL \cite{aadl}, one of the most broadly-used ADLs, is specifically designed for
embedded systems. AADL defines  three groups of components, one for software
architectures (including thread, process, and subprogram), the second one is for hardware architectures (such as processor, memory), and the last group is for specifying composite types. In AADL, ports and subprogram-calls are used to define interaction between components. PRISMA \cite{prisma}, another recent ADL, was designed to address aspect-oriented software engineering. Similar to Wright,  
PRISMA defines first-class connector elements, which are specified with a set of roles (i.e. components) and the behaviour of the roles is defined by aspects.
The semantics of PRISMA is defined with modal logic and $\pi$-calculus. A recent attempt of architectures specification towards automation is proposed in the project, called CONNECT \cite{connect}. The semantics of this ADL is based on the FSP (finite state process) algebra \cite{fsp}, which allows automation and stochastic analyses of architectures.  Finally, UML has also been used to specify architectures in practice, however, it is more high-level and lacks formal semantics. We note that none of these ADLs support the specification of  data protection and privacy properties.  


\subsection{Comparison with our work}  The main differences between the policy languages above and our work is that, for instance, P3P, APPEL, XPref and even PPL  are mainly designed for web applications/services, and the policies are defined in a XML-based language, with restricted options for the 
users, while ours is designed for any type of services.   
In addition, our policy language variant is defined on data types (data type centred language), and supports a more systematic and fine-grained policy specification, as its syntax and semantics cover seven sub-policies capturing a representative data life-cycle (from the point the data is collected until its deletion). Our language variant is inspired by the ones proposed in \cite{TaButin15, ButinFM14}, which were  proposed for biometrics surveillance systems and log design. We modified and extend those to specify different data protection requirements.  

Unlike the ADLs above, our architecture language variant is designed to capture the data protection and privacy properties, and also supports cryptographic primitives. Our language is data type centred, and its semantics does not rely on process algebra like most above mentioned ADLs but instead is based on the state of all the defined data types in a system. This concept was applied in some of our previous works, such as in \cite{TaAntignac14, Antignac14}. The language variants in \cite{TaAntignac14, Antignac14} mainly focus on the computation and integrity verification of data based on trust relations. Unlike \cite{TaAntignac14, Antignac14}, the language variant in this paper focuses primarily on data protection and privacy properties, rather than the data integrity perspective.

Finally, to the best of our knowledge, this is the first work that studies and proposes a fully automated conformance check between the policy and architecture levels. Our verification engine is based on the syntax of our policy and architecture language variants, and logic resolution based proofs.

\section{The Specification of a Data Protection Policy}
\label{sec:polgen} 
A policy is defined from the perspective of a data controller. Here, we assume that the data controllers are service providers who collect, store, use or transfer the personal data of the data subjects. The data subjects in our case are system users whose personal data is/will be collected and used by the data controller for some purposes.

\subsection{Proposed Policy Syntax}
\label{sec:syntaxpp}
A policy of a service provider, \textit{sp}, is defined on a finite set of different entities \textit{EntitySet}$^{sp}_{pol}$ = \{$E_{i_1}$,\dots, $E_{i_n}$\}, and a finite set of data types \textit{DataTypes}$^{sp}_{pol}$ = \{$\theta_1$,\dots, $\theta_m$\}, used by the service. An entity can be  a data subject, data controller, organisations, hardware/software components. 

\begin{ttd}
\label{ttd:1}
(Data Protection Policy). The syntax of the data protection policies is defined as the collection of seven sub-policies on a given data type, namely:   

\begin{figure}[htbp]
\small
\centering
\fbox{\begin{minipage}{11.87 cm}
\begin{tabbing}    
    \=1\=1\=1\=1\= \kill
    \> \textit{POL}$_{\textit{DataTypes}^{sp}_{pol}}$ $=$ Pol$_{Col}$ $\times$ Pol$_{Use}$ $\times$ Pol$_{Str}$ $\times$ Pol$_{Del}$ $\times$ Pol$_{Fw}$ $\times$ Pol$_{Has}$ $\times$ Pol$_{Link}$.\\\\ 
 \=1\=1\=1\=1\= \kill
    \> where: \\
 \=1\=1\=1\=1\= \kill
    \> 1. Pol$_{Col}$ $=$ Cons$_{col}$ $\times$ CPurp. \ \ \ \ \ \ \ \ \  \ \ \ \ \ \ \ \ \ \ \ \ \ (Data Collection Sub-policy)\\\\ 
 \=1\=1\=1\=1\= \kill
    \> 2. Pol$_{Use}$ $=$ Cons$_{use}$ $\times$ UPurp. \ \ \ \ \ \ \ \ \ \ \ \ \ \ \ \ \ \ \ \ \ (Data Usage Sub-policy)\\\\ 
 \=1\=1\=1\=1\= \kill
    \> 3. Pol$_{Str}$ $=$ Cons$_{str}$ $\times$ Where$_{str}$. \ \ \ \ \ \ \ \ \  \ \ \ \ \ \ \ \ \ \  (Data Storage Sub-policy)\\\\ 
 \=1\=1\=1\=1\= \kill
    \> 4. Pol$_{Del}$ $=$  FromWhere$_{del}$ $\times$ Del$_{delay}$. \ \ \ \ \ \ \ \ \  \ \ (Data Retention Sub-policy)\\\\ 
 \=1\=1\=1\=1\= \kill
    \> 5. Pol$_{Fw}$ $=$   Cons$_{fw}$ $\times$ FwTo $\times$ FwPurp.  \ \ \ \ \ \ \ \ \ (Data Transfer Sub-policy) \\\\ 
   \=1\=1\=1\=1\= \kill
 	\> 6. Pol$_{Has}$ $=$   Who$_{canhave}$. \ \ \ \ \ \ \ \ \ \ \ \ \ \ \ \ \ \ \ \ \ \ \ \ \ \ \ \ \ (Data Possession Sub-policy)\\\\ 
    \> 7. Pol$_{Link}$ $=$   Who$_{canlink}$. \ \ \ \ \ \ \ \ \ \ \ \ \ \ \ \ \ \ \ \ \ \ \ \ \ \ \ \ \ (Data Connection Sub-policy) 
\end{tabbing}
\end{minipage}
}
\label{fig:dcpol}
\end{figure}
\end{ttd}
\normalsize

\begin{enumerate}
\item The data collection sub-policy specifies whether a collection consent is required (\textit{Cons}$_{col}$) and a set of collection purposes (\textit{CPurp}). These aim to capture the consent and purposes limitation requirements in  \textit{Article 6} \cite{Gdpr6} and \textit{Article 5(1)(b)} \cite{Gdpr5} of the GDPR.

\item The data usage sub-policy specifies whether a usage consent is required (\textit{Cons}$_{use}$) for using a type of data, besides the set of purposes of the data usage (\textit{UPurp}). These capture the \textit{Article 6} \cite{Gdpr6} and \textit{Article 30(1)(b))} \cite{Gdpr30} of the GDPR, respectively.

\item The data storage sub-policy specifies whether a storage consent is required (\textit{Cons}$_{str}$) for storing a piece of data, and where the data can be stored (\textit{Where}$_{str}$). These elements partly capture the storage limitation principle in \textit{Article 5(1)(e)} \cite{Gdpr5} of the GDPR.

\item The data deletion sub-policy specifies from where the data can be deleted (\textit{FromWhere}$_{del}$), alongside the corresponding deletion delay (\textit{Del}$_{del}$). These elements partly capture the  \textit{Article 5(1)(e)} and  \textit{Article 17(1)(a)} \cite{Gdpr17}  of the GDPR. 

\item The data transfer sub-policy defines whether a transfer consent is required (\textit{Cons}$_{fw}$), and all the entities (\textit{FwTo}) to which the data can be transferred with the given purposes  (\textit{FwPurp}). These partly capture the requirement of transferring data the third-party organisations in \textit{Article 46(1)} \cite{Gdpr46}, GDPR.

\item The data possession sub-policy determines who has the right to possess a piece of data of given type. 

\item The data connection sub-policy determines who has the right to link two types of data.   
\end{enumerate}

\begin{table}
\centering
\small
\begin{tabular}{ |c|l| } 
 \hline
 $\pi_{\theta}$ & A policy on a data type $\theta$, where $\pi_{\theta}$ $=$ ($\pi_{col}$, $\pi_{use}$, $\pi_{str}$, $\pi_{del}$, $\pi_{fw}$, $\pi_{has}$, $\pi_{link}$),\\  
 & and  $\pi_{\theta}$ $\in$ \textit{POL}$_{\textit{DataTypes}^{sp}_{pol}}$.\\ 
  \hline
 $\pi_{col}$ & A data collection sub-policy, $\pi_{col}$ $\in$ Pol$_{Col}$. \\ 
  \hline
 $\pi_{use}$ & A usage sub-policy, $\pi_{use}$ $\in$ Pol$_{Use}$. \\ 
  \hline
 $\pi_{str}$ & A storage sub-policy, $\pi_{str}$ $\in$ Pol$_{Str}$. \\ 
   \hline
 $\pi_{del}$ & A retention sub-policy, $\pi_{del}$ $\in$ Pol$_{Del}$. \\
   \hline
 $\pi_{fw}$ & A transfer sub-policy, $\pi_{fw}$ $\in$ Pol$_{Fw}$. \\
   \hline
 $\pi_{has}$ & A data possession sub-policy, $\pi_{has}$ $\in$ Pol$_{Has}$.\\
   \hline
 $\pi_{link}$ & A data connection sub-policy, $\pi_{link}$ $\in$ Pol$_{Link}$. \\
   \hline
 $\pi_{\theta}$.$\pi_{*}$ & A sub-policy $\pi_{*}$ of $\pi_{\theta}$, where $*$ $\in$ \{\textit{col}, \textit{use}, \textit{str}, \textit{del}, \textit{fw},\textit{has}, \textit{link}\}. \\
   \hline
 $\pi_{*}$.\textit{arg} & A reference to an argument \textit{arg} of a sub-policy $\pi_{*}$.\\ 
   \hline
 \textit{cons} & Specify if a consent is required ($Y$ for Yes, $N$ for No), where\\ & \textit{cons} $\in$ \textit{Cons}$_{col}$, or \textit{cons} $\in$ \textit{Cons}$_{use}$, or \textit{cons} $\in$ \textit{Cons}$_{str}$.\\
   \hline
 \textit{upurp, cpurp, fwpurp} & A set of usage, collection, and forward purposes, respectively,\\
 &  where each set is of the form \{\textit{act}$_1$:$\theta_1$, \dots, \textit{act}$_n$:$\theta_n$\}, and \\ 
   \hline
 & \textit{upurp} $\in$ \textit{UPurp}, \textit{cpurp} $\in$ \textit{CPurp}, \textit{fwpurp} $\in$ \textit{FwPurp}.\\
 \textit{act}$_i$:$\theta_i$ & A purpose defined for a policy  $\pi_{\theta}$. It specifies that a piece of data of type $\theta$ is\\ 
   \hline
 & used for an action \textit{act}$_i$, and as a result we get a piece of data of type $\theta_i$.\\
 \textit{where} & A set of places where a piece of data of type $\theta$ can be stored (\textit{where} $\in$ Where$_{str}$). \\ 
   \hline
 \textit{fromwhere} & A set of places from where a piece of data of type $\theta$ can be deleted \\ 
 & (\textit{fromwhere} $\in$ FromWhere$_{del}$). \\
   \hline
 \textit{deld} & A deletion delay value, that can be \textit{tt} or \textit{dd} (\textit{deld} $\in$ Del$_{delay}$). \\
   \hline
    \textit{tt}, \textit{dd}  & A non-specific time value, and a numerical time value, respectively. \\
   \hline
  \textit{fwto} & A set of entities to which a piece of data can be transferred (\textit{fwto} $\in$ FwTo). \\
    \hline
  \textit{whocanhave} & A set of entities who has the right to have a type of data\\ 
  & (\textit{whocanhave} $\in$ Who$_{canhave}$).\\  
    \hline
   \textit{whocanlink} & A set that contains which entity has the right to link which pairs of types of data\\ 
  & (\textit{whocanlink} $\in$ Who$_{canlink}$).\\ 
 \hline
\end{tabular}
\caption{\label{tab:notation1}The notations used in the policy syntax.}
\end{table}
\normalsize

A policy is defined on a data type ($\theta$), specifically, let $\pi_{\theta}$, $\pi_{\theta}$ $\in$ \textit{POL}$_{\textit{DataTypes}^{sp}_{pol}}$, be a policy defined on a data type $\theta$, and on the seven sub-policies $\pi_{col}$ $\in$ \textit{Pol}$_{Col}$, $\pi_{use}$ $\in$ \textit{Pol}$_{Use}$, $\pi_{str}$ $\in$ \textit{Pol}$_{Str}$, $\pi_{del}$ $\in$ \textit{Pol}$_{Del}$, $\pi_{fw}$ $\in$ \textit{Pol}$_{Fw}$,  $\pi_{has}$ $\in$ \textit{Pol}$_{Has}$, $\pi_{link}$ $\in$ \textit{Pol}$_{Link}$, where  

\begin{center}
$\pi_{\theta}$ $=$ ($\pi_{col}$, $\pi_{use}$, $\pi_{str}$, $\pi_{del}$, $\pi_{fw}$, $\pi_{has}$, $\pi_{link}$).   
\end{center} 
 
Each sub-policy of $\pi_\theta$ is defined as follows: 
 
\begin{enumerate}
\item $\pi_{col}$ = (\textit{cons}, \textit{cpurp}), where \textit{cons} $\in$ \{$Y$, $N$\} that specifies whether consent is required to be collected from the data subjects (Y) or not (N) for a data type $\theta$, and \textit{cpurp} is a set of collection purposes. A purpose has the form \textit{act}$_i$:$\theta_i$, which specifies that a piece of data of type $\theta$ is collected by the service provider to perform an action \textit{act}$_i$ in order to get some data of type $\theta_i$ (e.g. $\theta$ = \textit{name} is collected for \textit{creating} and \textit{account}, i.e. the purpose is \textit{creating}:\textit{account}).

\item $\pi_{use}$ = (\textit{cons}, \textit{upurp}), with a usage consent  requirement, \textit{cons} $\in$ \{$Y$, $N$\}, and \textit{upurp}, a set of usage purposes.  

\item $\pi_{str}$ = (\textit{cons}, \textit{where}), in which \textit{where} is a set of places where 
a piece of data of type $\theta$ can be stored, for instance, in a client's machine (\textit{where} = \{\textit{clientpc}\}), at a  third party cloud service,  or in the service provider's main or backup storage places (denoted by \textit{mainstorage}, \textit{backupstorage}).  

\item $\pi_{del}$ = (\textit{fromwhere}, \textit{deld}), where 
\begin{itemize}
\item \textit{fromwhere} contains the locations from where a piece of data can be deleted. This strongly depends on the storage locations, \textit{where}, defined in the storage policy (point 3).   

\item \textit{deld} is the delay value for deletion. This value can be either $tt$, which refers to a ``\textit{non specific} time'', or a specific ``numerical" time value (e.g. 1 day, 10 mins, 5 years, etc.).        

\end{itemize}

\item $\pi_{fw}$ = (\textit{cons}, \textit{fwto}, \textit{fwpurp}), where \textit{cons} specifies the requirements for the data transfer consent,  and \textit{fwto} specifies a set of entities to whom the data can be transferred. Finally, \textit{fwpurp} is a set of purposes for the data transfer.  

\item $\pi_{has}$ = \textit{whocanhave},  where \textit{whocanhave} = \{$E_1$,\dots, $E_k$\} is a set of entities in the service that have the right to have or possess a piece of data of type $\theta$. If we forbid for a given entity to be able to have a given data type, then the entity must not have it (by any means, e.g. by intercepting, eavesdropping, or calculating, etc.). 

\item $\pi_{link}$ = \textit{whocanlink},  where \textit{whocanlink} = \{($E_1$,$\theta_1$),\dots, ($E_k$,$\theta_k$)\}, is a set of pairs of entities and data types defined in the service. Each pair ($E_i$, $\theta_i$) specifies that $E_i$ has the right to link two pieces of data of types $\theta$ and $\theta_i$. For instance, whether a service provider has the right to link a piece of information about someone's disease with their work place.  
\end{enumerate}

Finally, let \{$\theta_1$, \dots, $\theta_m$\} be a set of all data types used by the service of a provider \textit{sp}, we have: 

\begin{figure}[htbp]
\centering
\fbox{\begin{minipage}{13 cm}
The data protection policy of a service provider \textit{sp} is defined by the set
\begin{center}
$\mathcal{P}\mathcal{L}$ $=$ $\{$$\pi_{\theta_1}$, \dots, $\pi_{\theta_m}$$\}$.
\end{center}
\end{minipage}
}
\label{fig:polsp}
\end{figure}

\subsection{Proposed Policy Semantics}
\label{sec:semantic0} 

\subsubsection{Events}
\label{sec:aevents} 

The semantics of the policy syntax can be defined using the events that capture the actions performed by different entities during an instance of a system operation. An event is defined by a tuple starting with an event name that denotes an action carried out by an entity, followed by the time of the event, and some further action-specific parameters. 

\begin{table}
\centering
\small
\begin{tabular}{ |c|l| } 
 \hline
 $\theta$ & A data type value, e.g. $\theta$ = \textit{name}. \\
   \hline
 $\theta'$ & A data type value that we get as a result of a \textit{service\_spec\_use\_event} \\ 
 & (e.g. createat or calculateat) on a piece of data of type $\theta$ and value $v$. \\
   \hline
 $v$ & The value of a piece of data of type $\theta$ (e.g. $v$ = \textit{Peter}, for $\theta$ = \textit{name}). \\
     \hline
 $t$ & The time value when an event takes place. \\
   \hline
 \textit{E}$_{\textit{to}}$ & An entity value to whom a piece of data is transferred/forwarded (e.g. \textit{E}$_{\textit{to}}$ = \textit{police}). \\
   \hline
 \textit{E}$_{\textit{from}}$ & An entity value from which a piece of data is originated (e.g. \textit{E}$_{\textit{from}}$ = \textit{clientpc}).\\
   \hline
 \textit{place} & A place where a piece of data of type $\theta$ and value $v$ is stored. It  can be \\
 & \textit{mainstorage}, \textit{backupstorage} of a serv. provider, or some other service spec. place.\\
 \hline
\end{tabular}
\caption{\label{tab:notation2}The notations used in the policy semantics.}
\normalsize
\end{table}

Our language includes the following ``built-in" events:  \textit{cconsentat}, \textit{collectat},   \textit{uconsentat},  \textit{sconsentat}, \textit{service\_spec\_use\_event}, \textit{storeat}, \textit{deleteat},  \textit{fwconsentat},  and \textit{forwardat},  defined as follows: 

\small
\begin{center}
\begin{list}
{\bfseries{}\textit{Ev}\textit{\arabic{qcounter}}:~}
{
\usecounter{qcounter}
}
\item (\textbf{\textit{cconsentat}}, $t$, \textit{E}$_{\textit{from}}$, $\theta$).  
This event specifies that a data collection consent is being collected at time $t$, by the service provider for a piece of data of type $\theta$ from an entity \textit{E}$_{\textit{from}}$.

\begin{center}
E.g. \textit{(\textbf{cconsentat}, 2020.01.21.11:18, \textit{client}, personalinfo)}    
\end{center}

\item (\textbf{\textit{collectat}}, $t$, \textit{E}$_{\textit{from}}$, $\theta$,  $v$). This event specifies when a piece of data of type $\theta$ and value $v$ is collected by the service provider from \textit{E}$_{\textit{from}}$ at time $t$.

\begin{center}
E.g. \textit{(\textbf{collectat}, 2020.01.21.11:20, client,  personalinfo, Peter)}
\end{center}

\item (\textbf{\textit{uconsentat}}, $t$,  \textit{E}$_{\textit{from}}$, $\theta$).  This event specifies that a data usage consent is collected by the service provider at time $t$ from \textit{E}$_{\textit{from}}$.

\begin{center}
 E.g. \textit{(\textbf{uconsentat}, 2020.01.21.11:18, client, energyconsumption)}    
\end{center}

\item (\textbf{\textit{service\_spec\_use\_event}}, $t$, \textit{E}$_{\textit{from}}$, $\theta'$, $\theta$, $v$).  This captures 
a service specific event,  specifically, a piece of data type $\theta$ from \textit{E}$_{\textit{from}}$ is used by the service provider to obtain  a piece of data type $\theta'$ after performing the action \textit{service\_spec\_use\_event}. 

\begin{center}
E.g. \textit{(\textbf{createat}, 2020.01.30.15:45, client, bill, energyconsumption, 20kWh)}   
\end{center}

\item (\textbf{\textit{sconsentat}}, $t$, \textit{E}$_{\textit{from}}$,  $\theta$). This event specifies that a data storage consent is being collected by the service provider for a piece of data of type $\theta$ from an entity \textit{E}$_{\textit{from}}$.

\begin{center}
E.g. \textit{(\textbf{sconsentat}, 2020.01.30.15:45, client, sickness)}   \end{center}

\item (\textbf{\textit{storeat}}, $t$,  \textit{E}$_{\textit{from}}$, $\theta$, $v$, \textit{place}). This event specifies that a piece of data of type $\theta$ and value $v$ is stored at a place \textit{place} at time $t$. We note that unlike the rest events, which all defines an action carried out by a service provider, this event can capture an action by  a different entity as well. For example, if \textit{place} = \textit{clientpc}, then event \textit{Ev6} can refer to a storage action done by a client PC. 

\begin{center} 
E.g. \textit{(\textbf{storeat}, 2020.01.30.15:45, client, sickness, leukemia, backupstorage)}
\end{center}

\item (\textbf{\textit{deleteat}}, $t$, \textit{E}$_{\textit{from}}$, $\theta$, $v$, \textit{place}). This event specifies that at some time $t$, a service provider  deletes a piece of data of type $\theta$ and value $v$ from a place \textit{place}. 

\begin{center} 
E.g. \textit{(\textbf{deleteat}, 2020.01.30.15:45, client, sickness, leukemia, mainstorage)}  
\end{center}

\item (\textbf{\textit{fwconsentat}}, $t$, \textit{E}$_{\textit{to}}$, \textit{E}$_{\textit{from}}$, $\theta$). This event specifies that a service provider  is collecting a data transfer consent on a piece of data of type $\theta$ originated from $E_{from}$.
 
\begin{center} 
E.g. \textit{(\textbf{fwconsentat}, 2020.01.21.11:18,  insurancecompany, client, personalinfo)}    
\end{center}

\item (\textbf{\textit{forwardat}}, $t$, \textit{E}$_{\textit{to}}$, \textit{E}$_{\textit{from}}$, $\theta$, $v$). This captures that at time $t$, \textit{E}$_{\textit{to}}$ receives a piece of data transferred by a service provider, which has a type $\theta$ and value $v$, and is originally from \textit{E}$_{\textit{from}}$.

\begin{center} 
E.g. \textit{(\textbf{forwardat}, 2020.01.21.11:18,  insurancecompany, client, personalinfo, Peter)}    
\end{center}
\end{list}
\end{center} 
\normalsize

\subsubsection{Policy-Compliant System Operation Trace}
\label{sec:compliance0}

\begin{figure}[htb!]
    \begin{center}
        \includegraphics[width=0.55\textwidth]{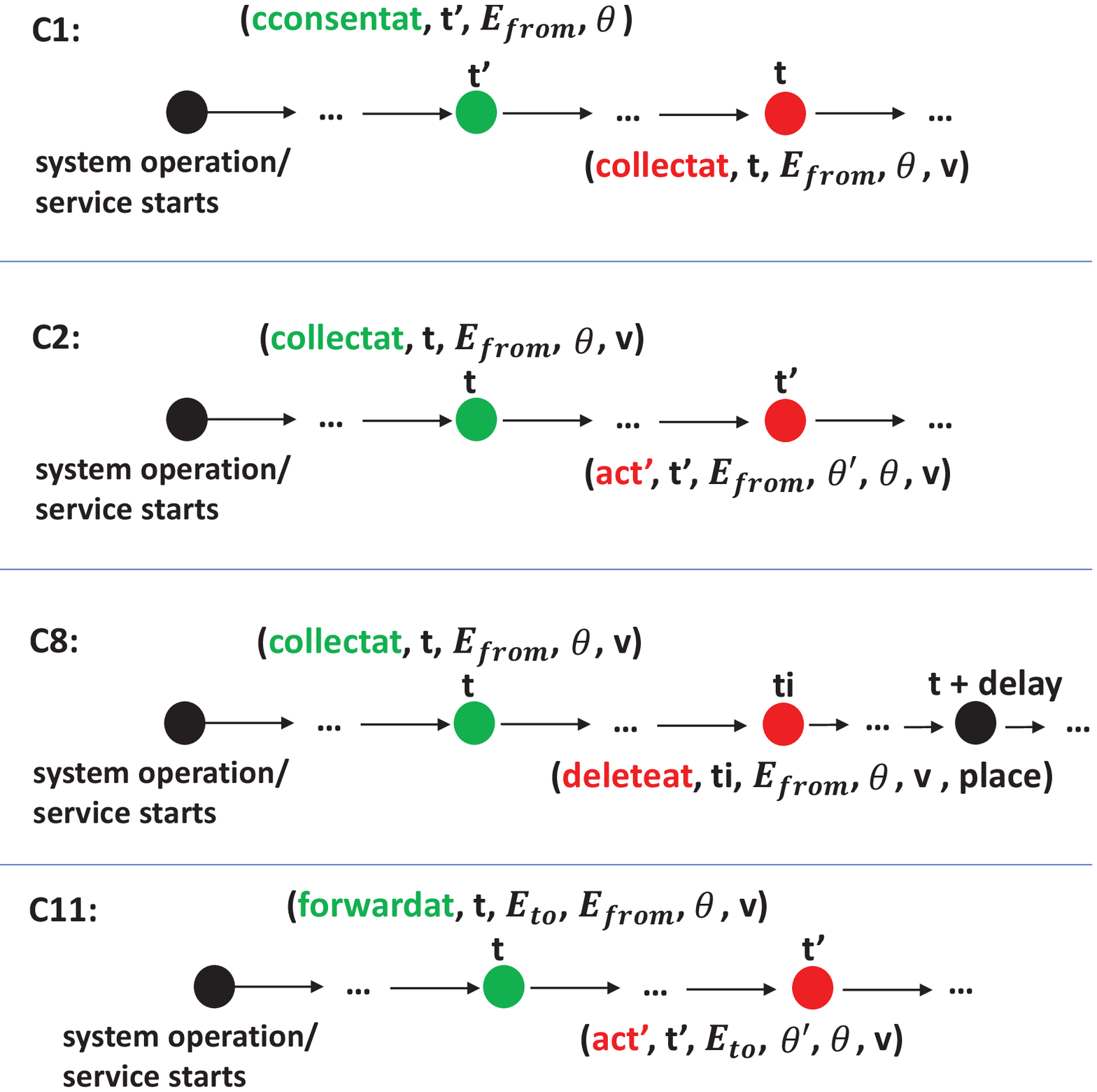}
    \end{center}
    \caption{The illustration of some policy compliance rules.}
    \label{fig:polsem}
\end{figure}

We discuss the policy compliant system operations based on the events defined in Section~\ref{sec:aevents}. Eleven rules ($C_1$-$C_{11}$) are defined, where each rule defines a system operation that respects a sub-policy in Definition~\ref{ttd:1} (see Figure~\ref{fig:polsem} for some illustration). In the sequel, we refer to each element $e$ of a tuple \textit{tup} as \textit{tup}.$e$, for example, we refer to $\pi_{str}$ in $\pi_{\theta}$ as $\pi_{\theta}$.$\pi_{str}$. Finally, in the following rules, we assume that a piece of data of type $\theta$ has not been deleted yet between any two actions. 

\begin{itemize}   
\item $C_1$ (collection consent): If in $\pi_{\theta}$.$\pi_{col}$, \textit{cons} = \textit{Y}, then a consent must be collected before the collection of the data itself. Formally:   

\begin{center}
\small
\begin{tabular}{ |c| } 
 \hline
 If during a system operation trace, $\exists$ \textit{Ev1} (\textbf{\textit{collectat}}, $t$, \textit{E}$_{\textit{from}}$, $\theta$,  $v$) for some time $t$,\\ then $\exists$ \textit{Ev2} (\textbf{\textit{cconsentat}}, $t'$, \textit{E}$_{\textit{from}}$, $\theta$) for some $t'$ in the trace, such that $t$ $\geq$ $t'$.\\    
 \hline
\end{tabular}
\end{center}
\normalsize
   
\item $C_2$ (collection purposes): If in $\pi_{\theta}$.$\pi_{col}$, \textit{cpurp} = \{\textit{act}$_1$:$\theta_1$, \dots, \textit{act}$_n$:$\theta_n$\}, then a piece of data of type $\theta$ must not be collected for any purpose that is not in \textit{cpurp}. Formally: 

\begin{center}
\small
\begin{tabular}{ |c| } 
 \hline
 If during a system operation trace, $\exists$ (\textbf{\textit{collectat}}, $t$,  \textit{E}$_{\textit{from}}$, $\theta$,  $v$) for some time $t$,\\ then for all instances of \textit{Ev4}, namely, each event (\textit{act}$'$, $t'$, \textit{E}$_{\textit{from}}$, $\theta'$, $\theta$, $v$)\\ in the trace,  where $t'$ $\geq$ $t$ we have \textit{act}$'$:$\theta'$ $\in$ \textit{cpurp}. \\(Note that \textit{act}$'$ is a service specific event, \textit{service\_spec\_use\_event} in \textit{Ev4}.)\\ 
 \hline
\end{tabular}
\end{center}
\normalsize

\item $C_3$ (usage consent): For $\pi_{\theta}$, if \textit{cons} = \textit{Y} in $\pi_{\theta}$.$\pi_{use}$, then consent must be collected before the usage of the data. Formally:   

\begin{center}
\small
\begin{tabular}{ |c| } 
 \hline
If during a system operation  trace, $\exists$ (\textbf{\textit{\textit{service\_spec\_use\_event}}}, $t$, \textit{E}$_{\textit{from}}$, $\theta'$, $\theta$, $v$) \\ for some time $t$,  then $\exists$  (\textbf{\textit{uconsentat}}, $t'$, \textit{E}$_{\textit{from}}$,  $\theta$) for some $t'$, such that $t$ $\geq$ $t'$.\\   
 \hline
\end{tabular}
\end{center}
\normalsize

\item $C_4$ (usage purposes): If in $\pi_{\theta}$.$\pi_{use}$, \textit{upurp} = \{\textit{act}$_1$:$\theta_1$, \dots, \textit{act}$_n$:$\theta_n$\}, then a piece of data of type $\theta$ must not be collected for any purpose not specified in \textit{upurp}. Formally: 

\begin{center}
\small
\begin{tabular}{ |c| } 
 \hline
If during a system operation trace, there is an instance of \textit{Ev4}, (\textit{act}$'$, $t$, \textit{E}$_{\textit{from}}$, $\theta'$, $\theta$, $v$)\\ for some time $t$, then \textit{act}$'$:$\theta'$ $\in$ \textit{upurp}.\\   
 \hline
\end{tabular}
\end{center}
\normalsize

\item $C_5$ (storage consent): If in $\pi_{\theta}$.$\pi_{str}$, \textit{cons} = \textit{Y}, then a consent must be collected before the storage of the data itself. Formally:   

\begin{center}
\small
\begin{tabular}{ |c| } 
 \hline
If during a system operation trace, $\exists$ (\textbf{\textit{storeat}}, $t$, \textit{E}$_{\textit{from}}$, $\theta$, $v$, \textit{places}) for some time $t$,\\ then $\exists$ (\textbf{\textit{sconsentat}}, $t'$, \textit{E}$_{\textit{from}}$, \textit{E},  $\theta$) for some $t'$ in the trace, such that $t$ $\geq$ $t'$.\\      
 \hline
\end{tabular}
\end{center}
\normalsize

\item $C_{6}$ (storage places): If in $\pi_{\theta}$.$\pi_{str}$, \textit{where} = \{\textit{place}$_1$, \dots, \textit{place}$_m$\}, then this data type must not be stored in any place that is not in \textit{where}. Formally:   

\begin{center}
\small
\begin{tabular}{ |c| } 
 \hline
If during a system operation trace, $\exists$ (\textbf{\textit{storeat}}, $t$, \textit{E}$_{\textit{from}}$, $\theta$, $v$, \textit{place}) for some time $t$,\\ then \textit{place} $\in$ \textit{where}.    
\\      
 \hline
\end{tabular}
\end{center}
\normalsize

\item $C_{7}$ (deletion places): If in $\pi_{\theta}$.$\pi_{del}$, \textit{fromwhere} = \{\textit{place}$_1$, \dots, \textit{place}$_m$\}, then this data type must be deleted from all the places defined in \textit{fromwhere}. Formally:   

\begin{center}
\small
\begin{tabular}{ |c| } 
 \hline
For all the events\\ (\textbf{\textit{deleteat}}, $t_1$, \textit{E}$_{\textit{from}}$, $\theta$, $v$, \textit{place}$_1$), \dots, (\textbf{\textit{deleteat}}, $t_n$, \textit{E}$_{\textit{from}}$, $\theta$, $v$, \textit{place}$_n$)\\ in a system operation trace,   \{\textit{place}$_1$, \dots, \textit{place}$_n$\}  $=$ \textit{fromwhere}.       
\\      
 \hline
\end{tabular}
\end{center}
\normalsize

\item $C_{8}$ (deletion delay): If in $\pi_{\theta}$.$\pi_{del}$, \textit{deld} = \textit{delay}, then this data type must be deleted up to the delay \textit{delay} from the time of its collection. Formally:   

\begin{center}
\small
\begin{tabular}{ |c| } 
 \hline
If during a system operation trace, $\exists$ (\textbf{\textit{collectat}}, $t$, \textit{E}$_{\textit{from}}$, $\theta$,  $v$) for some time $t$, and\\ $\exists$ events (\textbf{\textit{deleteat}}, $t_1$, \textit{E}$_{\textit{from}}$, $\theta$, $v$, \textit{places}$_1$), \dots, (\textbf{\textit{deleteat}}, $t_n$, \textit{E}$_{\textit{from}}$, $\theta$, $v$, \textit{places}$_n$),\\ for some $n$, then $t$ + \textit{delay} $\geq$ $t_1$  $\geq$ $t$, \dots, $t$ + \textit{delay} $\geq$ $t_n$  $\geq$ $t$.       
\\     
 \hline
\end{tabular}
\end{center}
\normalsize

\item $C_9$ (transfer consent): If in $\pi_{\theta}$.$\pi_{fw}$, \textit{cons} = \textit{Y}, then a consent must be collected before the transfer of a piece of data of type $\theta$. Formally:   

\begin{center}
\small
\begin{tabular}{ |c| } 
 \hline
If during a system operation trace, $\exists$ (\textbf{\textit{forwardat}}, $t$, \textit{E}$_{\textit{to}}$, \textit{E}$_{\textit{from}}$, $\theta$, $v$) for some time $t$,\\ then $\exists$  (\textbf{\textit{fwconsentat}}, $t'$,  \textit{E}$_{\textit{to}}$, \textit{E}$_{\textit{from}}$, $\theta$), such that $t$ $\geq$ $t'$.    
\\      
 \hline
\end{tabular}
\end{center}
\normalsize
   
\item $C_{10}$ (transfer to): If in $\pi_{\theta}$.$\pi_{fw}$, \textit{fwto} = \{\textit{E}$_1$, \dots, \textit{E}$_n$\}, then a piece of data of type $\theta$ must not be transferred to any entity not defined  in \textit{fwto}. Formally: 

\begin{center}
\small
\begin{tabular}{ |c| } 
 \hline
If during a system operation trace, $\exists$ (\textbf{\textit{forwardat}}, $t$, \textit{E}$_{\textit{to}}$, \textit{E}$_{\textit{from}}$, $\theta$, $v$) for some time $t$,\\ then \textit{E}$_{\textit{to}}$ $\in$ \textit{fwto}.
\\      
 \hline
\end{tabular}
\end{center}
\normalsize

\item $C_{11}$ (transfer purposes): If in $\pi_{\theta}$.$\pi_{fw}$, \textit{fwpurp} = \{\textit{act}$_1$:$\theta_1$, \dots, \textit{act}$_n$:$\theta_n$\}, then a piece of data of type $\theta$ must not be transferred for any purpose not defined in \textit{fwpurp}. Formally: 

\begin{center}
\small
\begin{tabular}{ |c| } 
 \hline
If during a system operation trace, $\exists$ (\textbf{\textit{forwardat}}, $t$,\textit{E}$_{\textit{to}}$, \textit{E}$_{\textit{from}}$, $\theta$, $v$) for some time $t$,\\ then for all instances of \textit{Ev4}, namely, each event (\textit{act}$'$, $t'$, \textit{E}$_{\textit{from}}$, $\theta'$, $\theta$, $v$)\\ in the trace,  where $t'$ $\geq$ $t$ we have \textit{act}$'$:$\theta'$ $\in$ \textit{fwpurp}.
\\      
 \hline
\end{tabular}
\end{center}   
\end{itemize}
\normalsize

\subsubsection{The semantics of the policy events}
\label{sec:semevent}

In this section, we discuss the semantics of the policy events defined in Section~\ref{sec:aevents}. For this purpose, we define the so-called \textit{data states}. We have the following assumptions before starting with the definitions:
\begin{itemize}
\item We assume a service provider \textit{sp} that provides a service \textit{serv}. 
\item We assume that the system that provides the service \textit{serv} involves the entities $E_1$, \dots, $E_m$ (namely, \textit{EntitySet}$^{sp}_{pol}$ = \{$E_1$, \dots, $E_m$\}.).   
\end{itemize}

The semantics of the policy events is defined based on the so-called \textit{data states of an entity} and the \textit{data states of a service}. 

\textbf{Data:} In this context, a piece of data is defined by a pair of  data type and the entity from which the data is originated, namely, \textit{data} = ($\theta$, $E_{\textit{from}}$), e.g. \textit{data} = (\textit{name}, \textit{clientpc}) or (\textit{disease}, \textit{healthapp}).

\textbf{Data value:} Each piece of data, \textit{data} = ($\theta$, $E_{\textit{from}}$), takes its value, $v$, during a system run. For example, during a system run the data (\textit{disease}, \textit{healthapp}) can take the value $v$ = \textit{coronavirus}. 

\textbf{Assigning a piece of data a value}: We denote that data ($\theta$, $E_{\textit{from}}$) takes the value $v$ by 
\begin{center}
($\theta$, $E_{\textit{from}}$) : $v$, e.g. (\textit{disease}, \textit{healthapp}) : \textit{coronavirus}.       
\end{center}


\textbf{The data state of an entity $E$}. Given a service provider \textit{sp} and the service \textit{serv}, the data state of an entity $E$, $E$ $\in$ \textit{EntitySet}$^{sp}_{pol}$,  captures the actual values of \textit{data},  \textit{data} = ($\theta$, $E_{\textit{from}}$), during a system run, for all $\theta$ $\in$ \textit{DataTypes}$^{sp}_{pol}$ from the perspective of $E$. 

Intuitively, the data state of an entity $E$ captures how the value of a piece of data, \textit{data} = ($\theta$, $E_{\textit{from}}$), changes from the perspective of $E$  during a system operation trace. 

Formally, the data state of $E$ is defined by the function $\textit{dstate}_E$ that assigns a value (including the undefined value $\bot$) to a piece of data \textit{data} = ($\theta$, $E_{\textit{from}}$).  

\begin{center}
\begin{tabular}{ |c| } 
 \hline
 \textbf{The data state of an entity $E$, $E$ $\in$ \textit{EntitySet}$^{sp}_{pol}$, is defined by a function:} \\\\
    \textit{dstate}$_{E}$ $:$ \textit{DataTuple} $\mapsto$ \textit{ValueTuple}$_{\bot}$, where\\\\  \textit{DataTuple} = (\textit{datadomain}$_1$, \dots, \textit{datadomain}$_n$) is the tuple of all defined \textit{data} in \textit{serv}, \\ where \textit{datadomain} is the set of all \textit{data} defined in \textit{serv}, besides \textit{data} = ($\theta$, $E_{\textit{from}}$).\\\\  \textit{ValueTuple}$_{\bot}$ = (\textit{valuedomain}$_1$, \dots, \textit{valuedomain}$_n$) is the tuple of values of the data in \textit{DataTuple},\\ where \textit{valuedomain} denotes the domain of the values, including the undefined value $\bot$.\\\\
    
    \textit{DataTuple} $\mapsto$ \textit{ValueTuple}$_{\bot}$ means that each \textit{datadomain}$_j$ is assigned a value \textit{valuedomain}$_j$.\\\\
    
    Finally, each $E$ $\in$ \textit{EntitySet}$^{sp}_{pol}$ has one data state, \textit{dstate}$_E$. \\
 \hline
\end{tabular}
\end{center}

For example, let $E$ = \textit{sp}, and \textit{DataTuple} = ((\textit{disease}, \textit{appOfTom}), (\textit{id}, \textit{appOfTom})), namely, there are two kinds of data defined in the service,  disease and an ID originated from \textit{appOfTom}. At the very start of the service, the value of both kinds of data is undefined. Specifically, the initial data state of \textit{sp} is \textit{dstate}$^{init}_{sp}$ = ($\bot$, $\bot$).  During the system run, when \textit{sp} receives the disease information from Tom's app, the data state can change to \textit{dstate}$_{sp}$ = (\textit{coronavirus}, $\bot$). Similarly,  when \textit{sp} receives the ID \textit{12345} from Tom's app, we can get the updated state as \textit{dstate}$_{sp}$ = (\textit{coronavirus}, \textit{12345}). Later, the value of (\textit{disease}, \textit{appOfTom}) can change to \textit{influenza}, and the state of \textit{sp} changes to 
\textit{dstate}$_{sp}$ = (\textit{influenza}, \textit{12345}).

\textbf{The data state of a service}. We assume a service \textit{serv} of a provider \textit{sp}, where  \textit{EntitySet}$^{sp}_{pol}$ = \{$E_{1}$, \dots, $E_{m}$\}. \textit{The data state of Service} is the tuple of the data states of entities $E_{1}$, \dots, $E_{m}$, and a time variable. 

\begin{center}
\begin{tabular}{ |c| } 
 \hline
 \textbf{The data state of a service \textit{serv}}\\\\
    $\delta_{serv}$ = (\textit{dstate}$_{E_1}$, \dots,  \textit{dstate}$_{E_m}$, \textit{TV}).
\\\\      
 \hline
\end{tabular}
\end{center}

The \textit{initial data state of a service} of a provider \textit{sp} is denoted by $\delta^{init}_{serv}$, which is the collection of the initial states of each defined entity in \textit{EntitySet}$^{sp}_{pol}$. Initially, at the start of a service, all the data have the undefined value, $\bot$. 

\begin{center}
\noindent\fbox{%
    \parbox{12cm}{%
    \textbf{The initial state of a service \textit{serv}:}
\begin{align*}
  \delta^{init}_{serv} & = ( \textit{dstate}^{init}_{E_1}, \ldots , \textit{dstate}^{init}_{E_m}, TV^{init})\ with\\ 
  \forall i \in [1,m],\ \textit{dstate}^{init}_{E_i} & = (\bot,\dots,\bot)  \\ 
  TV^{init} & = \bot  \textit{ (* denoting an undefined time value *)}.
\end{align*}
}%
}
\end{center}

\textbf{Event trace and state updates:}  An event trace of the operation of a service is denoted by $\tau$, and contains a finite sequence of events defined in Figure~\ref{sec:aevents}, happening during a corresponding system operation. Below, we define the semantics function, denoted by $S^{pol}_{Trace}$, which defines how a trace $\tau$ changes the state of a service  (Figure~\ref{fig:semeventpol}). 

$S^{pol}_{Trace}$ relies on the function $S^{pol}_{Ev}$ that defines how an event in $\tau$ changes the current global state of $\mathcal{P}\mathcal{L}$. 

\begin{center}
\small
\begin{tabular}{ |c| } 
 \hline
 \textbf{Semantics function (Policy)}\\\\
    $S^{pol}_{Trace}$ $:$ \textit{EventTrace} $\times$ \textit{ServDataStates}  $\mapsto$   \textit{ServDataStates}\\\\
    $S^{pol}_{Ev}$ $:$ \textit{Event} $\times$ \textit{ServDataStates} $\mapsto$  \textit{ServDataStates}\\\\

    where \textit{ServDataStates} denotes the ``domain" of the states of services,\\ 
    \textit{EventTrace} denotes the ``domain" of the event traces in a  system run.\\
    \textit{Event} denotes the ``domain" of the events. 
\\     
 \hline
\end{tabular}
\end{center}
\normalsize

\begin{figure}[htbp]
\centering
\fbox{\begin{minipage}{11.87 cm}
\small
\begin{tabbing}
\=1\=1\=1\=1\= \kill	
$S^{pol}_{Trace}$(\textit{emptytrace}, $\delta_{serv}$)  $=$ $\delta_{serv}$\\\\
\=1\=1\=1\=1\= \kill
$S^{pol}_{Trace}$(\textit{event}\textbf{.}$\tau$, $\delta_{serv}$) $=$  $S^{pol}_{Trace}$($\tau$, $S^{pol}_{Ev}$(\textit{event}, $\delta_{serv}$)) \\\\ 
\=1\=1\=1\=1\= \kill
$S^{pol}_{Ev}$((\textbf{\textit{cconsentat}}, $t$, \textit{E}$_{\textit{from}}$, $\theta$), $\delta_{serv}$) $=$ $\delta_{serv} [dstate_{sp} / dstate_{sp}[(\textit{cconsenttype}, \textit{E}_{\textit{from}}) : v_{cconsent}], TV/ t]$, \\ 
\ \ \ \ \ \ \ \ \ \ \ \ \ \ \ \ \ \ \ \ \ \ \ \ \ \ \ \ \ \ \ \ \ \ \ \ \ \ \ \ \ \ \ \ \ \ \ \ where $v_{cconsent}$ is the value of a collection consent.\\\\
\=1\=1\=1\=1\= \kill
$S^{pol}_{Ev}$((\textbf{\textit{collectat}}, $t$, \textit{E}$_{\textit{from}}$, $\theta$,  $v$), $\delta_{serv}$) $=$ $\delta_{serv}$ [$dstate_{sp}$/$dstate_{sp}$$[(\theta, \textit{E}_{\textit{from}}) : v]$, $TV$/$t$]\\\\ 

\=1\=1\=1\=1\= \kill
$S^{pol}_{Ev}$((\textbf{\textit{uconsentat}}, $t$,  \textit{E}$_{\textit{from}}$, $\theta$), $\delta_{serv}$) $=$ $\delta_{serv}$ [$dstate_{sp}$/$dstate_{sp}$$[(\textit{uconsenttype}, \textit{E}_{\textit{from}}) : v_{uconsent}]$, $TV$/$t$] \\ 
\ \ \ \ \ \ \ \ \ \ \ \ \ \ \ \ \ \ \ \ \ \ \ \ \ \ \ \ \ \ \ \ \ \ \ \ \ \ \ \ \ \ \ \ \ \ \ \ where $v_{uconsent}$ is the value of a usage consent.\\\\ 
\=1\=1\=1\=1\= \kill
$S^{pol}_{Ev}$((\textbf{\textit{createat}}, $t$, \textit{E}$_{\textit{from}}$, $\theta'$, $\theta$, $v$), $\delta_{serv}$) $=$ $\delta_{serv}$ [$dstate_{sp}$/$dstate_{sp}$$[(\theta', \textit{E}_{\textit{from}}) : v]$, $TV$/$t$]\\\\ 

\=1\=1\=1\=1\= \kill
$S^{pol}_{Ev}$((\textbf{\textit{calculateat}}, $t$, \textit{E}$_{\textit{from}}$, $\theta'$, $\theta$, $v$), $\delta_{serv}$) $=$ $\delta_{serv}$ [$dstate_{sp}$/$dstate_{sp}$$[(\theta', \textit{E}_{\textit{from}}) : v]$, $TV$/$t$]\\\\ 

\=1\=1\=1\=1\= \kill
$S^{pol}_{Ev}$((\textbf{\textit{sconsentat}}, $t$, \textit{E}$_{\textit{from}}$,  $\theta$), $\delta_{serv}$) $=$ $\delta_{serv} [dstate_{sp} / dstate_{sp}$$[(\textit{sconsenttype}, \textit{E}_{\textit{from}}) : v_{sconsent}]$, $TV$/$t$] \\ 
\ \ \ \ \ \ \ \ \ \ \ \ \ \ \ \ \ \ \ \ \ \ \ \ \ \ \ \ \ \ \ \ \ \ \ \ \ \ \ \ \ \ \ \ \ \ \ \ where $v_{sconsent}$ is the value of a storage consent.\\\\ 
\=1\=1\=1\=1\= \kill
$S^{pol}_{Ev}$((\textbf{\textit{storeat}}, $t$,  \textit{E}$_{\textit{from}}$, $\theta$, $v$, \textit{place}), $\delta_{serv}$) $=$ $\delta_{serv}$ [$dstate_{\textit{place}}$/$dstate_{\textit{place}}$$[(\theta, \textit{E}_{\textit{from}}) : v]$, $TV$/$t$]\\\\  
\=1\=1\=1\=1\= \kill
$S^{pol}_{Ev}$((\textbf{\textit{deleteat}}, $t$, \textit{E}$_{\textit{from}}$, $\theta$, $v$, \textit{place}), $\delta_{serv}$)\\
\=1\=1\=1\=1\= \kill
\>\> $=$  $\delta_{serv}$ [$dstate_{place}$/$dstate_{place}$[($\theta$, \textit{E}$_{\textit{from}}$) : $\bot$, 
(\textit{cconsenttype}, \textit{E}$_{\textit{from}}$) : $\bot$,
(\textit{uconsenttype}, \textit{E}$_{\textit{from}}$) : $\bot$,\\ 
\ \ \ \ \ \ \ \ \ \ \ \ \ \ \ \ \ \ \ \ \ \ \ \ \ \ \ \ \ \ \ \ \ \ \ \ \ \ \ \ \ \ \    (\textit{sconsenttype}, \textit{E}$_{\textit{from}}$) : $\bot$,  
(\textit{fwconsenttype}, \textit{E}$_{\textit{from}}$) : $\bot$], $TV$/$t$)].\\\\
\=1\=1\=1\=1\= \kill
$S^{pol}_{Ev}$((\textbf{\textit{fwconsentat}}, $t$, \textit{E}$_{\textit{to}}$, \textit{E}$_{\textit{from}}$, $\theta$), $\delta_{serv}$) $=$ $\delta_{serv}$ [$dstate_{sp}$/$dstate_{sp}$ $[(\textit{fwconsenttype}, \textit{E}_{\textit{from}}) : v_{fwconsent}]$, $TV$/$t$] \\ 
\ \ \ \ \ \ \ \ \ \ \ \ \ \ \ \ \ \ \ \ \ \ \ \ \ \ \ \ \ \ \ \ \ \ \ \ \ \ \ \ \ \ \ \ \ \ \ \ \ \ \ \ \ \ \ \ \ \ \ \ \ \ \ \ where $v_{fwconsent}$ is the value of a transfer consent.\\\\   
\=1\=1\=1\=1\= \kill
$S^{pol}_{Ev}$((\textbf{\textit{forwardat}}, $t$, \textit{E}$_{\textit{to}}$, \textit{E}$_{\textit{from}}$, $\theta$, $v$), $\delta_{serv}$) $=$ $\delta_{serv}$ [$dstate_{E_{\textit{to}}}$/$dstate_{E_{\textit{to}}}$$[(\theta, \textit{E}_{\textit{from}}) : v]$, $TV$/$t$] 
   \end{tabbing}
   \normalsize
\end{minipage}
}
\caption{The semantics of the policy events, where \textit{createat} and \textit{calculateat} are the two instances of \textit{service\_spec\_use\_event} (\textit{Ev4}).}\label{fig:semeventpol}
\end{figure}

\begin{table}
\centering
\begin{tabular}{ |c|l| } 
 \hline
  $\bot$ & An undefined data and time value.\\
 \hline 
 \textit{serv} & A service.\\ 
 \hline
 \textit{dstate}$_{E}$ & The data state of an entity $E$.\\
 \hline
 $\delta_{serv}$ & The data state of a service \textit{serv}.\\
 \hline
 \textit{dstate}$^{init}_{E}$ & The initial data state of an entity $E$.\\
 \hline
 $\delta^{init}_{serv}$ & The initial data state of a service \textit{serv}.\\
 \hline
 $\tau$ & An event trace.\\
 \hline 
 \textit{TV} & A time variable.\\
 \hline 
 $\delta_{serv} [dstate_E / dstate_E[(\theta, E_{\textit{from}}) : v], TV / t]$ & A change made on the state of a service \textit{serv} as\\ 
 & a result of an event. Here, inside $\delta_{serv}$, the data $(\theta, E_{\textit{from}})$\\ 
 & is assigned a value $v$ inside the 
 data state of $E$ ($dstate_E$). \\ 
 & Besides, \textit{TV} is assigned the time value $t$.\\
 \hline 
 $dstate_E / dstate_E[(\theta, E_{\textit{from}}) : v$] & The current data state of $E$ is updated with a new one\\ 
 & in which the data $(\theta, E_{\textit{from}})$ is assigned a value $v$.\\
 \hline
\end{tabular}
\caption{\label{tab:notationsemev}The notations used in the semantics of the policy events.}
\end{table}

Figure~\ref{fig:semeventpol} summarises the semantics of the events defined in Section~\ref{sec:aevents}. Each event can either change the global state or leave it unchanged. To capture the modification made by an event at time $t$ on the state of the variable ($\theta$, $E_{\textit{from}}$) from the perspective of an entity $E$ we write $\delta_{serv} [dstate_E / dstate_E[(\theta, E_{\textit{from}}) : v], TV / t]$ (or $\delta_{serv} [dstate_E / dstate_E[(\theta, E_{\textit{from}}) : \bot], TV / t]$ in case of the undefined value, e.g. when a piece of data has been deleted). Intuitively, this notation captures that the old state $dstate_E$ is replaced with the new state $dstate_E[(\theta, E_{\textit{from}}) : v]$ ($dstate_E[(\theta, E_{\textit{from}}) : \bot]$), in which the variable $(\theta, E_{\textit{from}})$ has been given the value $v$ (or the undefined value $\bot$) as a result of the event, the time variable $TV$ is given the time value $t$. 

\subsection{Well-formed Policies}
\label{wellformedpol} 

A policy $\mathcal{P}$$\mathcal{L}$, $\mathcal{P}\mathcal{L}$ $=$ $\{$$\pi_{\theta_1}$, \dots, $\pi_{\theta_m}$$\}$, is well-formed if for each data type $\theta_1$, \dots, $\theta_n$ (where $\pi_{\theta_j}$ $=$ ($\pi_{col}$, $\pi_{use}$, $\pi_{str}$, $\pi_{del}$, $\pi_{fw}$, $\pi_{has}$, $\pi_{link}$)), there is not any pair of sub-policies which are conflicting. For example, the pair  ($\pi_{col}$, $\pi_{has}$) or ($\pi_{str}$, $\pi_{has}$) is conflicting if the collection ($\pi_{col}$) or storage ($\pi_{str}$) sub-policy specifies that an entity $E$ can collect or store a type of data ($\theta_j$), but in $\pi_{has}$, $E$ does not have the right to have/posses this type of data.   

\section{The Corresponding Architecture Level}
\label{sec:arch0}
System architectures describe how a system is composed of components and how these components relate to each other (which is abstracted away from the policy), however, they abstract away from the implementation details, such as the cryptographic algorithms, the specific order and timing of the messages (e.g. we only define that \textit{sp} can receive a sickness record from a \textit{health app}, but we do not specify the  authentication, key exchange or communication protocol behind that). 

\subsection{Proposed Architecture Syntax}
\label{sec:syntax0}

In line with the policy specification, a system architecture is defined on a set of entities (components) and data types. For a service provider \textit{sp}, we define a finite set of entities, \textit{EntitySet}$^{sp}_{arch}$ = \{$E_{i_1}$, \dots, $E_{i_n}$\}.  Let \textit{DataTypes}$^{sp}_{arch}$ = \{$\theta_1$,\dots, $\theta_m$\}  be the set of all the data types defined in an architecture. We assume the finite sets of data variables \textit{Var}, ($X_{\theta}$ $\in$ \textit{Var}), time variables ($TT$ $\in$ \textit{TVar}), data values \textit{Val} ($V_\theta$ $\in$ \textit{Val}), respectively. Finally, we define the finite sets of the time and deletion delay values (\textit{t} $\in$ \textit{TVal}, and \textit{dd} $\in$ \textit{DVal}), respectively.   

\textbf{Terms:} As shown in Figure~\ref{fig:archterms},  a term, denoted by $T$, can be: 
\begin{itemize}
\item A variable ($X_\theta$) that represents some data of type $\theta$, and a data constant or value ($V_\theta$) of type $\theta$.

\item  A special term \textit{ds} that specifies the real identity value of a data subject (this will be used for modelling pseudonyms). 

\item A term can be an entity $E$ that specifies any software or hardware component, organisations, a data controller, or a data subject. 

\item A special function (\textit{SpecFunc}) that specifies the time, pseudonyms and four types of consents. 

\item Finally, a term can be a time value ($Ti$). 
\end{itemize}

A variable $X_\theta$ $\in$ \textit{Var} represents a piece of data of type $\theta$ supported by \textit{sp}, such as the users' personal information, photos, videos,  energy data, insurance number, etc. $X_\theta$ can be a  non-function/simple data $D_\theta$ of type $\theta$, a cryptographic or meta function (\textit{CryptoFunc}), and finally, any other service specific function.   

\begin{figure}[htbp]
\centering
\fbox{\begin{minipage}{11.87 cm}
\small
\begin{tabbing}    
    \=123456\=1\=1\=1\= \kill
		\>\> \underline{Terms}: \\\\    
    \=123456\=1\=1\=1\= \kill
    \>\> $T$ ::= $X_{\theta}$ $|$ $V_{\theta}$ $|$ \textit{ds} $|$ $E$ $|$ \textit{SpecFunc} $|$ $Ti$.\\\\
        \=123456\=1\=1\=1\= \kill
		\>\> $X_{\theta}$ ::=   $D_{\theta}$  $|$ \textit{CryptoFunc} $|$ \textit{Service\_spec\_fun}($X_{\theta_1}$, \dots, $X_{\theta_n}$). \\ 
		 \=123456\=1\=1\=1\= \kill
		\>\> (where TYPE(\textit{CryptoFunc}) = $\theta$, TYPE(\textit{Service\_spec\_fun}($X_{\theta_1}$, \dots, $X_{\theta_n}$)) = $\theta$).\\\\ 
		\=123456\=1\=1\=1\= \kill
		\>\> $Ti$ ::=   \textit{dd}  $|$ $TT$.\\\\ 
		\=123456\=1\=1\=1\= \kill
		\>\> \textit{SpecFunc} ::= \textbf{\textit{Time}}(\textit{Ti}) $|$ \textbf{\textit{P}}(\textit{ds}) $|$   \textbf{\textit{Cconsent}}(\textit{Data}) $|$ \textbf{\textit{Uconsent}}(\textit{Data}). \\
		\=123456\=1\=1\=1\= \kill
		\>\> \ \ \ \ \ \ \ \ \ \ \ \ \  \ \ \ \ \ \ $|$   \textbf{\textit{Sconsent}}(\textit{Data}) $|$   \textbf{\textit{Fwconsent}}(\textit{Data}, $E_{to}$)  \\\\
		\>\> \textit{Data} ::= ($X_{\theta}$, $E_{from}$)  where $E_{from}$ is an entity who originally sent the data $X_{\theta}$.\\\\
		\>\> \textit{CryptoFunc} ::= \textbf{\textit{Sk}}($X_{pkeytype}$) $|$ \textbf{\textit{Senc}}($X_{\theta}$, $X_{keytype}$) $|$ \textbf{\textit{Aenc}}($X_{\theta}$, $X_{pkeytype}$).  \\ 
		\=123456\=1\=1\=1\= \kill
		\>\>\ \ \ \ \ \ \ \ \ \ \ \ \ \ \ \ \ \ \ \ \ $|$ \textbf{\textit{Hash}}($X_{\theta}$) $|$ \textbf{\textit{Mac}}($X_{\theta}$, $X_{keytype}$) $|$   \textbf{\textit{Meta}}($X_{\theta}$).\\\\  
		\=123456\=1\=1\=1\= \kill
		\>\> \underline{Destructor application on terms}: \\\\
		\=123456\=1\=1\=1\= \kill
		\>\>  $G$($T_1$, \dots, $T_n$) $\rightarrow$ $T$\\\\ 
		\=123456\=1\=1\=1\= \kill
		\>\> \underline{Function that returns a type of a term $T$}: \\\\
\=123456\=1\=1\=1\= \kill
		\>\> TYPE($T$) = $\theta$,  where $\theta$ $\in$ \textit{DataTypes}$^{sp}_{arch}$.\\\\
		 \=123456\=1\=1\=1\= \kill
		\>\> \underline{Function \textit{HasAccessTo}}: \\\\
		\>\> \textit{HasAccessTo}: $E_i$ $\in$ \textit{EntitySet}$^{sp}_{arch}$ $\rightarrow \{$$E_j$ $\in$ \textit{EntitySet}$^{sp}_{arch}$\}.
\end{tabbing}
\end{minipage}
}
\caption{Terms, Destructors and Types.}\label{fig:archterms}
\end{figure}
\normalsize

\textbf{Functions:} The two groups of functions \textit{SpecFunc} and \textit{CryptoFunc} are defined as follows:  
\begin{itemize}

\item Function  \textbf{\textit{Time}}(\textit{Ti}) specifies the time with either a non-specific time value \textit{TT} or a numerical delay value, \textit{dd}. While \textit{dd} captures a numerical time value such as 3 years, 2 months, etc., the value \textit{TT} is not numerical, and is used to express the informal term ``at some point/time". Function \textbf{P}(\textit{ds}) specifies a pseudonym of a real identity \textit{ds}.

\item \textbf{\textit{Cconsent}}(\textit{Data}),  \textbf{\textit{Uconsent}}(\textit{Data}) and  \textbf{\textit{Sconsent}}(\textit{Data}), besides \textit{Data} = ($X_\theta$, $E_{\textit{from}}$), specify a piece of data of type collection, usage, and storage consent, respectively, on a piece of data $X_\theta$ that is originally sent by $E_{\textit{from}}$. Finally, \textbf{\textit{Fwconsent}}(\textit{Data}, $E_{to}$) specifies a transfer consent on \textit{Data}, alongside an entity to whom the data can be transferred ($E_{to}$). 

\item \textbf{Meta}($X_\theta$) defines the metadata (information about other data), or information located in the header of the packets (e.g. IP address). For simplicity, they are both modelled by \textbf{Meta}. 


\item The basic cryptographic functions:    
\begin{itemize}
\item \textbf{Sk}($X_{pkeytype}$): This function defines a type of private key used in asymmetric key encryption algorithms. Its argument has a type of public key (pkeytype).    

\item \textbf{Senc}($X_\theta$, $X_{keytype}$): This defines a type of symmetric key  encryption, and has two arguments, a piece of data (of type $\theta$) and a symmetric key (of type keytype). 

\item \textbf{Aenc}($X_\theta$, $X_{pkeytype}$): This defines a type of the cipher text resulted from an asymmetric key encryption, and has two arguments, a piece of data and a public key (pkeytype).

\item \textbf{Mac}($X_\theta$, $X_{keytype}$): This defines a type of the message authentication code that has two arguments, a piece of data  and a symmetric key. 

\item \textbf{Hash}($X_\theta$): This defines a type of the cryptographic hash that has one argument, a piece of data of type $\theta$. 

\end{itemize}
\end{itemize} 
 
\textbf{Values}: A variable $X_{\theta}$  will be given a specific data value $V_{\theta}$ during an instance of a system run (see Section~\ref{sec:archsem}). $V_{\theta}$ can be the value of both a simple (non-function) data or a function, and it can also be $\bot$, which denotes an undefined value (every data variable $X_\theta$ has the value $\bot$ at the start of a service).  

\textbf{Destructor:} This represents an evaluation of a function, used to model a verification procedure. For instance, if $X_{enc}$ $=$ \textit{Senc}($X_{name}$, $X_{Skey}$) that represents the encryption of data $X$ with the server key $X_{Skey}$, and $X_{Skey}$ represents a symmetric key, then $G(X_{enc}, X_{Skey})$ $\rightarrow$ $X$ is \textit{Dec}(\textit{Senc}($X_{name}$, $X_{Skey}$), $X_{Skey}$) $\rightarrow$ $X_{name}$.  Note that not all functions have a corresponding destructor, e.g., in case $X_{hash}$ is a one-way cryptographic hash function, $X_{hash}$ $=$ \textit{Hash}($X_{password}$), then due to the one-way property there is no destructor (reverse procedure) that returns 
$X_{password}$ from the hash $X_{hash}$.

\textbf{HasAccessTo}: This is a function that expects an entity as input and returns a set of other entities defined in the same architecture. It specifies which entity can have access to the data handled/stores/collected by other entities. For example, if $E_m$ and $E_{p}$ represent a smart meter, and a digital panel, respectively, and we want to specify that the service provider, \textit{sp}, can have access to the panel and the meter, then, we define the relation \textit{HasAccessTo}(\textit{sp}) = \{$E_{m}$, $E_{p}$\}. It is used for verifying the data possesion and link policies.

\subsubsection{System Architecture}

\textbf{The definition of a system architecture:} An architecture $\mathcal{P}\mathcal{A}$ is defined as a set of \textit{actions} 
(denoted by $\{\mathcal{F}\}$). The formal definition is given as follows: 

\begin{figure}[htbp]
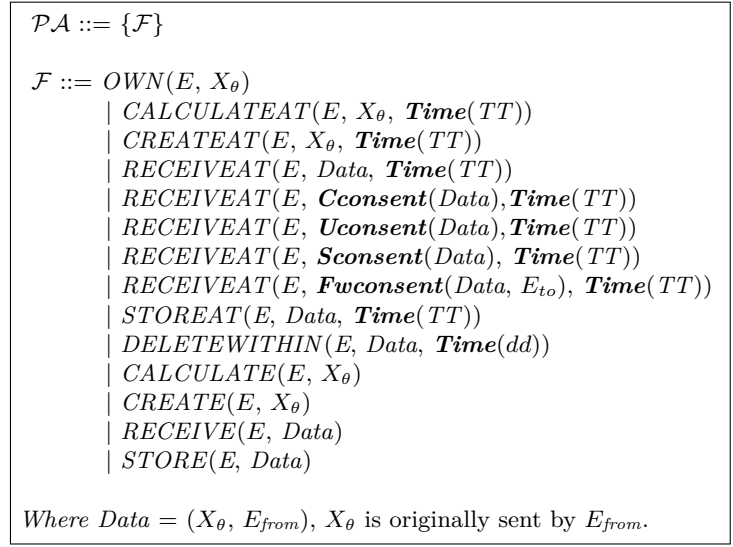

\centering
\fbox{\begin{minipage}{11.87 cm}
\small
\begin{tabbing}  
    \=1\=1\=1\= \kill
    \>\> $\mathcal{P}\mathcal{A}$ ::= $\{\mathcal{F}\}$\\\\
    \=1\=1\=1\=1\= \kill
    \>\> $\mathcal{F}$ ::= \textit{OWN}($E$, $X_{\theta}$)\\
    \=1\=1\=1\=1\= \kill
		\>\>\ \ \ \ \ \ \ \ \ $|$ 
    \textit{CALCULATEAT}($E$, $X_{\theta}$, \textbf{\textit{Time}}(\textit{TT}))\\
    \=1\=1\=1\=1\= \kill
		\>\>\ \ \ \ \ \ \ \ \ $|$ 
    \textit{CREATEAT}($E$, $X_{\theta}$, \textbf{\textit{Time}}(\textit{TT}))\\
    \=1\=1\=1\=1\= \kill
		\>\>\ \ \ \ \ \ \ \ \ $|$ 
    \textit{RECEIVEAT}($E$, \textit{Data}, \textbf{\textit{Time}}(\textit{TT}))\\
    \=1\=1\=1\=1\= \kill
		\>\>\ \ \ \ \ \ \ \ \ $|$ 
    \textit{RECEIVEAT}($E$, \textbf{\textit{Cconsent}}(\textit{Data}),\textbf{\textit{Time}}(\textit{TT}))\\
    \=1\=1\=1\=1\= \kill
		\>\>\ \ \ \ \ \ \ \ \ $|$ 
    \textit{RECEIVEAT}($E$, \textbf{\textit{Uconsent}}(\textit{Data}),\textbf{\textit{Time}}(\textit{TT}))\\
    \=1\=1\=1\=1\= \kill
		\>\>\ \ \ \ \ \ \ \ \ $|$ 
    \textit{RECEIVEAT}($E$, \textbf{\textit{Sconsent}}(\textit{Data}), \textbf{\textit{Time}}(\textit{TT}))\\
    \=1\=1\=1\=1\= \kill
		\>\>\ \ \ \ \ \ \ \ \ $|$ 
    \textit{RECEIVEAT}($E$, \textbf{\textit{Fwconsent}}(\textit{Data}, $E_{to}$),  \textbf{\textit{Time}}(\textit{TT}))\\
	\=1\=1\=1\=1\= \kill
		\>\>\ \ \ \ \ \ \ \ \ $|$ 
    \textit{STOREAT}(\textit{E}, \textit{Data}, \textbf{\textit{Time}}(\textit{TT}))\\
		\=1\=1\=1\=1\= \kill
		\>\>\ \ \ \ \ \ \ \ \ $|$ \textit{DELETEWITHIN}(\textit{E}, \textit{Data}, \textbf{\textit{Time}}($dd$))\\
		\=1\=1\=1\=1\= \kill
		\>\>\ \ \ \ \ \ \ \ \ $|$ 
    \textit{CALCULATE}($E$, $X_{\theta}$)\\
    \=1\=1\=1\=1\= \kill
		\>\>\ \ \ \ \ \ \ \ \ $|$ 
    \textit{CREATE}($E$, $X_{\theta}$)\\
\=1\=1\=1\=1\= \kill
		\>\>\ \ \ \ \ \ \ \ \ $|$ 
    \textit{RECEIVE}($E$, \textit{Data})\\
    \>\>\ \ \ \ \ \ \ \ \ $|$ 
    \textit{STORE}(\textit{E}, \textit{Data})\\
		\\
		\=1\=1\=1\=1\= \kill
		\> \textit{Where} \textit{Data} = ($X_{\theta}$, $E_{\textit{from}}$),  $X_{\theta}$ is originally sent by $E_{\textit{from}}$. 
\end{tabbing}
\end{minipage}
}
\caption{The table shows the syntax of a system architecture with the defined actions between components/entities.}\label{fig:sysarch}
\end{figure}
\normalsize

\begin{itemize}
\item Action \textit{OWN}($E$, $X_{\theta}$) captures that $E$ can own the data variable $X$ of type $\theta$ (during a service regardless of time). Note that  $X_{\theta}$ is  the originally owned data (not the data obtained/received by $E$).

\item \textit{CALCULATEAT}($E$, $X_{\theta}$, \textbf{\textit{Time}}(\textit{TT})) specifies that an entity $E$ can calculate the variable $X_{\theta}$ based on an equation $X_{\theta}$ $=$ $T$, for some term $T$ at non-specific time $TT$ (e.g. $\theta$ $=$ bill, and $X_\theta$ = \textit{Bill$($energyconsumption, tariff$)$}). 

\item \textit{CREATEAT}($E$, $X_{\theta}$, \textbf{\textit{Time}}(\textit{TT})) specifies that $E$ can create a piece of data of type $\theta$, based on an equation $X_{\theta}$ $=$ $T$ (e.g. $\theta$ $=$ account, and $X_{\theta}$ $=$ \textit{Account(name, address)}). The actions \textit{create} and \textit{calculate} merely differ in the nature of $T$, for example, we calculate a bill, while create an account. 

\item \textit{RECEIVEAT}($E$, \textit{Data}, \textbf{\textit{Time}}(\textit{TT})) means that $E$ can receive \textit{Data} (i.e. ($X_{\theta}$, $E_{\textit{from}}$)) at some  non-specific time \textit{TT}.

\item \textit{RECEIVEAT}($E$, \textbf{\textit{Cconsent}}(\textit{Data}), \textbf{\textit{Time}}(\textit{TT})), \textit{RECEIVEAT}($E$, \textbf{\textit{Uconsent}}(\textit{Data}), \textbf{\textit{Time}}(\textit{TT})), and \textit{RECEIVEAT}($E$, \textbf{\textit{Sconsent}}(\textit{Data}),  \textbf{\textit{Time}}(\textit{TT})) specify that a collection, usage and  storage consent on \textit{Data}, \textit{Data}=($X_{\theta}$, $E_{\textit{from}}$), respectively, can be received by $E$ at time \textit{TT}. 

\item \textit{RECEIVEAT}($E$, \textbf{\textit{Fwconsent}}(\textit{Data}, $E_{to}$), \textit{Time}(\textit{TT})) specifies that a transfer consent on \textit{Data} and $E_{to}$ can be received by $E$ at time \textit{TT}. 

\item \textit{STOREAT}(\textit{E}, \textit{Data}, \textbf{\textit{Time}}(\textit{TT}))  specifies that \textit{Data} can be stored at some non-specific time \textit{TT} in a place \textit{E}.  A place can be  \textbf{\textit{mainstorage}} and \textbf{\textit{backupstorage}}, which represent a collection of main storage places such as main servers, and a collection of backup storage places (e.g. backup servers) of a service provider, respectively, or any service specific place (e.g., \textit{clientPC}).

\item \textit{DELETEWITHIN}(\textit{E}, \textit{Data}, \textbf{\textit{Time}}(\textit{dd})) specifies  that \textit{Data} must be deleted from a place $E$ within a certain time delay \textit{dd} (where \textit{dd} is a numerical time value, e.g. 10 years).   

\item The last four CALCULATE, CREATE, RECEIVE and STORE actions at the end are the corresponding versions of the previous four but without the \textbf{Time}() construct.  They capture the correspinding actions regardless of time, and are  defined for convenient purposes, offering the user an option to specify the simpler actions if they only want to reason about privacy properties. The actions with the \textbf{Time}() construct are mainly used for reasoning about data protection properties and requirements (e.g. whether a consent has been collected before collection, usage, or transfer).  The semantics of these four actions are the same as the previous four.
\end{itemize}

\begin{figure}[htb!]
    \begin{center}
        \includegraphics[width=0.7\textwidth]{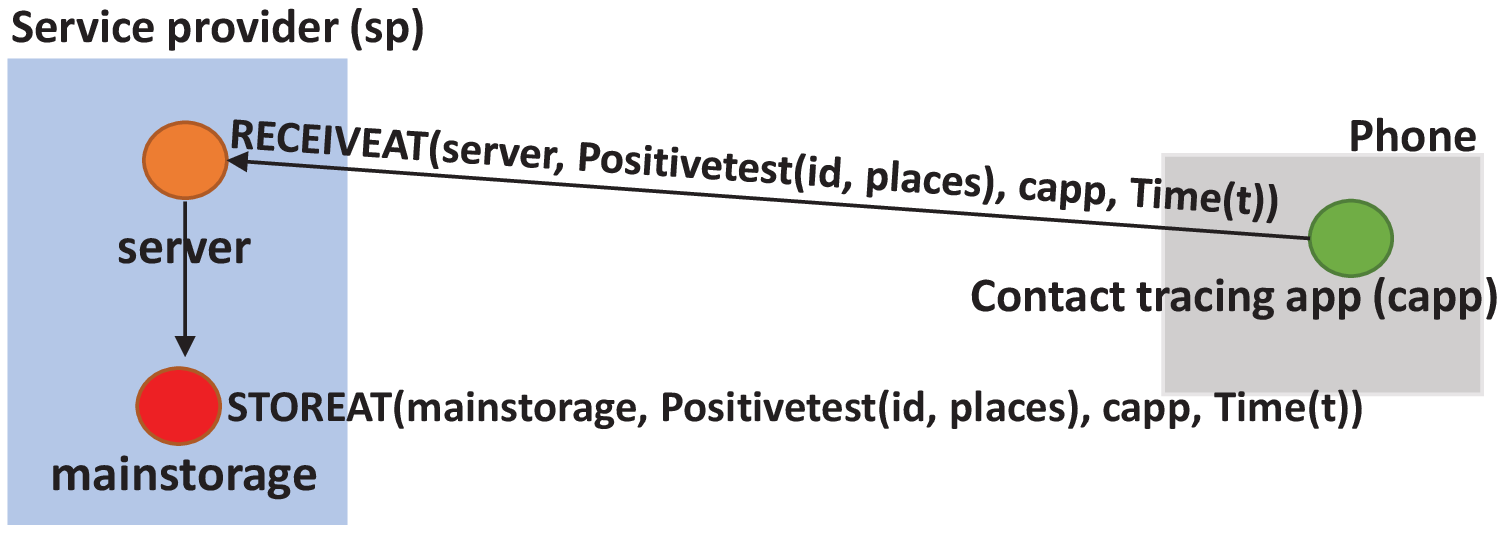}
    \end{center}
    \caption{A simple example architecture, where \textit{Data} = ($X_{\theta}$, $E_{\textit{from}}$) = (\textit{Positivetest(id, places)}, \textit{capp}).}
    \label{fig:archexample1}
\end{figure}

An example architecture is shown in Figure~\ref{fig:archexample1}, where a service provider collects positive (virus) test records sent by contact tracing apps. A record contains an unique ID and a set of places where the phone has been brought to, and the record is stored in the main storage place(s) of \textit{sp}. Here, we also define HasAccessTo(\textit{sp}) = \{\textit{server}, \textit{mainstorage}\} so that \textit{sp} can have access to \textit{server} and \textit{mainstorage}.

\subsection{Proposed Architecture Semantics}
\label{sec:archsem}
Like the policy case, the semantics of an architecture is based on events and system run traces. A trace $\Gamma$ is a sequence of high-level events 
\textit{Seq}($\epsilon$) taking place in during a service, as presented in Figure~\ref{fig:eventarch}.  

\begin{figure}[htbp]
\centering
\small
\fbox{\begin{minipage}{11.87 cm}
\begin{tabbing}
    \=1\=1\=1\=1\= \kill
		\> $\Gamma$ ::= \textit{Seq}($\epsilon$)\\
    \>$\epsilon$ ::=
    \textit{own}($E$, $X_{\theta}$$:$$V_{\theta}$, $t$), for all $t$ in any traces during a service\\ 
    \=1\=1\=1\=1\= \kill
     \=1\=1\=1\=1\= \kill
		\>\ \ \ \ \ \ \ 
	 $|$ \textit{calculateat}($E$, $X_{\theta}$$:$$T$, $t$)\\
	 \=1\=1\=1\=1\= \kill
		\>\ \ \ \ \ \ \ 
	 $|$ \textit{createat}($E$, $X_{\theta}$$:$$T$, $t$)\\
	 \=1\=1\=1\=1\= \kill
		\>\ \ \ \ \ \ \  $|$  \textit{receiveat}($E$, \textit{Data}$:$$V_{\textit{TYPE(Data)}}$, $t$)\\ 
		\=1\=1\=1\=1\= \kill
		\>\ \ \ \ \ \ \  $|$  \textit{receiveat}($E$, \textbf{\textit{Cconsent}}(\textit{Data}):$V_{cconsent}$, $t$)\\ 
		\=1\=1\=1\=1\= \kill
		\>\ \ \ \ \ \ \  $|$  \textit{receiveat}($E$, \textbf{\textit{Uconsent}}(\textit{Data}):$V_{uconsent}$, $t$)\\ 
		\=1\=1\=1\=1\= \kill
		\>\ \ \ \ \ \ \  $|$  \textit{receiveat}($E$, \textbf{\textit{Sconsent}}(\textit{Data}):$V_{sconsent}$, $t$)\\ 
		\=1\=1\=1\=1\= \kill
		\>\ \ \ \ \ \ \  $|$  \textit{receiveat}($E$, \textbf{\textit{Fwconsent}}(\textit{Data}):$V_{fwconsent}$, $t$)\\ 
    \=1\=1\=1\=1\= \kill
		\>\ \ \ \ \ \ \  $|$  \textit{storeat}(\textit{E}, \textit{Data}$:$$V_{\textit{TYPE(Data)}}$, $t$)\\ 
		\=1\=1\=1\=1\=\kill
		\>\ \ \ \ \ \ \
		$|$ \textit{deletewithin}(\textit{E}, \textit{Data}$:$$V_{\textit{TYPE(Data)}}$, $dd$, $t$).\\\\
		\=1\=1\=1\=1\= \kill
		\> \textit{Where} \textit{Data} = ($X_{\theta}$, $E_{\textit{from}}$),  $X_{\theta}$ is originally sent by $E_{\textit{from}}$.  
		\end{tabbing}
\end{minipage}
}
\caption{Events defined for architectures.}\label{fig:eventarch}
\end{figure}
\normalsize

 An event can be seen as an instance of an action defined in Figure~\ref{fig:sysarch} that happens at some specific time  $t$ (e.g. \textit{2020.01.30.15:45}) during a system run trace. Events are given the same names as the corresponding actions but in lower-case letters in order to avoid confusion.  

\begin{itemize} 
\item Event \textit{own}($E$, $X_{\theta}$$:$$V_{\theta}$, $t_{\textit{all}}$) captures that $E$ owns $X_{\theta}$ with a value $V_{\theta}$ at time $t_{\textit{all}}$ (where $t_{\textit{all}}$ denotes ``all the time" during a service). $X_{\theta}$$:$$V_{\theta}$ means that the variable $X_{\theta}$ is assigned a value $V_{\theta}$\footnote{$V_{\theta}$ can be a name, e.g. Peter, that is assigned to the $X_{\theta}$ during a service/system operation.}. 

\item \textit{calculateat}($E$, $X_{\theta}$$:$$T$, $t$) captures that at some time $t$, $E$ calculates a piece of data of type $\theta$ that is equal to a term $T$ (based on the equation $X_{\theta}$$=$$T$, e.g. $X_{hash}$ $=$  \textit{Hash}$(X_{password})$.).  

\item  \textit{createat}($E$, $X_{\theta}$$:$$T$, $t$) captures that at some time $t$, $E$ creates a piece of data of type $\theta$ that is equal to a term $T$ (e.g. $X_{\theta}$ $=$ \textit{Account}($X_{name}$, $X_{address}$)).   

\item  \textit{receiveat}($E$, \textit{Data}$:$$V_{\textit{TYPE(Data)}}$, $t$) specifies that $E$ receives a piece of data of type \textit{TYPE(Data)} and value $V_{\textit{TYPE(Data)}}$ at some specific time $t$.

\item Events \textit{receiveat}($E$, \textbf{\textit{Cconsent}}(\textit{Data}):$V_{cconsent}$, $t$), \textit{receiveat}($E$, \textbf{\textit{Uconsent}}(\textit{Data}):$V_{uconsent}$, $t$), \textit{receiveat}($E$, \textbf{\textit{Sconsent}}(\textit{Data}):$V_{sconsent}$, $t$), and    \textit{receiveat}($E$, \textbf{\textit{Fwconsent}}(\textit{Data}):$V_{fwconsent}$, $t$) specify that $E$ receives a (collection, usage, storage, or transfer) consent on \textit{Data} with a value $V_{\theta}$, where $\theta$ is a corresponding type of consent ($\theta$ $\in$ \{\textit{cconsent}, \textit{uconsent}, \textit{sconsent}, \textit{fwconsent}\}). 

\item  \textit{storeat}(\textit{E}, \textit{Data}$:$$V_{\textit{TYPE(Data)}}$, $t$) says that a piece of data of type \textit{TYPE(Data)} is stored in a place \textit{E}. 

\item   \textit{deletewithin}(\textit{E}, \textit{Data}$:$$V_{\textit{TYPE(Data)}}$, $dd$, $t$) specifies that at time $t$, a piece of data of type \textit{TYPE(Data)} is deleted from a place \textit{E}, where $t$ $\leq$ $t_{\textit{collect}}$ + $dd$, where the data was collected at $t_{\textit{collect}}$\footnote{This can be extended to the time of any other action (e.g. the time when the data is stored).}.

\end{itemize}

The semantics of each architecture event above is defined by the semantics function, $S_T$, which specifies the impact made by each event on the states of the data variables (i.e. how the values of $X_{\theta}$, for all $\theta$ $\in$ \textit{DataTypes}$^{sp}_{arch}$, changes after an event takes place). For example, let  \textit{DataTypes}$^{sp}_{arch}$ =\{\textit{name}, \textit{bill}\}, the two types supported by \textit{sp}, and \textit{Entity}$^{sp}_{arch}$ =\{\textit{sp}, \textit{client}\}. At the start of the service, the variable states of both \textit{sp} and \textit{client} are ($X_{name}$ = $\bot$, $X_{bill}$ = $\bot$), where $\bot$ is an undefined (initial) value. As a result of an event \textit{own}(\textit{client}, $X_{name}$$:$\textit{Peter}, $t_{\textit{all}}$), the variable state of \textit{sp} remains unchanged, while the state of \textit{client} has changed to ($X_{name}$ = \textit{Peter}, $X_{bill}$ = $\bot$).

\subsection{The Semantics of Architecture Events}
\label{app:2}

\textbf{States:} The semantics of events is defined based on \textit{local states} and the \textit{global state} of the data types defined in a system. 
Given a service provider \textit{sp}, a local state captures the values of (a data variable) $X_{\theta}$, for all $\theta$ $\in$ \textit{DataTypes}$^{sp}_{arch}$ from the perspective of an entity (component) $E$. Intuitively, a local state of $E$ captures how the value of $X_{\theta}$, $\theta$ $\in$ \textit{DataTypes}$^{sp}_{arch}$, changes from the perspective of an $E$  during a system operation.

Formally, a local state of $E$ is a function $\textit{State}_V$ that assigns a value (including the undefined value $\bot$) to each variable.

\begin{center}
\small
\begin{tabular}{ |c| } 
 \hline
 \textbf{Local state of $E$ (denoted by $\mu_E$)}\\\\
    \textit{State}$_E$ $:$ \textit{Var} $\mapsto$ \textit{Val}$_{\bot}$, where \textit{Var} is a set of all possible data variables and\\  \textit{Val}$_{\bot}$ a set of all possible values, including the undefined value $\bot$.\\ 
 \hline
\end{tabular}
\end{center}
\normalsize

Assume that there are $m$ entities $E_{1}$, \dots, $E_{m}$ defined in an architecture. The \textit{global state} of an architecture is the collection of all the local states in a system. A global state is denoted by $\mu$, where $\mu$ $=$ ($\mu_{E_1}$, \dots, $\mu_{E_m}$, $TT$). 

\begin{center}
\small
\begin{tabular}{ |c| } 
 \hline
 \textbf{Global state of an architecture (denoted by $\mu$)}\\\\
    \textit{State} : \textit{State}$_{E}^m$ $\times$ \textit{TVar}.
\\      
 \hline
\end{tabular}
\end{center}
\normalsize

The \textit{initial $($global$)$ state} for an architecture $\mathcal{P}\mathcal{A}$ is denoted by $\sigma^{init}$, and is the collection of the initial states of each defined entity. Initially the values of all the variables defined in the architecture (including the time variable) have the undefined value, $\bot$. 

\begin{center}
\noindent\fbox{%
    \parbox{9cm}{%
    \underline{\textbf{$\mu^{init}$: Initial Global State}}
\begin{align*}
  \mu^{init} & = ( \mu^{init}_{E_1}, \ldots , \mu^{init}_{E_m}, TT^{init})\ with\\ 
  \forall i \in [1,m],\ \mu^{init}_{E_i} & = (\bot,\dots,\bot)  \\ 
  TT^{init} & = \bot.
\end{align*}
}%
}
\end{center}

\textbf{Event trace and state updates:}  An event trace of an architecture $\mathcal{P}\mathcal{A}$ is denoted by $\tau_{\mathcal{P}\mathcal{A}}$, and contains a finite sequence of events defined in Figure~\ref{fig:eventarch}, happening during a system operation. Below we define the semantics function, denoted by $S_T$, which defines how a trace $\tau_{\mathcal{P}\mathcal{A}}$ changes the global state of an architecture (Figure~\ref{fig:semevent}). 

$S_T$ makes use of the function $S_E$, which defines how each event in $\tau_{\mathcal{P}\mathcal{A}}$ changes the current global state of $\mathcal{P}\mathcal{A}$. 

\begin{center}
\small
\begin{tabular}{ |c| } 
 \hline
 \textbf{Semantics function}\\\\
    $S_T$ $:$ \textit{EventTrace} $\times$ \textit{State}  $\mapsto$   \textit{State}\\
    $S_{Ev}$ $:$ \textit{Event} $\times$ \textit{State} $\mapsto$  \textit{State}
\\      
 \hline
\end{tabular}
\end{center}
\normalsize

\begin{ttd}[The semantics of architectures]
The semantics of an architecture $\mathcal{P}\mathcal{A}$ is defined as a set of global states that can be reached from the initial global state :
\begin{center}
$\{\mu \in \textit{State} \, | \, \exists\ \tau_{\mathcal{P}\mathcal{A}}, S_T(\tau_{\mathcal{P}\mathcal{A}}, \mu^{init}) = \mu \}$.
\end{center}
\end{ttd}

\begin{figure}[htbp]
\centering
\fbox{\begin{minipage}{11.87 cm}
\small
\begin{tabbing}
\=1\=1\=1\=1\= \kill	
$S_T$(\textit{emptytrace}, $\mu$)  $=$ $\mu$\ \ \ \ \ \ \ \ \ \ \ \ \ \ \ \ \ \ \ \ \ \ \ \ \ \ \ \ \ \ \ \ \ \ \  $S_T$(\textit{event}\textbf{.}$\tau_{\mathcal{P}\mathcal{A}}$, $\mu$) $=$  $S_T$($\tau_{\mathcal{P}\mathcal{A}}$, $S_{Ev}$(\textit{event}, $\mu$)) \\\\ 
\=1\=1\=1\=1\= \kill
$S_{Ev}$(\textit{own}($E$, $X_{\theta}$$:$$V_{\theta}$, $t$), $\mu$) $=$ $\mu [\mu_E / \mu_E[X_\theta/V_{\theta}], TT / t]$\\\\
\=1\=1\=1\=1\= \kill
$S_{Ev}$(\textit{calculateat}($E$, $X_{\theta}$$:$$T$, $t$), $\mu$) $=$ $\mu$ [$\mu_E$/$\mu_E$[$X_{\theta}$/\textit{eval}($T$, $\mu_E$)], $TT$/$t$]\\\\ 
\=1\=1\=1\=1\= \kill
$S_{Ev}$(\textit{createat}($E$, $X_{\theta}$$:$$T$, $t$), $\mu$) $=$ $\mu$ [$\mu_E$/$\mu_E$[$X_{\theta}$/\textit{eval}($T$, $\mu_E$)], $TT$/$t$]\\\\ 
\=1\=1\=1\=1\= \kill
$S_{Ev}$(\textit{receiveat}($E$, \textit{Data}$:$$V_{\textit{TYPE(Data)}}$, $t$), $\mu$) $=$ $\mu [\mu_{E} / \mu_{E}[\textit{Data}/V_{\textit{TYPE(Data)}}],\ TT / t]$\\\\ 
\=1\=1\=1\=1\= \kill
$S_{Ev}$(\textit{receiveat}($E$, \textbf{\textit{Cconsent}}(\textit{Data}):$V_{cconsent}$, $t$), $\mu$) $=$ $\mu[\mu_{E}[\textbf{\textit{Cconsent}}(\textit{Data})/V_{cconsent}], TT / t]$\\\\ 
\=1\=1\=1\=1\= \kill
$S_{Ev}$(\textit{receiveat}($E$, \textbf{\textit{Uconsent}}(\textit{Data}):$V_{uconsent}$, $t$),  $\mu$) $=$ $\mu[\mu_{E}[\textbf{\textit{Uconsent}}(\textit{Data})/V_{uconsent}], TT / t]$\\\\ 
\=1\=1\=1\=1\= \kill
$S_{Ev}$(\textit{receiveat}($E$, \textbf{\textit{Sconsent}}(\textit{Data})/$V_{sconsent}$, $t$),  $\mu$) $=$ $\mu[\mu_{E}[\textbf{\textit{Sconsent}}(\textit{Data})/V_{sconsent}], TT / t]$\\\\ 
\=1\=1\=1\=1\= \kill
$S_{Ev}$(\textit{receiveat}($E$, \textbf{\textit{Fwconsent}}(\textit{Data})/$V_{fwconsent}$, $t$),  $\mu$) $=$ $\mu[\mu_{E}[\textbf{\textit{Fwconsent}}(\textit{Data})/V_{fwconsent}], TT / t]$\\\\ 
\=1\=1\=1\=1\= \kill
$S_{Ev}$(\textit{storeat}(\textit{E}, \textit{Data}$:$$V_{\textit{TYPE(Data)}}$, $t$), $\mu$) $=$ $\mu$ [$\mu_E$/$\mu_{E}$[$X_\theta$/$V_\theta$], $TT$/$t$]\\\\  
\=1\=1\=1\=1\= \kill
$S_{Ev}$(\textit{deletewithin}(\textit{E}, \textit{Data}$:$$V_{\textit{TYPE(Data)}}$, $dd$, $t$), $\mu$)\\
\=1\=1\=1\=1\= \kill
\>\> $=$  $\mu$ [$\mu_E$/$\mu_E$[$X_\theta$/$\bot$, \textbf{\textit{Cconsent}}(\textit{Data})/$\bot$, \textbf{\textit{Uconsent}}(\textit{Data})/$\bot$, \textbf{\textit{Sconsent}}(\textit{Data})/$\bot$,\\ \ \ \ \ \ \ \ \ \ \textbf{\textit{Fwconsent}}(\textit{Data})/$\bot$], $TT$/$t$)].
   \end{tabbing}
   \normalsize
\end{minipage}
}
\caption{The semantics of architectural events.}\label{fig:semevent}
\end{figure}

Each event can either change the global state or leave it unchanged. To capture the modification made by an event at time $t$ on (only) the variable state of an entity $E$ we write $\mu [\mu_E / \mu_E[X_\theta/V_\theta], TT / t]$ (or $\mu [\mu_E / \mu_E[X_\theta/\bot], TT / t]$ in case of the undefined value, e.g., when a variable has been deleted). Intuitively, this denotation captures that the old state $\mu_E$ is replaced with the new state $\mu_E[X_\theta/V_\theta]$ ($\mu_E[X_\theta/\bot]$), in which the variable $X_\theta$ has been given the value $V_\theta$ (or the undefined value $\bot$) as a result of the event, the time variable $TT$ is given the value $t$. \textit{eval}($T$, $\mu_E$) is a function that evaluates the variables in $T$ with $\mu_E$. 

\subsection{Well-formed Architectures}
\label{wellformed} 

An architecture $\mathcal{P}$$\mathcal{A}$ is well-formed if: 
\begin{itemize}
\item  Whenever \textit{STORE}($E$, \textit{Data}) $\in$ $\mathcal{P}$$\mathcal{A}$ or \textit{STOREAT}($E$, \textit{Data}, Time(TT)) $\in$ $\mathcal{P}$$\mathcal{A}$, we have 
\begin{itemize}
\item \textit{OWN}($E$, $\theta$) $\in$ $\mathcal{P}$$\mathcal{A}$, \textit{RECEIVE}($E$, \textit{Data}) $\in$ $\mathcal{P}$$\mathcal{A}$ or \textit{RECEIVEAT}($E$, \textit{Data}, \textbf{Time}(TT)) $\in$ $\mathcal{P}$$\mathcal{A}$, or 
\item \textit{CREATE}($E$, \textit{Data}) $\in$ $\mathcal{P}$$\mathcal{A}$ or \textit{CREATEAT}($E$, \textit{Data}, \textbf{Time}(TT)) $\in$ $\mathcal{P}$$\mathcal{A}$, or
\item \textit{CALCULATE}($E$, \textit{Data}) $\in$ $\mathcal{P}$$\mathcal{A}$ or \textit{CALCULATEAT}($E$, \textit{Data}, \textbf{Time}(TT)) $\in$ $\mathcal{P}$$\mathcal{A}$.
\end{itemize}

In case of \textit{OWN}, \textit{CREATE/AT} or \textit{CALCULATE/AT}, instead of \textit{Data}, there can be any function of $\theta$ in \textit{Data} (where \textit{Data} = ($\theta$, $E_{\textit{from}}$)), except for \textit{Cconsent}, \textit{UConsent}, \textit{Sconsent}, or \textit{Fwconsent}. In case of \textit{RECEIVE/AT}, instead of \textit{Data}, there can be any function of $\theta$ except for cryptographic functions and the ``consent functions". 

\item  Whenever \textit{DELETEWITHIN}($E$, \textit{Data}, \textbf{Time}(dd)) $\in$ $\mathcal{P}$$\mathcal{A}$, we have 
\begin{itemize}
\item \textit{STORE}($E$, \textit{Data}) $\in$ $\mathcal{P}$$\mathcal{A}$ or \textit{STOREAT}($E$, \textit{Data}, Time(TT)) $\in$ $\mathcal{P}$$\mathcal{A}$. 
\end{itemize}
\end{itemize}

\section{The Conformance Between Policies and Architectures}
\label{conformance} 
We propose three types of conformance relation: (i) privacy conformance, (ii) conformance with regards to data protection properties (which we refer to as DPR conformance in this paper), and (iii) functional conformance. Privacy conformance compares a policy and an architecture based on the privacy properties. Specifically, if we do not give an entity the right to have or link certain types of data, then in the architecture this entity cannot have or link those types of data. 

\begin{ttd} 
\label{def:priv}
(Proposed privacy conformance definition)
\begin{enumerate}
\item If in a policy $\pi_{\theta}$ an entity $E$ does not have the right to have any data of type $\theta$, then $E$ cannot have this type of data in the corresponding architecture. 
\item If in a policy $\pi_{\theta}$ an entity $E$ does not have the right to link two types of data, $\theta_1$ and $\theta_2$, then $E$ cannot link these types of data in the corresponding architecture.      
\end{enumerate}
\end{ttd}

The DPR conformance relation deals with the data protection requirements (specified in the sub-policies), such as appropriate consent collection, satisfaction of the defined deletion/retention delay, appropriate storage and  transfer of a given type of data. 

\begin{ttd}
\label{def:dpr} 
(Proposed DPR conformance definition): 
\begin{enumerate}
\item If in a policy $\pi_{\theta}$, the collection of a (collection, usage, storage, or transfer) consent is required for a piece of data of a given type, then in the architecture the reception of a consent can happen before or at the same time with the reception of the data itself. 

\item If in an architecture there is an action \textbf{act} (\textit{createat} or \textit{calculateat}) defined on a data type $\theta$, then in the policy $\pi_{\theta}$, there is a (collection, usage, storage, or transfer) purpose \textbf{act}:$\theta'$ defined for the type $\theta$ (besides some $\theta'$).  

\item If in an architecture a piece of data of type $\theta$ can be stored in some storage place, \textbf{strplace}, then in the policy $\pi_{\theta}$, \textbf{strplace} $\in$ $\pi_{str}$.\textbf{where} (see Table~\ref{tab:notation1} for notations). 

\item If in the policy $\pi_{\theta}$, \textbf{delplace} $\in$  $\pi_{del}$.\textbf{fromwhere}, then in the corresponding architecture the same data type can be deleted from the place \textbf{delplace}.

\item If in an architecture, a piece of data of type $\theta$ can be deleted within a delay \textbf{dd} (from collection), then in the corresponding policy $\pi_{\theta}$, \textbf{dd} $\leq$ $\pi_{del}$.\textbf{deld}.  In other words, the retention delay defined in the policy must be respected in the architecture.

\item If in an architecture, a piece of data of type $\theta$ can be  
transferred to an entity $E$, then in the policy $\pi_{\theta}$, 
$E$ $\in$ $\pi_{fw}$.\textbf{fwto} (again, see Table~\ref{tab:notation1} for notations). 
\end{enumerate}
\end{ttd}

Finally, functional conformance compares a policy and an architecture from the perspective of functionality or effectiveness. 
This conformance can help a system designer to find an appropriate trade-off between functionality and privacy as in real life, a system is expected to be able to provide certain services. 

\begin{ttd} 
\label{def:func}
(Proposed functional conformance definition)
\begin{enumerate}
\item   If in a policy $\pi_{\theta}$, an entity $E$ has the right to have a type of data,  $\theta$, then $E$ can have this type of data in the corresponding architecture. 

\item  If in a policy $\pi_{\theta}$, an entity $E$ has the right to link two types of data, $\theta_1$ and $\theta_2$, then $E$ can link these types of data in the corresponding architecture.   

\item If in a policy $\pi_{\theta}$, the collection of a (collection, usage, storage, or transfer) consent  is \textbf{not} required, then \textbf{no} corresponding consent can be received in the corresponding architecture.  

\item If in a policy $\pi_{\theta}$, there is a (collection, usage, storage, or transfer) purpose \textbf{act}:$\theta'$ defined, then in the corresponding  architecture there is an action \textbf{act} defined on a data type $\theta$ (besides some $\theta'$).

\item If in a policy $\pi_{\theta}$, (\textbf{strplace} $\in$ $\pi_{str}$.\textbf{where}) for some storage place \textbf{strplace}, then in the corresponding architecture this type of data can be stored in \textbf{strplace}.    

\item If in an architecture a piece of data of type $\theta$ can be deleted from a  storage place, \textbf{delplace}, then in the corresponding policy $\pi_{\theta}$, we have (\textit{delplace} $=$ $\pi_{del}$.\textbf{fromwhere}). 

\item If in the policy $\pi_{\theta}$, $E$ $\in$ $\pi_{fw}$.\textbf{fwto}, then  
in the corresponding architecture, the same type of data can be  
transferred to the same entity $E$.    
\end{enumerate}
\end{ttd}

\section{The proposed automated verification engine}
\label{sec:aut} 

The verification engine is based on logic and resolution based proofs. Below, we define the inference rules that will be used in the proof process in Algorithm~\ref{alg1}. See Table~\ref{tab:notation3} for the notations used in this section.

\begin{ttd}
An inference rule $R$ is denoted by $R$ $=$ $H$ $\vdash$ $T_1$, \dots, $T_n$, where $H$ is the head of the rule and $T_1$, \dots, $T_n$ is the tail of the rule. Each element $T_i$ of the tail is called a fact (or condition), and a head is called a ``consequence". The rule $R$ reads as ``if $T_1$, \dots, $T_n$, then $H$".
\end{ttd}

\begin{table}
\centering
\small
\begin{tabular}{ |c|l| } 
 \hline
 \textit{An inference rule} & $H$ $\vdash$ $T_1$, \dots, $T_n$. \\
 \textit{The head of a rule} & $H$  (in $H$ $\vdash$ $T_1$, \dots, $T_n$).\\
 \textit{The tail of a rule} & $T_1$, \dots, $T_n$ (in $H$ $\vdash$ $T_1$, \dots, $T_n$). A $T_i$ is called as a (sub-)goal in a proof.\\
 \textit{A fact} & Any of $H$, $T_1$, \dots, $T_n$.\\
 \textit{A predicate} & Each fact has the form of PREDICATE(\textit{Argument}$_1$,\dots, \textit{Argument}$_m$).\\
  \hline
 $\theta V$ & A variable that can be mapped to a data type $\theta$ in the policy/architecture. \\
 \textit{EV} & A variable that can be mapped to an entity $E$ in the policy/architecture. \\
 \textit{DD} & A variable that can be mapped to a deletion delay \textit{dd} in the policy/arch. \\
 \textit{TV} & A variable that can be mapped to a non-specific time value \textit{TT} in the arch. \\
 \textit{TT}, \textit{dd} & A non-specific time value (\textit{TT}), and a numerical time value (\textit{dd}). \\
  \textit{K}, \textit{PK} & The variables that can be mapped to a type of symmetric and public key. \\
 ds & A value that specifies a real identity of a living individual. \\
 P(ds) & The pseudonym of the real identity ds.\\
  \hline
 \textit{GoalsToBeProved} & The set of goals (sub-goals) to be proved.\\
 \textit{initgoal} & A goal to be proved, which is generated from/captures a sub-policy. \\
 \textit{nextgoal} & The (next) goal in the set \textit{GoalsToBeProved} to be proved. \\
 \textit{previousgoal} & The goal in \textit{GoalsToBeProved} was proved right before \textit{nextgoal}. \\
 $\mathbb{A}\mathbb{G}$ & A set of all possible \textit{initgoal}s (covers all the seven sub-policies). \\
 \textit{C/U/FwPurpSet} & A set of facts that capture the collection/usage/transfer purposes, respectively.\\
 \textit{UniqueTypes} & A set of facts that capture the unique data types (UNIQUE($\theta$)).\\
 \textit{TrivialHASLINKFacts} & A set of ``trivial" HAS, LINK, LINKUNIQUE facts generated from\\ 
 & architectural actions (See Figures~\ref{fig:trivi}-\ref{fig:trivi2}).\\ 
 \hline
 Anytype(arg1,\dots, argn) & A piece of (compound) data Anytype that has $n$ arguments\\  & (Anytype $\notin$ \{\textit{Senc}, \textit{Aenc}, \textit{Mac}, \textit{Hash}\}). \\
 Anytype\textbf{[$\theta V$]} & A piece of data Anytype that contains a piece of data of type $\theta V$.\\
  & (Anytype $\notin$ \{\textit{Senc}, \textit{Aenc}, \textit{Mac}, \textit{Hash}\}). $\theta V$ may be an argument of another \\ 
  & compound data inside  Anytype, and so on.\\ 
 Anytypeinccrypto\textbf{[$\theta V$]} & Similar to above, but Anytypeinccrypto can also be \textit{Senc}, \textit{Aenc}, \textit{Mac}, or \textit{Hash}.\\
 & $\theta V$ may be an argument of another compound data inside  Anytypeincrypto,\\ 
 & and so on.\\
 \hline
 $\sigma$ & A unifier or mapping, e.g. $\sigma$ = \{\textit{EV} $\mapsto$ \textit{E}, $\theta V$ $\mapsto$ $\theta$, \textit{DD} $\mapsto$ \textit{dd}, \textit{TV} $\mapsto$ \textit{TT}\},\\
 &  where \textit{E} is an entity value (e.g. client), $\theta$ is a type value (e.g. name),\\ 
 & \textit{dd} (e.g. 6 years). \\
 $T\sigma$ & Apply the mapping $\sigma$ to the variables in $T$.\\ 
 \textit{nextgoal}$\sigma$ & Apply the mapping $\sigma$ to the variables in \textit{nextgoal}.\\ 
 \textit{Data} & ($\theta V$, \textit{EV}$_{\textit{from}}$), $\theta V$ is a type of a piece of data, \textit{EV}$_{\textit{from}}$ is who originally sent\\ 
 & this data.\\ 
 \textit{isSuccessful}[(rule, goal)] & A dictionary used in e.g. the Python language, with (rule, goal) as the key. \\
 \hline
\end{tabular}
\caption{\label{tab:notation3} The notations used in the automated verification engine.}
\normalsize
\end{table}

Figure~\ref{fig:inf} presents the proposed  rules used in the verification of the DPR conformance relations. For instance: 
\begin{itemize}
\item \textit{D1} specifies that if an entity \textit{EV} can receive a transfer consent on \textit{Data}, \textit{Data} =($\theta V$, \textit{EV}$_{\textit{from}}$),  to \textit{EV}$_{\textit{to}}$ at some non-specific time $TV$, and \textit{EV}$_{\textit{to}}$ can receive this at the same time (or later\footnote{This is modelled in an abstract way by using the same non-specific time value $TV$.}), then we say that \textit{EV} can collect the transfer consent on $\theta V$ to \textit{EV}$_{\textit{to}}$. 

\item Rule \textit{D2} is defined for data collection consent, rules  \textit{D3}-\textit{D4} are for usage consent collection, and \textit{D5} is for the storage consent.      

\item Rules \textit{D6} and \textit{D7} are the corresponding version of \textit{D1} and \textit{D2}, respectively, where $\theta V$ is inside another compound data type\footnote{For example,  Anytypeinccrypto[$\theta V$] can be \textit{Sicknessrec}($\theta V$,\dots), \textit{Sicknessrec}(Anytypeinccrypto1[$\theta V$],\dots), \textbf{Senc}($\theta V$, \textit{K}), or  \textbf{Senc}(Anytypeinccrypto1[$\theta V$], \textit{K}), etc}.  
\end{itemize}

Figure~\ref{fig:inf2} shows the proposed rules used in the verification of the privacy conformance relation (i.e. a HAS/HASUPTO data possession property). For instance: 
\begin{itemize}
\item Rule \textit{P1} says that if an entity \textit{EV} can store  \textit{Data}, \textit{Data} = ($\theta V$, \textit{EV}$_{\textit{from}}$), and can delete \textit{Data} within a time delay \textit{DD}, then the entity can have this data\footnote{More precisely, it can have the corresponding type of data ($\theta V$) in \textit{Data}  = ($\theta V$,  \textit{EV}$_{\textit{from}}$).} up to \textit{DD} time. 

\item Rule \textit{P2} says that if a trusted authority/organisation has any data that contains a pseudonym (\textbf{P}(\textit{ds})),  alongside some other data, then the trusted authority can also have the same data that contains the ``real" identity \textit{ds}. 

\item \textit{P3} says that if \textit{EV} can own a type of data (regardless of time), then it can have this type of data. 

\item Rule \textit{P4} says that if \textit{EV} can receive \textit{Data} at some non-specific time \textit{TV}, then it can have this data. The rest rules can be interpreted in a similar way. 

\item Finally, rules \textit{P8}-\textit{P10} capture the decryption of the cryptographic data types. \textit{P8} says that if \textit{EV} can have an encryption of \textit{Data} using a symmetric key $K$, and it can also have $K$, then it can have \textit{Data}. Similarly, \textit{P9}-\textit{P10} deal with the decryption of a message authentication code, and  the asymmetric decryption process, respectively.    
\end{itemize}

\begin{figure}[htbp!]
\centering
\fbox{\begin{minipage}{11.87 cm}
\small
\begin{tabbing}    
    \=1\kill
    D1. FWCONSENTCOLLECTED(\textit{EV}, $\theta V$, \textit{EV}$_{\textit{to}}$) $\vdash$\\  
    12345\=1 \kill
    \> RECEIVEAT(\textit{EV}, \textbf{Fwconsent}(\textit{Data}, \textit{EV}$_{\textit{to}}$),\textbf{Time}(\textit{TV})), RECEIVEAT(\textit{EV}$_{\textit{to}}$, \textit{Data}, \textbf{Time}(\textit{TV})) \\\\
    
    \=1\kill
    D2. CCONSENTCOLLECTED(\textit{EV}, $\theta V$) $\vdash$\\  
    12345\=1 \kill
    \>RECEIVEAT(\textit{EV}, \textbf{Cconsent}(\textit{Data}), \textbf{Time}(\textit{TV})), RECEIVEAT(\textit{EV}, \textit{Data}, \textbf{Time}(\textit{TV}))\\\\
    
    \=1\kill
    D3. UCONSENTCOLLECTED(\textit{EV}, $\theta V$) $\vdash$\\  
    12345\=1 \kill
    \>RECEIVEAT(\textit{EV}, \textbf{Uconsent}(\textit{Data}),\textbf{Time}(\textit{TV})), CREATEAT(\textit{EV}, Anytype[$\theta V$], \textit{EV}$_{\textit{from}}$, \textbf{Time}(\textit{TV}))\\\\
    
     \=1\kill
    D4. UCONSENTCOLLECTED(\textit{EV}, $\theta V$) $\vdash$\\  
    12345\=1 \kill
    \>RECEIVEAT(\textit{EV}, \textbf{Uconsent}(\textit{Data}), \textbf{Time}(\textit{TV})), CALCULATEAT(\textit{EV}, Anytype[$\theta V$], \textit{EV}$_{\textit{from}}$, \textbf{Time}(\textit{TV}))\\\\
    
    \=1\kill
    D5. STRCONSENTCOLLECTED(\textit{EV}, $\theta V$) $\vdash$\\  
    12345\=1 \kill
    \>RECEIVEAT(\textit{EV}, \textbf{Sconsent}(\textit{Data}), \textbf{Time}(\textit{TV})), STOREAT(\textit{EV}, \textit{Data}, \textbf{Time}(\textit{TV}))\\\\ 
    
    \=1\kill
    D6.  FWCONSENTCOLLECTED(\textit{EV}, $\theta V$, \textit{EV}$_{\textit{to}}$) $\vdash$ \\ 
    1234567890123\=1 \kill 
    \> RECEIVEAT(\textit{EV}, \textbf{Fwconsent}($\theta V$, \textit{EV}$_{\textit{from}}$, \textit{EV}$_{\textit{to}}$),\textbf{Time}(\textit{TV})),\\  
    1234567890123\=1 \kill
    \> RECEIVEAT(\textit{EV}$_{\textit{to}}$, Anytypeinccrypto[$\theta V$], \textit{EV}$_{\textit{from}}$, \textbf{Time}(\textit{TV}))\\\\
    
    \=1\kill
    D7.  CCONSENTCOLLECTED(\textit{EV}, $\theta V$) $\vdash$ \\ 
     1234567890123\=1 \kill 
    \> RECEIVEAT(\textit{EV}, \textbf{Cconsent}($\theta V$, \textit{EV}$_{\textit{from}}$), \textbf{Time}(\textit{TV})),\\  
     1234567890123\=1 \kill
    \> RECEIVEAT(\textit{EV}, Anytypeinccrypto[$\theta V$], \textit{EV}$_{\textit{from}}$, \textbf{Time}(\textit{TV}))\\\\

    \=1\kill   
    Where Data = ($\theta V$, \textit{EV}$_{\textit{from}}$) ($\theta V$ represents a data type, and \textit{EV}$_{\textit{from}}$, an entity that originally sent this data).  
\end{tabbing}
\end{minipage}
}
\caption{\small The proposed inference rules for DPR conformance check. The predicates and arguments of the heads and tails in the rules are in line with the architecture syntax in Figure~\ref{fig:sysarch}. \normalsize}\label{fig:inf}
\end{figure}
\normalsize

\begin{figure}[htbp!]
\centering
\fbox{\begin{minipage}{11.87 cm}
\small
\begin{tabbing}    
    \=1\kill
 	P1. HASUPTO(\textit{EV}, $\theta V$, \textbf{Time}(\textit{DD})) $\vdash$\\  
    123456\=1 \kill
    \> STOREAT(\textit{EV}, \textit{Data}, \textbf{Time}(\textit{TV})), DELETEWITHIN(\textit{EV}, \textit{Data}, \textbf{Time}(\textit{DD})) \\\\
    
    P2. HAS(\textbf{trusted}, Anytype(\textit{ds}, $\theta V$)) $\vdash$  HAS(\textbf{trusted}, Anytype($\theta V$, \textbf{P}(\textit{ds}))), \\ \ \ \ \ \ where Anytype is not a crypto function (Anytype $\notin$ \{\textit{Senc}, \textit{Aenc}, \textit{Mac}, \textit{Hash}\}). \\\\  

    P3. HAS(\textit{EV}, $\theta V$) $\vdash$ OWN(\textit{EV}, $\theta V$)\\\\ 
    
    P4. HAS(\textit{EV}, $\theta V$) $\vdash$ RECEIVEAT(\textit{EV}, \textit{Data}, \textbf{Time}(\textit{TV}))\\\\ 
    
    P5. HAS(\textit{EV}, $\theta V$) $\vdash$ STOREAT(\textit{EV},  \textit{Data}, \textbf{Time}(\textit{TV}))\\\\ 
    
    P6. HAS(\textit{EV}, $\theta V$) $\vdash$ CREATEAT(\textit{EV}, $\theta V$, \textbf{Time}(\textit{TV}))\\\\ 
    
    P7. HAS(\textit{EV}, $\theta V$) $\vdash$ CALCULATEAT(\textit{EV}, $\theta V$, \textbf{Time}(\textit{TV}))\\\\ 
     
    P8. HAS(\textit{EV}, $\theta V$) $\vdash$ HAS(\textit{EV}, \textbf{Senc}($\theta V$, \textit{K})), HAS(\textit{EV}, \textit{K})\\\\ 
     
    P9. HAS(\textit{EV}, $\theta V$) $\vdash$ HAS(\textit{EV}, \textbf{Mac}($\theta V$,\textit{K})), HAS(\textit{EV}, \textit{K}) \\\\ 
    
    P10. HAS(\textit{EV}, $\theta V$) $\vdash$ HAS(\textit{EV}, \textbf{Aenc}($\theta V$, \textit{PK})), HAS(\textit{EV}, \textbf{Sk}(\textit{PK}))\\\\
    \=1\kill
 	P11. HASUPTO(EV, $\theta V$,\textbf{Time}(DD)) $\vdash$\\  
    123456\=1 \kill
    \> STORE(EV, Data), DELETEWITHIN(EV, Data, \textbf{Time}(DD)) \\\\
    
     P12. HAS(\textbf{trusted}, Anytype(ds, $\theta V$)) $\vdash$  HAS(\textbf{trusted}, Anytype(\textbf{P}(ds), $\theta V$))\\\\ 
    
    P13. HAS(\textbf{trusted}, Anytype($\theta V$, ds)) $\vdash$  HAS(\textbf{trusted},  Anytype($\theta V$, \textbf{P}(ds)))\\\\  
    
    P14. HAS(\textbf{trusted}, Anytype($\theta V$, ds)) $\vdash$  HAS(\textbf{trusted}, Anytype(\textbf{P}(ds), $\theta V$)) \\\\
    
    P15. HAS(EV, $\theta V$) $\vdash$ RECEIVE(EV, Data)\\\\ 
    
    P16. HAS(EV, $\theta V$) $\vdash$ STORE(EV, Data)\\\\ 
         
    P17. HAS(EV, $\theta V$) $\vdash$ CREATE(EV, $\theta V$)\\\\ 
    
    P18. HAS(EV, $\theta V$) $\vdash$ CALCULATE(EV, $\theta V$). 
    \footnote{No rule is defined for the trivial HAS, LINK, LINKUNIQUE properties (e.g. if \textit{sp} can receive \textit{Bill}(\textit{name}, \textit{address}), then it can have \textit{name}, \textit{address}, and can link them, but the facts HAS(sp, name),\dots, LINK(sp, name, address), LINKUNIQUE(sp, name, address) are generated directly from the architectural actions/facts). See Figure~\ref{fig:trivi} for details on how these facts are generated.}   
\end{tabbing}
\end{minipage}
}
\caption{\small Inference rules for privacy conformance check (HAS and HASUPTO property). \textit{P8}-\textit{P10} capture the cryptographic verification/decryption process, i.e. the destructor application defined in Figure~\ref{fig:archterms}. \normalsize}\label{fig:inf2}
\end{figure}
\normalsize

\begin{figure}[htbp!]
\centering
\fbox{\begin{minipage}{11.87 cm}
\small
\begin{tabbing}    
	L0. LINK(\textit{EV}, $\theta V_1$, $\theta V_2$) $\vdash$ \\ 
    1234567\=1 \kill
    \>  HAS(\textit{EV}, Anytype1($\theta V_1$, $\theta V$, \textbf{Meta}($\theta V_3$))), HAS(\textit{EV}, Anytype2($\theta V_2$, $\theta V'$, \textbf{Meta}($\theta V_3$)))\\\\ 
    
	\=1\kill
   	L1. LINK(\textit{EV}, $\theta V_1$, $\theta V_2$) $\vdash$\\ 
	1234567\=1 \kill   	
   	\> HAS(\textit{EV}, Anytype1($\theta V_1$, $\theta V$, $\theta V_3$)), HAS(\textit{EV}, Anytype2($\theta V_2$, $\theta V'$, $\theta V_3$)).\\\\ 

     \=1\kill
   	L2. LINK(\textit{EV}, $\theta V_1$, $\theta V_2$) $\vdash$\\  
    1234567\=1 \kill
     \> HAS(\textit{EV},Anytype1($\theta V_1$, $\theta V$, $\theta V_3$)), HAS(\textit{EV}, Anytype2($\theta V_3$, $\theta V'$, $\theta V_2$)) \\\\
     
    \=1\kill
   	L3. LINK(\textit{EV}, $\theta V_2$, $\theta V_1$) $\vdash$\\  
    1234567\=1 \kill
    \> HAS(\textit{EV}, Anytype1($\theta V_1$, $\theta V$, $\theta V_3$)), HAS(\textit{EV}, Anytype2($\theta V_2$, $\theta V'$, $\theta V_3$)) \\\\		
	 
 	\=1\kill
   	L4. LINK(\textit{EV}, $\theta V_2$, $\theta V_1$) $\vdash$\\  
    1234567\=1 \kill
    \> HAS(\textit{EV}, Anytype1($\theta V_1$, $\theta V$, $\theta V_3$)), HAS(\textit{EV}, Anytype2($\theta V_3$, $\theta V'$, $\theta V_2$)) \\\\
    
    L5-L8 are similar to L1-L4, respectively, but with HAS(\textit{EV}, Anytype1($\theta V_3$, $\theta V$, $\theta V_1$))\\ 
    123456\=1 \kill
    \>instead of HAS(\textit{EV}, Anytype1($\theta V_1$, $\theta V$, $\theta V_3$)) to capture the different order of the\\ 
    123456\=1 \kill
    \>data types.\\\\
     U1. LINKUNIQUE(\textit{EV}, $\theta V_1$, $\theta V_2$) $\vdash$\\  
    123456\=1 \kill
    \> HAS(\textit{EV}, Anytype1($\theta V_1$, $\theta V$, $\theta V_3$)), HAS(\textit{EV}, Anytype2($\theta V_2$, $\theta V'$, $\theta V_3$)), UNIQUE($\theta V_3$)\\
    \ \ \ \ \ \ where Anytype1 and Anytype2 are not crypto functions.\\\\ 
     \=1\kill
   	U2. LINKUNIQUE(\textit{EV}, $\theta V_1$, $\theta V_2$) $\vdash$\\  
    123456\=1 \kill
   \> HAS(\textit{EV}, Anytype1($\theta V_1$,$\theta V$,$\theta V_3$)), HAS(\textit{EV}, Anytype2($\theta V_3$,$\theta V'$,$\theta V_2$)), UNIQUE($\theta V_3$) \\\\	
 	
    \=1\kill
   	U3. LINKUNIQUE(\textit{EV}, $\theta V_2$,$\theta V_1$) $\vdash$\\  
    123456\=1 \kill
    \> HAS(\textit{EV}, Anytype1($\theta V_1$, $\theta V$, $\theta V_3$)), HAS(\textit{EV}, Anytype2($\theta V_2$, $\theta V'$,$\theta V_3$)), UNIQUE($\theta V_3$) \\\\		
	 
 	\=1\kill
   	U4. LINKUNIQUE(\textit{EV}, $\theta V_2$, $\theta V_1$) $\vdash$\\  
    123456\=1 \kill
    \> HAS(\textit{EV}, Anytype1($\theta V_1$, $\theta V$, $\theta V_3$)), HAS(\textit{EV}, Anytype2($\theta V_2$, $\theta V'$, $\theta V_3$)), UNIQUE($\theta V_3$)\\\\
    
   U5-U8 are similar to U1-U4, respectively, but with HAS(\textit{EV}, Anytype1($\theta V_3$, $\theta V$, $\theta V_1$))\\ 
    123456\=1 \kill
    \>instead of HAS(\textit{EV}, Anytype1($\theta V_1$, $\theta V$, $\theta V_3$)) to capture the different order of the\\ 
    123456\=1 \kill
    \>data types.  
\end{tabbing}
\end{minipage}
}
\caption{ \small Inference rules for the privacy conformance check (basic linkability and unique linkability).}\label{fig:inf3}
\normalsize
\end{figure}
\normalsize

\begin{figure}[htbp!]
\centering
\fbox{\begin{minipage}{11.87 cm}
\small
\begin{tabbing}    
    \=1\kill
    C1. CRYPTHAS(\textit{EV}, $\theta V$) $\vdash$ HAS(\textit{EV}, \textbf{Senc}($\theta V$, \textit{K})), HAS(\textit{EV}, \textit{K})\\\\ 
     
    C2. CRYPTHAS(\textit{EV}, $\theta V$) $\vdash$ HAS(\textit{EV}, \textbf{Mac}($\theta V$,\textit{K})), HAS(\textit{EV}, \textit{K}) \\\\ 
    
    C3. CRYPTHAS(\textit{EV}, $\theta V$) $\vdash$ HAS(\textit{EV}, \textbf{Aenc}($\theta V$, \textit{PK})), HAS(\textit{EV}, \textbf{Sk}(\textit{PK}))
\end{tabbing}
\end{minipage}
}
\caption{\small Inference rules for CRYPTHAS check. The three rules \textit{C1}-\textit{C3} in Figure~\ref{fig:infcrypt} are similar to \textit{P8}-\textit{P10}, and we define them with verify whether a piece of data of type $\theta V_1$  can be obtained by a decryption step. We intentionally differentiate between CRYPTHAS and HAS to deal with the linkability rules in Figure~\ref{fig:inf3b}.  \normalsize}\label{fig:infcrypt}
\end{figure}
\normalsize

\begin{figure}[htbp!]
\centering
\fbox{\begin{minipage}{11.87 cm}
\small
\begin{tabbing}    
    	\=1\kill
   	L1/b. LINK(\textit{EV}, $\theta V_1$, $\theta V_2$) $\vdash$\\ 
	1234567\=1 \kill   	
   	\> HAS(\textit{EV}, Anytype1($\theta V_1$, $\theta V$, $\theta V_3$)), HAS(\textit{EV}, Anytype2($\theta V_2$, $\theta V'$, Anytypeincrypto[$\theta V_3$])), \\
   		1234567\=1 \kill   	
   	\> CRYPTHAS(\textit{EV}, $\theta V_3$)
   	\\\\
   	
   	\=1\kill
   	L1/c. LINK(\textit{EV}, $\theta V_1$, $\theta V_2$) $\vdash$\\ 
	1234567\=1 \kill   	
   	\> HAS(\textit{EV}, Anytype1($\theta V_1$, $\theta V$, $\theta V_3$)), HAS(\textit{EV}, Anytype2(Anytypeincrypto[$\theta V_2$], $\theta V'$, $\theta V_3$)), \\
   		1234567\=1 \kill   	
   	\> CRYPTHAS(\textit{EV}, $\theta V_2$)
   	\\\\
   	
   	   	\=1\kill
   	L1/d. LINK(\textit{EV}, $\theta V_1$, $\theta V_2$) $\vdash$\\ 
	1234567\=1 \kill   	
   	\> HAS(\textit{EV}, Anytype1($\theta V_1$, $\theta V$, $\theta V_3$)),\\
   		1234567\=1 \kill   	
   	\> HAS(\textit{EV}, Anytype2(Anytypeincrypto1[$\theta V_2$], $\theta V'$, Anytypeincrypto2[$\theta V_3$])), \\
   		1234567\=1 \kill   	
   	\> CRYPTHAS(\textit{EV}, $\theta V_2$), CRYPTHAS(\textit{EV}, $\theta V_3$)
   	\\\\

     \=1\kill
   	L2/b. LINK(\textit{EV}, $\theta V_1$, $\theta V_2$) $\vdash$\\  
    1234567\=1 \kill
     \> HAS(\textit{EV},Anytype1($\theta V_1$, $\theta V$, $\theta V_3$)), HAS(\textit{EV}, Anytype2($\theta V_3$, $\theta V'$, Anytypeincrypto[$\theta V_2$])))\\
   		1234567\=1 \kill   	
   	\>  CRYPTHAS(\textit{EV}, $\theta V_2$) \\\\ 
     
     \=1\kill
   	L2/c. LINK(\textit{EV}, $\theta V_1$, $\theta V_2$) $\vdash$\\  
    1234567\=1 \kill
     \> HAS(\textit{EV},Anytype1($\theta V_1$, $\theta V$, $\theta V_3$)), HAS(EV, Anytype2(Anytypeincrypto[$\theta V_3$], $\theta V'$, $\theta V_2$))) \\
   		1234567\=1 \kill   	
   	\> CRYPTHAS(\textit{EV}, $\theta V_3$)\\\\ 
     
     \=1\kill
   	L2/d. LINK(\textit{EV}, $\theta V_1$, $\theta V_2$) $\vdash$\\  
    1234567\=1 \kill
     \> HAS(\textit{EV},Anytype1($\theta V_1$, $\theta V$, $\theta V_3$)),\\ 
     1234567\=1 \kill
     \> HAS(\textit{EV}, Anytype2(Anytypeincrypto1[$\theta V_3$], $\theta V'$, Anytypeincrypto2[$\theta V_2$]))) \\
   		1234567\=1 \kill   	
   	\> CRYPTHAS(\textit{EV}, $\theta V_3$), CRYPTHAS(\textit{EV}, $\theta V_2$)\\\\

    \=1\kill
   	L3/b. LINK(\textit{EV}, $\theta V_2$, $\theta V_1$) $\vdash$\\  
    1234567\=1 \kill
    \> HAS(\textit{EV}, Anytype1($\theta V_1$, $\theta V$, $\theta V_3$)), HAS(\textit{EV}, Anytype2($\theta V_2$, $\theta V'$, Anytypeincrypto[$\theta V_3$])), \\
   		1234567\=1 \kill   	
   	\> CRYPTHAS(\textit{EV}, $\theta V_3$)\\\\
    
    \=1\kill
   	L3/c. LINK(\textit{EV}, $\theta V_2$, $\theta V_1$) $\vdash$\\  
    1234567\=1 \kill
    \> HAS(\textit{EV}, Anytype1($\theta V_1$, $\theta V$, $\theta V_3$)), HAS(\textit{EV}, Anytype2(Anytypeincrypto[$\theta V_2$], $\theta V'$, $\theta V_3$))\\
   		1234567\=1 \kill   	
   	\> CRYPTHAS(\textit{EV}, $\theta V_3$)\\\\
    
    \=1\kill
   	L3/d. LINK(\textit{EV}, $\theta V_2$, $\theta V_1$) $\vdash$\\  
    1234567\=1 \kill
    \> HAS(\textit{EV}, Anytype1($\theta V_1$, $\theta V$, $\theta V_3$)), \\
   		1234567\=1 \kill
   		\> HAS(\textit{EV}, Anytype2(Anytypeincrypto1[$\theta V_2$], $\theta V'$, Anytypeincrypto2[$\theta V_3$])), \\
   		1234567\=1 \kill   	
   	\> CRYPTHAS(\textit{EV}, $\theta V_3$), CRYPTHAS(\textit{EV}, $\theta V_2$) \\\\
	 
 	\=1\kill
   	L4/b. LINK(\textit{EV}, $\theta V_2$, $\theta V_1$) $\vdash$\\  
    1234567\=1 \kill
    \> HAS(\textit{EV}, Anytype1($\theta V_1$, $\theta V$, $\theta V_3$)), HAS(EV, Anytype2($\theta V_3$, $\theta V'$, Anytypeincrypto[$\theta V_2$])),\\
   		1234567\=1 \kill   	
   	\>  CRYPTHAS(\textit{EV}, $\theta V_2$)\\\\
    
    \=1\kill
   	L4/c. LINK(\textit{EV}, $\theta V_2$, $\theta V_1$) $\vdash$\\  
    1234567\=1 \kill
    \> HAS(\textit{EV}, Anytype1($\theta V_1$, $\theta V$, $\theta V_3$)), HAS(EV, Anytype2(Anytypeincrypto[$\theta V_3$], $\theta V'$, $\theta V_2$)), \\
   		1234567\=1 \kill   	
   	\> CRYPTHAS(\textit{EV}, $\theta V_3$)\\\\
    
    \=1\kill
   	L4/d. LINK(\textit{EV}, $\theta V_2$, $\theta V_1$) $\vdash$\\  
    1234567\=1 \kill
    \> HAS(\textit{EV}, Anytype1($\theta V_1$, $\theta V$, $\theta V_3$)),  \\
   		1234567\=1 \kill
   		\>  HAS(\textit{EV}, Anytype2(Anytypeincrypto1[$\theta V_3$], $\theta V'$, Anytypeincrypto2[$\theta V_2$])), \\
   		1234567\=1 \kill   	
   	\> CRYPTHAS(\textit{EV}, $\theta V_3$), CRYPTHAS(\textit{EV}, $\theta V_2$).\\\\
   	
   	For each rule in \textbf{L5-L8} and \textbf{U1-U8} we define a corresponding three rules (/b, /c, /d) with \\ Anytypeincrypto[$\theta V_2$] and Anytypeincrypto1[$\theta V_3$] in the same manner, respectively.   
    
\end{tabbing}
\end{minipage}
}
\caption{\small Inference rules for linkability and unique linkability (part 2). These rules are the extension to the basic rules in Figure~\ref{fig:inf3}, in order to capture the case of compound data types for the completeness property. E.g.  Anytypeincrypto[$\theta V_3$] represents a piece of compound data type (which may be a crypto function) that contains a type $\theta V_3$ (note Anytypeincrypto[$\theta V_3$] means that $\theta V_3$ can be an argument of another compound data inside  Anytypeincrypto, and so on).}\label{fig:inf3b} 
\normalsize
\end{figure}
\normalsize

Figure~\ref{fig:inf3} includes the proposed rules used in the verification of the privacy conformance relation (for the LINK property). For instance: 
\begin{itemize}
\item  Rule \textit{L0} says that if an entity (specified by the variable) \textit{EV} can have two pieces of data of types $\theta V_1$ and $\theta V_2$, inside any compound data types with the same metadata, then this entity can link $\theta V_1$ and $\theta V_2$.       
\item Rule \textit{L1} says that if the entity \textit{EV} can have any data that contains two pieces of data of types $\theta V_1$, and $\theta V_3$ besides any other data (denoted by $\theta V$ and $\theta V'$), and any data that contains two pieces of data of types $\theta V_2$ and $\theta V_3$, then \textit{EV} can link $\theta V_1$ and $\theta V_2$. Note that this is not ``unique" linkability, meaning that \textit{EV} cannot be sure that the data of types $\theta V_1$ and $\theta V_2$ belong to the same individual (although it can narrow down the set of possible individuals to some extent). 

\item Extending \textit{L1}, rule \textit{U1} also says that if  a type $\theta V_3$ is unique (e.g. public IP addresses), then \textit{EV} can ``uniquely" link the data of types $\theta V_1$ and $\theta V_2$, namely, it can also be sure that they belong to the same individual. 
\end{itemize}


In Figure~\ref{fig:inf3b},  for example, rule L1/b says that if an entity  \textit{EV} can have $\theta V_1$ inside a compound data type (Anytype1), and  $\theta V_2$ in some compound data type, then it can link $\theta V_1$ and $\theta V_2$.  The main difference between L1/b and L1 is that in the first case, $\theta V_3$ is also inside a compound data type (which can be a type of a cryptographic function). We use the fact CRYPTHAS(\textit{EV}, $\theta V_3$) to capture that \textit{EV} can have $\theta V_3$ by decrypting the cryptographic function that contains $\theta V_3$\footnote{Namely,  CRYPTHAS(\textit{EV}, $\theta V_3$) is defined to deal with the case when  $\theta V_3$ is inside a cryptographic function in the second HAS fact.}. 

\textbf{The forward search strategy}: 
In order to speed up the verification process and avoid an infinite loop of resolution steps during a proof, let us consider the ``trivial" HAS, LINK and LINKUNIQUE properties, namely, if an entity \textit{EV} can have Anytype($\theta V_1$, \dots, $\theta V_n$) for some $n$, then:
\begin{itemize}
\item \textit{EV} can have each of $\theta V_1$, \dots, $\theta V_n$.    
\item \textit{EV} can link and uniquely link all the possible pairs among $\theta V_1$, \dots, $\theta V_n$.
\end{itemize}

\noindent For these ``trivial" properties, instead of defining the inference rules  

\begin{itemize}
\item HAS(\textit{EV}, $\theta V_1$) $\vdash$ HAS(\textit{EV}, Anytype($\theta V_1$, \dots, $\theta V_n$)), \dots, 
\item LINK(\textit{EV}, $\theta V_1$, $\theta V_2$) $\vdash$ HAS(\textit{EV}, Anytype($\theta V_1$, \dots, $\theta V_n$)), \dots,  
\item LINKUNIQUE(\textit{EV}, $\theta V_{n-1}$, $\theta V_n$) $\vdash$ HAS(\textit{EV}, Anytype($\theta V_1$, \dots, $\theta V_n$)), 
\end{itemize}
we generate the HAS, LINK and LINKUNIQUE facts directly from the architectural actions. 

\begin{figure}[htbp!]
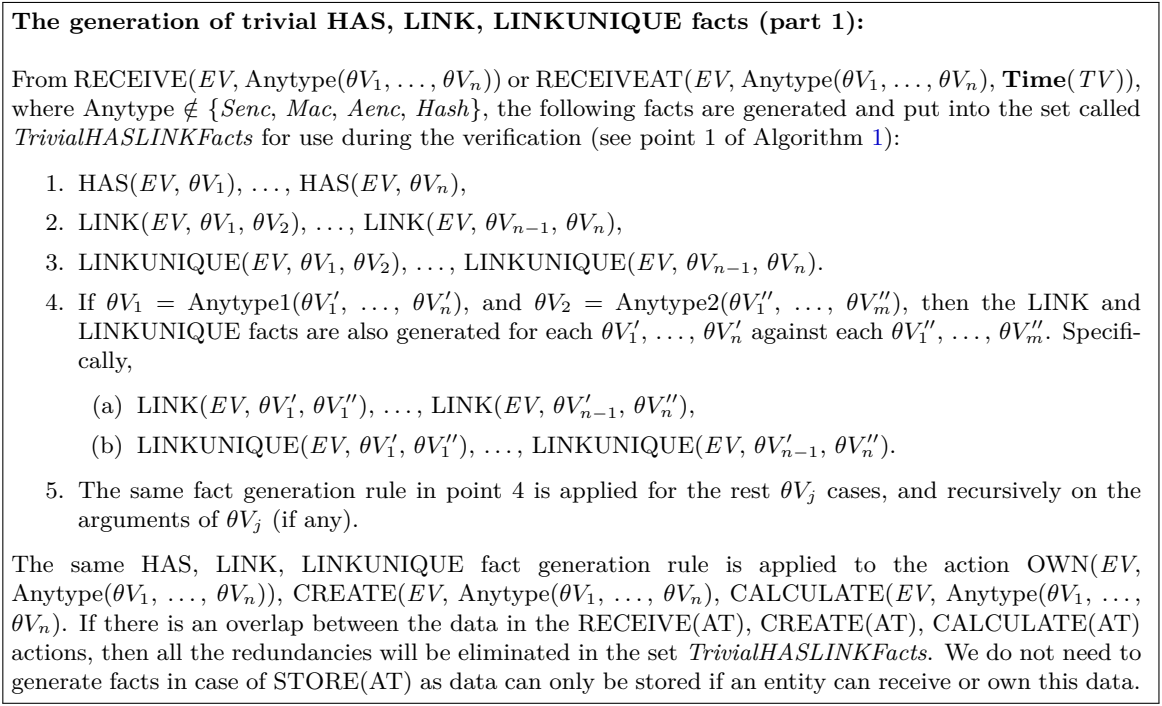

\centering
\fbox{\begin{minipage}{14.87 cm}
\small
\textbf{The generation of trivial HAS, LINK, LINKUNIQUE facts (part 1):}\\

From RECEIVE(\textit{EV}, Anytype($\theta V_1$, \dots, $\theta V_n$)) or RECEIVEAT(\textit{EV}, Anytype($\theta V_1$, \dots, $\theta V_n$), \textbf{Time}(\textit{TV})), where Anytype $\notin$ \{\textit{Senc}, \textit{Mac}, \textit{Aenc}, \textit{Hash}\}, the following facts are generated and put into the set called \textit{TrivialHASLINKFacts} for use during the verification (see point 1 of Algorithm~\ref{alg1}): 

\begin{enumerate}
\item HAS(\textit{EV}, $\theta V_1$), \dots, HAS(\textit{EV}, $\theta V_n$), 

\item LINK(\textit{EV}, $\theta V_1$, $\theta V_2$), \dots, LINK(\textit{EV}, $\theta V_{n-1}$, $\theta V_n$), 

\item LINKUNIQUE(\textit{EV}, $\theta V_1$, $\theta V_2$), \dots, LINKUNIQUE(\textit{EV}, $\theta V_{n-1}$, $\theta V_n$). 

\item If $\theta V_1$ = Anytype1($\theta V'_1$, \dots, $\theta V'_n$), and $\theta V_2$ = Anytype2($\theta V''_1$, \dots, $\theta V''_m$), then the LINK and LINKUNIQUE facts are also generated for each  $\theta V'_1$, \dots, $\theta V'_n$ against each $\theta V''_1$, \dots, $\theta V''_m$. Specifically,  
\begin{enumerate}
\item LINK(\textit{EV}, $\theta V'_1$, $\theta V''_1$), \dots, LINK(\textit{EV}, $\theta V'_{n-1}$, $\theta V''_n$),

\item LINKUNIQUE(\textit{EV}, $\theta V'_1$, $\theta V''_1$), \dots, LINKUNIQUE(\textit{EV}, $\theta V'_{n-1}$, $\theta V''_n$).
\end{enumerate}

\item The same fact generation rule in point 4 is applied for the rest $\theta V_j$ cases, and recursively on the arguments of $\theta V_j$ (if any).  
\end{enumerate}

The same HAS, LINK, LINKUNIQUE fact generation rule is applied to the action OWN(\textit{EV}, Anytype($\theta V_1$, \dots, $\theta V_n$)), CREATE(\textit{EV}, Anytype($\theta V_1$, \dots, $\theta V_n$), CALCULATE(\textit{EV}, Anytype($\theta V_1$, \dots, $\theta V_n$). If there is an overlap between the data in the RECEIVE(AT), CREATE(AT), CALCULATE(AT) actions,  then all the redundancies will be eliminated in the set \textit{TrivialHASLINKFacts}. We do not need to generate facts in case of STORE(AT) as data can only be stored if an entity can receive or own this data.       
 
\end{minipage}
}
\caption{Generate ``trivial" HAS, LINK, LINKUNIQUE facts from the arch. actions.}\label{fig:trivi}
\end{figure}

Formally, the rule for generating the ``trivial" HAS, LINK and LINKUNIQUE facts is given in Figure~\ref{fig:trivi}.   

\normalsize

If a resulted HAS fact from point 1 in Figure \ref{fig:trivi} still contains a compound data type (e.g. in HAS(\textit{EV}, $\theta V_1$), $\theta V_1$ = Anytype$'$($\theta V'_1$, \dots, $\theta V'_n$)), then we recursively generate the ``trivial" HAS, LINK and LINKUNIQUE facts from it. 

\begin{figure}[htbp!]
\centering
\fbox{\begin{minipage}{14.87 cm}
\small
\textbf{The (recursive) generation of trivial HAS, LINK, LINKUNIQUE facts (part 2):}\\

From HAS(\textit{EV}, Anytype$'$($\theta V'_1$, \dots, $\theta V'_n$)), where Anytype$'$ $\notin$ \{\textit{Senc}, \textit{Mac}, \textit{Aenc}, \textit{Hash}\}, the following facts are generated and added into the set \textit{TrivialHASLINKFacts} for use during the verification (see point 1 of Algorithm~\ref{alg1}): 

\begin{enumerate}
\item HAS(\textit{EV}, $\theta V'_1$), \dots, HAS(\textit{EV}, $\theta V'_n$), 

\item LINK(\textit{EV}, $\theta V'_1$, $\theta V'_2$), \dots, LINK(\textit{EV}, $\theta V'_{n-1}$, $\theta V'_n$), 

\item LINKUNIQUE(\textit{EV}, $\theta V'_1$, $\theta V'_2$), \dots, LINKUNIQUE(\textit{EV}, $\theta V'_{n-1}$, $\theta V'_n$). 

\item If $\theta V'_1$ = Anytype1($\theta V^1_1$, \dots, $\theta V^1_n$), and $\theta V'_2$ = Anytype2($\theta V^2_1$, \dots, $\theta V^2_m$), then the LINK and LINKUNIQUE facts are also generated for each  $\theta V^1_1$, \dots, $\theta V^1_n$ against each $\theta V^2_1$, \dots, $\theta V^2_m$. Specifically,  
\begin{enumerate}
\item LINK(\textit{EV}, $\theta V^1_1$, $\theta V^2_1$), \dots, LINK(\textit{EV}, $\theta V^1_{n-1}$, $\theta V^2_n$),

\item LINKUNIQUE(\textit{EV}, $\theta V^1_1$, $\theta V^2_1$), \dots, LINKUNIQUE(\textit{EV}, $\theta V^1_{n-1}$, $\theta V^2_n$).
\end{enumerate}

\item The same fact generation rule in point 4 is applied for the rest $\theta V_j$ cases, and recursively on the arguments of $\theta V_j$ (if any).  
\end{enumerate} 
\end{minipage}
}
\caption{Generate HAS, LINK, LINKUNIQUE facts from a  HAS fact).}\label{fig:trivi2}
\end{figure}
\normalsize

\textbf{Facts generation example}: If RECEIVE(\textit{sp}, \textit{Sicknessrec}(\textit{Personalinfo}(name,address),disease)\footnote{We intentionally do not use any space character between the arguments as this is the way how we need to specify an action in the software tool} $\in$ $\mathcal{P}\mathcal{A}$ (the same is valid for RECEIVEAT), then the following set of HAS, LINK and LINKUNIQUE facts will be generated: 

\begin{center}
\textbf{TrivialHASLINKFacts} =  \{HAS(\textit{sp}, \textit{Personalinfo}(name,address)), HAS(\textit{sp}, disease),	HAS(\textit{sp}, name), HAS(\textit{sp}, address), LINK(\textit{sp}, \textit{Personalinfo}(name, address), disease), LINK(\textit{sp}, name, address), LINK(\textit{sp}, name, disease), LINK(\textit{sp}, address, disease)\} 
\end{center}

\subsection{Proposed Automated Conformance Check Algorithm}

The automated conformance verification is based on the execution of \textit{resolution} steps and backward search.  
Resolution is well-known in logic programming and is widely supported in logic programming languages.  The formal definition of resolution is based on the so-called substitution and unification steps. A substitution binds some value to some variable, and we denote it by $\sigma$ in this paper. 

\begin{ttd}
A substitution $\sigma$ is the most general unifier of a set of facts $\mathbb{F}$ if it unifies $\mathbb{F}$,and for any unifier $\mu$ of $\mathbb{F}$, 
there is a unifier $\lambda$ such that $\mu$ = $\lambda$$\sigma$.  
\end{ttd}


\begin{ttd}
\label{def:res}
Given a goal (fact) $F$, and a rule $R$ $=$ $H$ $\vdash$ $T_1$,\dots, $T_n$ , where $F$ is  unifiable  with $H$ with  the  most  
general  unifier $\sigma$,  then  the  resolution $F\circ_{(F,H)} R$ results in  $T_1$$\sigma$, \dots, $T_n$$\sigma$. 
\end{ttd}


\begin{ttd} \label{def:goalgen}
The function that generates initial (verification) goals is defined as: 
\begin{center}
$\mathbb{G}$ : $\textit{Policy}_{DataType^{sp}_{pol}}$ $\rightarrow$ $\{$\textit{ColG} $\cup$ 
\textit{UseG} $\cup$ \textit{StoreG} $\cup$ \textit{DelG} $\cup$ \textit{TransfG} $\cup$ \textit{HasG} $\cup$ \textit{LinkG}$\}$.
\end{center}
\end{ttd}

$\mathbb{G}$ expects a policy as input and returns a set of seven subsets of goals to be proved in a conformance check. Each subset contains the goals capturing  each sub-policy in Section~\ref{sec:syntaxpp}.   
For a data type $\theta$, we have:  

\begin{center} 
$\mathbb{G}$($\pi_\theta$) $=$ \{$\mathcal{G}^{\theta}_{\textit{col}}$ $\cup$ $\mathcal{G}^{\theta}_{\textit{use}}$ $\cup$ $\mathcal{G}^{\theta}_{\textit{str}}$ $\cup$ $\mathcal{G}^{\theta}_{\textit{del}}$ $\cup$  $\mathcal{G}^{\theta}_{\textit{fw}}$ $\cup$ $\mathcal{G}^{\theta}_{\textit{has}}$ $\cup$ $\mathcal{G}^{\theta}_{\textit{link}}$\} 
\end{center} 

\textbf{The goals generation rules}: In the following, we provide the rules for goals generation based on the specific values of the sub-policies inside $\pi_{\theta}$, namely, ($\pi_{col}$, $\pi_{use}$, $\pi_{str}$, $\pi_{del}$, $\pi_{fw}$, $\pi_{has}$, $\pi_{link}$):  

\small
\begin{tabbing}  
    \=1\=1\=1\=1\= \kill
    1. For $\pi_{\textbf{col}}$ with the collection purpose values \{$cp_1$:$\theta_1'$,\dots, $cp_n$:$\theta_n'$\}, the following verification goals are generated:  \\\\
    \=123\=1\=1\=1\= \kill
		\>\> $\mathcal{G}^{\theta}_{\textbf{col}}$ $=$ $\mathcal{G}^{\theta}_{\textbf{ccons}}$ $\cup$ $\mathcal{G}^{\theta}_{\textbf{cpurp}}$, where $\mathcal{G}^{\theta}_{\textbf{ccons}}$ = \{CCONSENTCOLLECTED(sp, $\theta$)\},\\ 
	\=1234567890123456789012345678901\=1\=1\=1\= \kill
		\>\>  $\mathcal{G}^{\theta}_{\textbf{cpurp}}$ = \{CPURPOSE($\theta_1'$, $cp_1$), \dots, CPURPOSE($\theta_n'$, $cp_n$)\}.\\\\ 
		\=1\=1\=1\=1\= \kill
	\>\>\>\> If \textit{cons} = $N$, then CCONSENTCOLLECTED(sp, $\theta$) $\notin$ $\mathcal{G}^{\theta}_{\textbf{use}}$.\\\\
		2. For $\pi_{\textbf{use}}$ with the usage purpose values \{$up_1$:$\theta_1'$,\dots, $up_n$:$\theta_n'$\}, the following verification goals are generated:\\\\
    \=123\=1\=1\=1\= \kill
		\>\> $\mathcal{G}^{\theta}_{\textbf{use}}$ $=$ $\mathcal{G}^{\theta}_{\textbf{ucons}}$ $\cup$ $\mathcal{G}^{\theta}_{\textbf{upurp}}$, where $\mathcal{G}^{\theta}_{\textbf{ucons}}$ = \{UCONSENTCOLLECTED(sp, $\theta$)\},\\ 
	\=1234567890123456789012345678901\=1\=1\=1\= \kill 
		\>\>  $\mathcal{G}^{\theta}_{\textbf{upurp}}$ $=$ \{UPURPOSE($\theta_1'$, $up_1$), \dots, UPURPOSE($\theta_n'$, $up_n$)\}.\\\\ 
		\=1\=1\=1\=1\= \kill
	\>\>\>\> Again, if the first argument of $\pi_{\textbf{use}}$ is \textit{cons} = $N$, then UCONSENTCOLLECTED(sp, $\theta$) $\notin$ $\mathcal{G}^{\theta}_{\textbf{use}}$.\\\\
		3. For $\pi_{\textbf{str}}$ with the storage place values \{\textit{E}$_1$,\dots, \textit{E}$_n$\}, the next verification goals are generated:  \\\\
    \=123\=1\=1\=1\= \kill
		\>\> $\mathcal{G}^{\theta}_{\textbf{str}}$ $=$ $\mathcal{G}^{\theta}_{\textbf{scons}}$ $\cup$ $\mathcal{G}^{\theta}_{\textbf{places}}$, where $\mathcal{G}^{\theta}_{\textbf{scons}}$ = \{STRCONSENTCOLLECTED(sp,$\theta$)\},\\ 
	\=123\=1\=1\=1\= \kill
		\>\> $\mathcal{G}^{\theta}_{\textbf{places}}$ = \{STORE(\textit{E}$_1$, $\theta$, \textit{EV}$_{\textit{from}}$), \dots, STORE(\textit{E}$_n$, $\theta$, \textit{EV}$_{\textit{from}}$), \dots, STOREAT(\textit{E}$_n$, $\theta$, \textit{EV}$_{\textit{from}}$,  \textbf{Time}(\textit{TT}))\}.\\\\ 
		4. If $\pi_{\textbf{del}}$ = (\{\textit{E}$_1$,\dots, \textit{E}$_n$\}, $dd$), where \textit{E}$_1$,\dots, \textit{E}$_n$ are the values of the deletion places,  and\\
\=1\=1\=1\=1\= \kill
	\>\>\>\> 
		$dd$ is the value of the deletion delay,  then:  \\\\
    \=123\=1\=1\=1\= \kill
		\>\> $\mathcal{G}^{\theta}_{\textbf{\textbf{del}}}$ $=$ $\mathcal{G}^{\theta}_{\textbf{hasupto}}$ $\cup$ $\mathcal{G}^{\theta}_{\textbf{within}}$, where\\
	\>\>  $\mathcal{G}^{\theta}_{\textbf{hasupto}}$ = \{HASUPTO(\textit{E}$_1$, $\theta$, \textbf{Time}(\textit{dd})), \dots, HASUPTO(\textit{E}$_n$, $\theta$, \textbf{Time}(\textit{dd}))\},\\ 
	\=123\=1\=1\=1\= \kill
		\>\> $\mathcal{G}^{\theta}_{\textbf{within}}$ $=$ \{DELETEWITHIN(\textit{E}$_1$, $\theta$, \textit{EV}$_{\textit{from}}$,  \textbf{Time}(\textit{dd})), \dots, DELETEWITHIN(\textit{E}$_n$, $\theta$, \textit{EV}$_{\textit{from}}$, \textbf{Time}(\textit{dd}))\}.\\\\ 
		5. If $\pi_{\textbf{fw}}$ = (\textit{cons}, \{$E_1$,\dots, $E_n$\}, \{\textit{fwp}$_1$:$\theta_1'$,\dots,\ \textit{fwp}$_m$:$\theta_m'$\}), where $E_1$,\dots, $E_n$ are the entities who can \\
		\=1\=1\=1\=1\= \kill
	\>\>\>\> receive the transferred data, and \textit{fwp}$_1$,\dots, \textit{fwp}$_m$ are the transfer purpose values, then:\\\\
   \=123\=1\=1\=1\= \kill
		\>\> $\mathcal{G}^{\theta}_{\textbf{fw}}$ $=$ $\mathcal{G}^{\theta}_{\textbf{fwcons}}$ $\cup$
		$\mathcal{G}^{\theta}_{\textbf{fwto}}$ $\cup$
		$\mathcal{G}^{\theta}_{\textbf{fwpurp}}$, where\\ 
		\=123\=1\=1\=1\= \kill
		\>\> $\mathcal{G}^{\theta}_{\textbf{fwto}}$ = \{RECEIVE($E_1$, $\theta$, \textit{EV}$_{\textit{from}}$), RECEIVE($E_n$, $\theta$,  \textit{EV}$_{\textit{from}}$), \dots, RECEIVEAT($E_n$, $\theta$,  \textit{EV}$_{\textit{from}}$,  \textbf{Time}(\textit{TT}))\},\\ 
		\=123\=1\=1\=1\= \kill
		\>\> $\mathcal{G}^{\theta}_{\textbf{fwcons}}$ = \{FWCONSENTCOLLECTED(sp, $\theta$, $E_1$),\dots, FWCONSENTCOLLECTED(sp, $\theta$, $E_n$)\},\\ 
	\=123\=1\=1\=1\= \kill
		\>\>  $\mathcal{G}^{\theta}_{\textbf{fwpurp}}$ = \{FWPURPOSE($\theta_1'$, \textit{fwp}$_1$), \dots, FWPURPOSE($\theta_n'$, \textit{fwp}$_m$)\}.\\\\ 
		6. For $\pi_{\textbf{has}}$, if \{$E_1$,\dots, $E_n$\} is the set of all defined entities in an architecture, then:\\\\
   \=123\=1\=1\=1\= \kill
		\>\> $\mathcal{G}^{\theta}_{\textbf{has}}$ $=$ \{HAS($E_1$, $\theta$),\dots,  HAS($E_n$, $\theta$)\}.\\\\ 		
		7. For $\pi_{\textbf{link}}$, if \{$E_1$,\dots, $E_n$\} is the set of all defined entities in an architecture, and \{$\theta_1$,\dots, $\theta_m$\} is\\ 
		\=123\=1\=1\=1\= \kill
		\>\> a set of all defined data types (different from $\theta$), then : \\\\ 
 
   \=123\=1\=1\=1\= \kill
		\>\> $\mathcal{G}^{\theta}_{\textbf{link}}$ $=$ \{LINK($E_1$, $\theta$,  $\theta_1$), LINK($E_1$, $\theta_1$, $\theta$), \dots, LINK($E_n$, $\theta$, $\theta_n$), \dots, LINKUNIQUE($E_n$, $\theta_m$, $\theta$)\}.\\	  	
\end{tabbing}
\normalsize

Finally, let us denote the set of all goals to be proved during a conformance verification by $\mathbb{A}\mathbb{G}$, namely:
\begin{center}
 $\mathbb{A}\mathbb{G}$ $=$ $\bigcup_{\forall \theta \in DataTypes^{sp}_{pol}}$ $\mathbb{G}$($\pi_\theta$), 
\end{center} 
where $DataTypes^{sp}_{pol}$ is a set of all data types defined in the policy for a service provider \textit{sp}.

\textbf{The generation of purpose-facts in architectures}: Besides the actions defined in Figure~\ref{fig:sysarch}, to verify the DPR conformance regarding the (collection, usage, or forward) purposes, the so-called purpose-facts are generated. This is based on the following purposes-fact generation rules, for a given architecture $\mathcal{P}$$\mathcal{A}$:

\begin{enumerate}
\item If \textit{CREATEAT}($E$, $X_{\theta}$, \textbf{Time}(\textit{TT})) $\in$ $\mathcal{P}$$\mathcal{A}$, then CPURPOSE($\theta$, createat) $\in$ \textit{CPurpSet}.  

\item If \textit{CALCULATEAT}($E$, $X_{\theta}$, \textbf{Time}(\textit{TT})) $\in$ $\mathcal{P}$$\mathcal{A}$, then UPURPOSE($\theta$, calculateat) $\in$ \textit{UPurpSet}. 

\item If \textit{RECEIVEAT}($E$, \textbf{Fwconsent}($X_{\theta}$,$E_{to}$), \textbf{Time}(\textit{TT})) $\in$ $\mathcal{P}$$\mathcal{A}$, and \textit{CREATEAT}($E_{to}$, $X_{\theta}$, \textbf{Time}(\textit{TT})) $\in$ $\mathcal{P}$$\mathcal{A}$, then FWPURPOSE($\theta$, createat) $\in$ \textit{FwPurpSet}.

\item If \textit{RECEIVEAT}($E$, \textbf{Fwconsent}($X_{\theta}$,$E_{to}$), \textbf{Time}(\textit{TT})) $\in$ $\mathcal{P}$$\mathcal{A}$, and \textit{CALCULATEAT}($E_{to}$, $X_{\theta}$, \textbf{Time}(\textit{TT})) $\in$ $\mathcal{P}$$\mathcal{A}$, then FWPURPOSE($\theta$, calculateat) $\in$ \textit{FwPurpSet}. 
\end{enumerate}

These rules define how the facts for the collection (point 1), usage (point 2), and transfer (points 3-4) purposes are generated from the architectural actions, and added into the sets \textit{CPurpSet}, \textit{UPurpSet}, and  \textit{FwPurpSet}, respectively, to be used in Algorithm~\ref{alg1}. 

To speed up the verification process, the actions defined in an architecture are divided into four subsets, specifically, \textit{ArchTime}, \textit{ArchPseudo}, \textit{ArchMeta}, and \textit{Arch}. \textit{ArchTime} includes the actions that contain the \textbf{Time}() construct, \textit{ArchPseudo} includes the actions that contain the \textbf{P}() construct for pseudonym, \textit{ArchMeta} includes the actions that contain the \textbf{Meta}() construct for metadata, and finally, \textit{Arch} is a set of actions without any specific construct above (see rules  \textit{P15}-\textit{P18}). 

Finally, if the set of unique data types\footnote{Unique data types are types that can be used to uniquely identify a living individual, e.g. passport numbers.} defined in the policy is \{$\theta_1$,\dots, $\theta_n$\},  \{$\theta_1$,\dots, $\theta_n$\} $\subseteq$ \textit{DataType}$^{sp}_{pol}$, then we have the corresponding set of facts, \textit{UniqueTypes}, which can be used to prove the unique linkability properties (see rule \textit{U1} in Figure~\ref{fig:inf3}): 

\begin{center}
\textit{UniqueTypes} = \{UNIQUE($\theta_1$), \dots, UNIQUE($\theta_n$)\}.
\end{center}    

Let us define the following rule sets that we will use in the inference algorithms, namely: 
\begin{itemize}
\item \textit{DPRRules} $=$ \{$D1$,\dots, $D5$, $D6$, $D7$\}, 
\item \textit{HasUpToRules} $=$ \{$P1$, $P2$\}, 
\item \textit{HasRules} $=$ \{$P3$, \dots, $P18$\},  
\item \textit{CryptHasRules} $=$ \{$C1$, $C2$, $C3$\},
\item \textit{LinkRules} $=$ \{$L0$, $L1$ (inc. L1/b-L1/d), \dots, $L8$ (inc. L8/b-L8/d)\} , and  
\item \textit{LinkUniqueRules} $=$ \{$U1$ (inc. U1/b-U1/d), \dots, $U8$ (inc. U8/b-U8/d)\}.
\end{itemize}

Algorithm~\ref{alg1} defines the process of checking whether the input architecture \textit{Architecture} is fulfilling the ``initial" goal, \textit{initgoal}, and  returns either $1$ if the proof is successful, or $0$ if failed.

\hfill \break

\small
\begin{algorithm}[H]
\SetAlgoLined
\tcc{(* Backward search strategy *)}
\tcc{\textbf{Note}: If the proof has failed for \textit{initgoal} for an (original) entity $E$, the algorithm will attempt the proof \textit{initgoal} in which $E$ is replaced with the entities in the set \textbf{HasAccessTo}($E$).}
\KwResult{Proof found (1) /Proof not found (0) \ \ \ \ \ \ \ \ \ \ \ \ \ \ \ \ \ \  (* see Table~\ref{tab:notation3} for the used notations *)}
 \textbf{Inputs}:  \\
 1. Rulesets = \\ \ \ \ \ \ \ \{\textit{DPRRules}, \textit{HasUpToRules},  \textit{HasRules}, \textit{CryptHasRules}, \textit{LinkRules}, \textit{LinkUniqueRules}, \textit{TrivialHASLINKFacts}\}.\\ 
 2. Architecture = \{\textit{ArchTime}, \textit{ArchPseudo}, \textit{ArchMeta}, \textit{Arch}\}. \\ 
 3. ArchPurposes = \{\textit{CPurpSet}, \textit{UPurpSet}, \textit{FwPurpSet}\}.\\
 4. UniqueTypes.\\
 6. Goal: \textit{initgoal}, where \textit{initgoal} $\in$ $\mathbb{A}\mathbb{G}$.\\
 7. Allowed layers of nested crypto functions: $N$.\\
 \eIf{\textit{initgoal} $\in$ $\mathcal{G^{\theta}}_{\textbf{places}}$ $\cup$ $\mathcal{G^{\theta}}_{\textbf{within}}$ $\cup$ $\mathcal{G^{\theta}}_{\textbf{fwto}}$}{
  	  		\For{\textit{arch} in \textit{Architecture}}{ 
  	  \If{(\textit{initgoal}\ $\circ_{(\textit{initgoal},\textit{arch})}$ \textit{arch}) is successful \textbf{or} (\textit{initgoal} == \textit{arch})}{
  	  		\textbf{return} 1
  	  }
  	  \textbf{return} 0
  	 }
  	  		}{
  	  		\eIf{the predicate of \textit{initgoal} matches the predicate of a purpose-fact in \textit{AP}, \textit{AP} $\in$ ArchPurposes}{
 \For{\textit{purp} in \textit{AP}}{ 
  	  \If{(\textit{initgoal}\ $\circ_{(\textit{initgoal},\textit{purp})}$ \textit{purp}) is successful \textbf{or} (\textit{initgoal} == \textit{purp})}{
  	  		\textbf{return} 1
  	  }
  	  \textbf{return} 0
  	 }
 
 }{
 \eIf{\textbf{VerifyAgainstRuleset}(initgoal, Architecture, UniqueTypes, Rulesets, N) == 1}{ 
   \textbf{return} 1
 }{
  \textbf{return} 0
 }
 } 
  	  		}
 
 \caption{\textbf{ConformanceCheck}(initgoal, Architecture, Rulesets, N)}
 \label{alg1}
\end{algorithm}
\normalsize


\small
\begin{algorithm}[H]
\SetAlgoLined
 \If{the predicate of \textbf{\textit{goal}} matches the predicate of a head of a rule in RS, RS $\in$ Rulesets}{ 
 \For{\textit{rule} in \textit{RS}}{
 \textit{isSuccessful}[(rule, goal)] = \textbf{VerifyRule}(\textit{rule, goal, Architecture, UniqueTypes, Rulesets, N})
 	}
 	\eIf{for all rule in RS: \textit{isSuccessful}[(rule, goal)] == 0}{
  \textbf{return} 0
  }{
  \textbf{return} 1
  } 
 }{
 
 }
 
 \caption{\textbf{VerifyAgainstRuleset}(goal, Architecture, UniqueTypes, Rulesets, N)}
 \label{alg3}
\end{algorithm}
\normalsize

\hfill \break

\small
\begin{algorithm}[H]
\SetAlgoLined
 
  	\For{\textit{unique} in \textit{UniqueTypes}}{ 
  	  \eIf{(\textit{goal}\ $\circ_{(\textit{goal},\ \textit{unique})}$ \textit{unique}) is successful \textbf{or} (\textit{goal} == \textit{unique})}{
  	  \tcc{A dictionary entry with the key of (\textit{rule}, \textit{goal}, \textit{unique}). The proof of goal  with arch was successful.}
  	  		\textit{Derivation\_Unique\_Successful[(rule,goal,unique)]} = 1
  	  }{
  	  \tcc{The proof of goal with unique and rule has failed.}  
  	 	 \textit{Derivation\_Unique\_Successful[(rule, goal, unique)]} = 0 
  	  } 
  	  
  	 }
  \eIf{for all unique in \textit{UniqueTypes}: \textit{Derivation\_Unique\_Successful[(rule, goal, unique)]} == 0}{
  \textbf{return} 0
  }{
  \textbf{return} 1
  }  
  
 \caption{\textbf{VerifyUniqueTypes}(\textit{rule, goal, UniqueTypes})}
 \label{algu}
\end{algorithm}
\normalsize

\small
\begin{algorithm}[H]
\SetAlgoLined
 
  	\For{\textit{arch} in \textit{AS}}{ 
  	  \eIf{(\textit{goal}\ $\circ_{(\textit{goal},\ \textit{arch})}$ \textit{arch}) is successful with $\sigma_{\textit{arch}}$ \textbf{or} (\textit{goal} == \textit{arch})}{
  	  	\tcc{A dictionary entry with the key of (\textit{rule}, \textit{goal}, \textit{arch}). The proof of goal  with arch was successful.}
  	  		\textit{Derivation\_Arch\_Successful[(rule, goal, arch)]} = 1   \\
  	  		\tcc{$\sigma_{\textit{arch}}$ is the mapping that proves \textit{goal} with \textit{arch}. Several \textit{arch}s can prove \textit{goal} with different $\sigma_{\textit{arch}}$-s. \textit{Mappings[goal]} contains all $\sigma_{\textit{arch}}$-s that prove goal.} 
  	  		\textbf{add} the mapping $\sigma_{\textit{arch}}$ into the set \textit{Mappings[goal]} \\
  	  }{
  	  \tcc{The proof of goal  with arch and rule has failed.}
  	 	 \textit{Derivation\_Arch\_Successful[(rule, goal, arch)]} = 0\\
  	  } 
  	  
  	 }
  
  \eIf{for all arch in AS: \textit{Derivation\_Arch\_Successful[(rule,goal,arch)]} == 0}{
  \textbf{return} 0
  }{
  \textbf{return} 1
  }  
  
 \caption{\textbf{VerifyAgainstArch}(\textit{rule, goal, AS})}
 \label{alg4}
\end{algorithm}
\normalsize


\hfill \break

 Algorithm~\ref{alg2} defines a  verification process of \textit{initgoal} via the sub-goals resulted from the resolution steps.  
 
 \hfill \break
 
\small
\begin{algorithm}[H]
\SetAlgoLined
\tcc{In this case, there is no \textit{previousgoal} right before \textit{nextgoal}. Hence, the proof will be attempted on \textit{nextgoal} instead of \textit{nextgoal}$\sigma$ (like in Algorithm~\ref{alg2}).}
  	  		\eIf{\textit{nextgoal} is an action, and matches the Time/P/Meta construct in AS, AS $\in$ Architecture}{
  	  		 \eIf{\textbf{VerifyAgainstArch}(\textit{rule, \textit{nextgoal}, AS}) ==1}{ 
  	  		 \tcc{The proof of \textit{nextgoal} with rule and the subset AS was successful.}
  	  		 \textit{isSuccessful}[(rule, nextgoal)] = 1
  	  		}{
  	  		 \tcc{The proof of \textit{nextgoal} with rule and the subset AS failed.}
  	  		\textit{isSuccessful}[(rule, nextgoal)] = 0
  	  		}
  	  		}{
  	  		\eIf{the predicate of \textbf{\textit{nextgoal}} matches a fact in UniqueTypes}{
  	  		 \eIf{\textbf{VerifyUniqueTypes}(\textit{rule, nextgoal, UniqueTypes}) ==1}{  \textit{isSuccessful}[(rule, nextgoal)] = 1
  	  		}{
  	  		\textit{isSuccessful}[(rule, nextgoal)] = 0
  	  		}
  	  		}{
  	  		\eIf{\textbf{VerifyAgainstRuleset}(\textit{nextgoal, Architecture, UniqueTypes, Rulesets, N}) == 1}{
  	  		\textit{isSuccessful}[(rule, nextgoal)] = 1
  	  		}{
  	  		\textit{isSuccessful}[(rule, nextgoal)] = 0
  	  		}
  	  		}
  	  		}

 \caption{ \textbf{CaseNoPreviousGoal}(rule, nextgoal, Architecture, UniqueTypes, Rulesets, N)}
 \label{alg6}
\end{algorithm}
\normalsize

\hfill \break

\small
\begin{algorithm}[H]
\SetAlgoLined
 \tcc{Note: The variable arguments in the inference rules are renamed before they are used in a resolution.}
  \textit{GoalsToBeProved} = \{\textit{goal}\}\;  
  	  \If{\textit{goal}\ $\circ_{(\textit{goal},\ head\ of\ rule)}$ \textit{rule} is successful}{
  	   \tcc{Check for the limit of nested layers of crypto functions.}
  	  	\eIf{$\exists$ fact in $($\textit{goal}\ $\circ_{(\textit{goal},\ head\ of\ rule)}$ \textit{rule}$)$ that contains more than $N$ nested layers of crypto functions \textbf{and} \textit{rule} $\in$ \{\textit{P8}, \textit{P9}, \textit{P10}\}}{
  	  	\textbf{return} 0\;
  	  	}{
  	  	\textbf{remove} \textit{goal} from \textit{GoalsToBeProved} \;
  	  		\textbf{add} \textit{the facts} in $($\textit{goal}\ $\circ_{(\textit{goal},\ head\ of\ rule)}$ \textit{rule}$)$ to the start of \textit{GoalsToBeProved}\;
  	  		\For{\textit{nextgoal} in \textit{GoalsToBeProved}}{ 
  	  		\tcc{For all mappings ($\sigma$) that can be used to prove \textit{previousgoal}  (Algorithm~\ref{alg4}).}
  	  		\eIf{there exists \textbf{\textit{previousgoal}} examined just before \textbf{\textit{nextgoal}} in \textit{GoalsToBeProved}}{
  	  		\For{$\sigma$ in \textit{Mappings[previousgoal]}}{    
  	  		\eIf{\textit{nextgoal}$\sigma$ is an action in AS, AS $\in$ Architecture}{ 
  	  		 \eIf{\textbf{VerifyAgainstArch}(\textit{rule, \textit{nextgoal}$\sigma$, AS}) ==1}{ 
  	  		 \tcc{The proof of \textit{nextgoal} with rule and $\sigma$ was successful.}
  	  		 \textit{isSuccessfulMapping}[(rule, nextgoal, $\sigma$)] = 1
  	  		}{
  	  		 \tcc{The proof of \textit{nextgoal} with rule and $\sigma$ was unsuccessful.}
  	  		\textit{isSuccessfulMapping}[(rule, nextgoal, $\sigma$)] = 0
  	  		}
  	  		}{
  	  		\eIf{the predicate of \textbf{\textit{nextgoal}$\sigma$} matches a fact in UniqueTypes}{
  	  		 \eIf{\textbf{VerifyUniqueTypes}(\textit{rule, nextgoal$\sigma$, UniqueTypes}) ==1}{  \textit{isSuccessfulMapping}[(rule, nextgoal, $\sigma$)] = 1
  	  		}{
  	  		\textit{isSuccessfulMapping}[(rule, nextgoal, $\sigma$)] = 0
  	  		}
  	  		}{
  	  		\eIf{\textbf{VerifyAgainstRuleset}(\textit{nextgoal$\sigma$, Architecture, UniqueTypes, Rulesets, N}) == 1}{
  	  		\textit{isSuccessfulMapping}[(rule, nextgoal, $\sigma$)] = 1
  	  		}{
  	  		\textit{isSuccessfulMapping}[(rule, nextgoal, $\sigma$)] = 0
  	  		}
  	  		}
  	  		}
  	  		}
  	  		}{\textbf{CaseNoPreviousGoal}(rule, nextgoal, Architecture, UniqueTypes, Rulesets, N)}
  	  		\eIf{for all $\sigma$ in \textit{Mappings[previousgoal]}: \textit{isSuccessfulMapping}[(rule, nextgoal, $\sigma$)] == 0}{
  \textit{isSuccessful}[(rule, nextgoal)] == 0
  }{
  \textit{isSuccessful}[(rule, nextgoal)] == 1
  } 
  	  		}
  	  		\eIf{for all \textit{nextgoal} in \textit{GoalsToBeProved}: \textit{isSuccessful}[(rule, nextgoal)] == 1}{
  \textbf{return} 1
  }{
  \textbf{return} 0
  } 
  	  		}
  	  } 
  
 \caption{ \textbf{VerifyRule}(\textit{rule, goal, Architecture, UniqueTypes, Rulesets, N})}
 \label{alg2}
\end{algorithm}
\normalsize

\hfill \break

\textbf{Algorithm Explanation}. Algorithm~\ref{alg1} expects as input the set of inference rules (\textit{Rulesets}, which also contains a set of ``trivial" HAS, LINK, LINKUNIQUE facts generated from the architectural actions (\textit{TrivialHASLINKFacts})), a set of facts that capture the actions in an architecture (\textit{Architecture}), a set of purposes defined in an architecture (\textit{ArchPurposes}), a set of unique data types (\textit{UniqueTypes}),  and a verification goal, \textit{initgoal}. $N$ is a defined number that denotes the maximum layers of nested cryptographic functions in a piece of data that the verification engine examines. A finite $N$ is used to ensure the termination of the proof process.

\begin{enumerate}
\item First of all, if \textit{initgoal} $\in$ $\mathcal{G^{\theta}}_{\textbf{places}}$ $\cup$ $\mathcal{G^{\theta}}_{\textbf{within}}$ $\cup$ $\mathcal{G^{\theta}}_{\textbf{fwto}}$   (see points 3-5 in Definition~\ref{def:goalgen}), then we check whether \textit{inigoal} can be unified with or equal to a fact in \textit{Architecture}. The algorithm returns $1$ if the proof was successful, and  $0$ otherwise.  

\item If \textit{initgoal} is not an action fact, then we check if the (collection, usage, or transfer) purposes in an architecture is in line with the policy, namely, whether \textit{initgoal} is in \textit{ArchPurposes}. The algorithm returns $1$ if the proof was successful, and  otherwise, $0$. 
 
\item If \textit{initgoal} is not a purpose-fact (e.g. \textit{initgoal} = HAS(sp, name)), then we try to prove it using the inference rule set and the given architecture. If a proof or a derivation was found for  \textit{initgoal}, then $1$ is returned, otherwise, $0$.  

\item In \textbf{VerifyRule}(\textit{rule, goal, Architecture, UniqueTypes, Rulesets, N}), inside  algorithm~\ref{alg3}, we attempt to carry out resolution steps between \textit{initgoal} and each rule in an appropriate \textit{RS}, \textit{RS} $\in$ \textit{Rulesets}. If the proof has failed for all rules in \textit{RS}, then $0$ is returned (proof failed). Otherwise, if at least one rule can be used to prove the goal, then $1$ is returned.     

\item In  algorithm~\ref{alg2}, a step \textit{goal} $\circ_{(\textit{goal},\   \textit{head of rule})}$ \textit{rule} can be successful or unsuccessful (in case there is no unifier $\sigma$ for \textit{goal} and the head of \textit{rule}). This step  results in  the new (sub-)goals to be proved. If there is a new (sub-)goal that contains more than $N$ layers of nested cryptographic functions (\textit{Senc}, \textit{Aenc}, \textit{Mac}, \textit{Hash}), then we return $0$, and this ``branch" of the proof was unsuccessful\footnote{A proof can be seen as a derivation tree, with \textit{initgoal} in the root and the facts in \textit{Architecture} are the leaves.}. If there is a new (sub-)goal which corresponds to an architectural action, then we attempt to prove it using the facts in \textit{Architecture}.  

\item Algorithm~\ref{alg4} specifies a proof attempt using the (action) facts in \textit{Architecture}. If there is no matching action for a goal, then this branch of the proof was unsuccessful. Otherwise, this branch of the proof has been successful.  

\item Finally, algorithm~\ref{algu} checks \textit{goal} against the set  \textit{UniqueTypes}. If there is no matching, then this branch of the proof was unsuccessful. Otherwise, this branch of the proof has been successful. 
\end{enumerate}
  

\textbf{Example 1}. Let \textit{Architecture} = \{RECEIVEAT(\textit{sp}, \textit{name}, \textit{client}, \textbf{Time}(\textit{TT}))\} and \textit{initgoal} = HAS(\textit{sp}, \textit{name}), namely, we want to prove that \textit{sp} can have \textit{name}. This can be proven with rule \textit{P4} in Figure~\ref{fig:inf2} and a resolution step in Definition~\ref{def:res}. 

\begin{figure}[htb!]
    \begin{center}
        \includegraphics[width=0.7\textwidth]{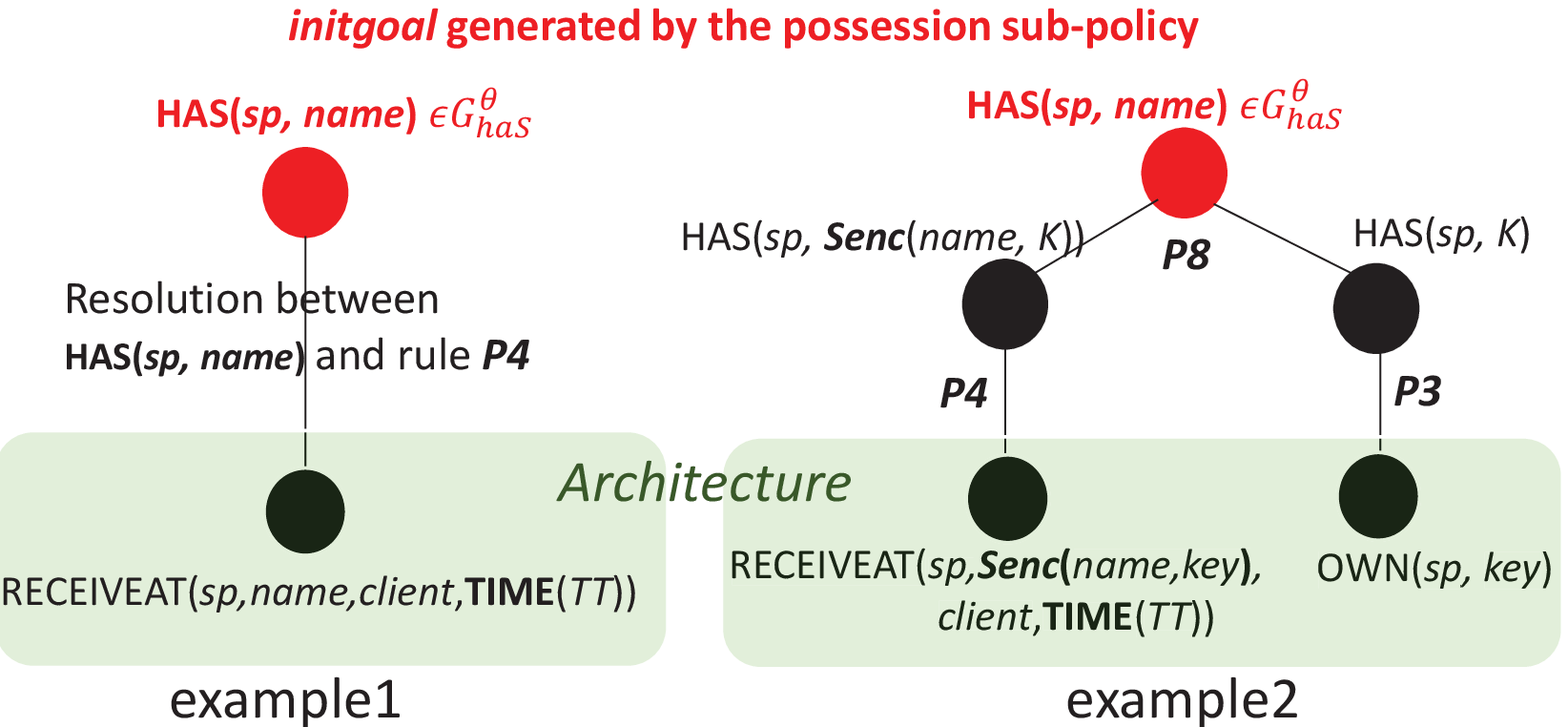}
    \end{center}
    \caption{Two example proofs (without and with encryption, respectively).}
    \label{fig:proof}
\end{figure}

\begin{itemize}
\item  \textbf{Step 1}: \textit{initgoal} $\circ_{(initgoal,\ HAS(\textit{EV}, \theta V))}$ \textit{P4} = RECEIVEAT(\textit{sp}, \textit{name}, \textit{client}, \textbf{Time}(\textit{TT}))), as \textit{initgoal} can be unified with HAS(\textit{EV}, $\theta V$), the head of rule \textit{P4}, with the unifier $\sigma$ = \{\textit{EV} $\mapsto$ \textit{sp}, $\theta V$  $\mapsto$ \textit{name}, \textit{EV}$_{\textit{from}}$ $\mapsto$ \textit{client}, \textit{TV} $\mapsto$ \textit{TT}\}. We have RECEIVEAT(\textit{EV}, $\theta V$, $E_{\textit{from}}$, \textbf{Time}(\textit{TV}))$\sigma$ as a result, which is equal to RECEIVEAT(\textit{sp}, \textit{name}, \textit{client}, \textbf{Time}(\textit{TT}))). 

\item \textbf{Step 2}: As RECEIVEAT(\textit{sp}, \textit{name}, \textit{client}, \textbf{Time}(\textit{TT}))) $\in$ \textit{Architecture}, therefore, we get \textbf{ConformanceCheck}(\textit{initgoal, Architecture, Rulesets, N}) == 1, for any natural $N$. 
\end{itemize}
 
\textbf{Example 2}. Let \textit{Architecture} = \{RECEIVEAT(\textit{sp}, \textbf{Senc}(\textit{name},key), \textit{client}, \textbf{Time}(\textit{TT})), OWN(sp, key)\} and \textit{initgoal} = HAS(\textit{sp}, \textit{name}). This can be proven with rules \textit{P8}, then \textit{P3}, \textit{P4} as shown in Figure~\ref{fig:proof}.

\subsubsection{Properties}

\begin{ttp} (Correctness)\\
\label{ttp:1}
We distinguish several cases based on the value of \text{initgoal}: 
\begin{enumerate}
\item If \textit{initgoal} $\in$ \{HAS(\textit{E}, $\theta$), HASUPTO(\textit{E}, $\theta$, \textbf{Time}(dd))\}, and $E$ $\in$ $\pi_{\theta}$.$\pi_{has}$ at the policy level, then whenever \textbf{ConformanceCheck}(\textit{initgoal, Architecture, Rulesets, N}) == 1, Architecture functionally conforms with this requirement of the policy. 

\item If \textit{initgoal} $\in$ \{HAS(E, $\theta$), HASUPTO(E, $\theta$, \textbf{Time}(dd))\}, and $E$ $\notin$ $\pi_{\theta}$.$\pi_{has}$, then whenever \textbf{ConformanceCheck}(\textit{initgoal, Architecture, Rulesets, N}) == 1, Architecture does not privacy conform with the policy. 

\item If \text{initgoal} $\in$ $\mathcal{G}^\theta_{link}$ and $(E_i,\theta_i)$ $\in$ $\pi_{\theta}$.$\pi_{link}$, then whenever \textbf{ConformanceCheck}(\textit{initgoal, Architecture, Rulesets, N}) == 1, the architecture functionally conforms with this link policy.  

\item If \text{initgoal} $\in$ $\mathcal{G}^\theta_{link}$ and $(E_i,\theta_i)$ $\notin$ $\pi_{\theta}$.$\pi_{link}$, then whenever \textbf{ConformanceCheck}(\textit{initgoal, Architecture, Rulesets, N}) == 1, the architecture does not  privacy conform with the policy. 

\item If \text{initgoal} $\in$ $\mathcal{G}^\theta_{ccons}$ $\cup$ $\mathcal{G}^\theta_{ucons}$ $\cup$ $\mathcal{G}^\theta_{scons}$ $\cup$ $\mathcal{G}^\theta_{fwcons}$, and $\pi_{col}$.\textit{cons} = $Y$, $\pi_{use}$.\textit{cons} = $Y$, $\pi_{str}$.\textit{cons} = $Y$, or $\pi_{fw}$.\textit{cons} = $Y$ in $\pi_{\theta}$, respectively, then the architecture DPR conforms with the actual sub-policy whenever \textbf{ConformanceCheck}(\textit{initgoal, Architecture, Rulesets, N}) == 1.   

\item If \text{initgoal} = CPURPOSE($\theta$, \textit{cp})  (i.e. \text{initgoal} $\in$ $\mathcal{G}^\theta_{cpurp}$), and (\textit{cp}:$\theta$ $\in$ $\pi_{use}$.\textit{cpurp}), then whenever \textbf{ConformanceCheck}(\textit{initgoal, Architecture, Rulesets, N}) == 1, the architecture functionally conforms with the policy.  

\item If \text{initgoal} = CPURPOSE($\theta$, \textit{cp})  (i.e. \text{initgoal} $\in$ $\mathcal{G}^\theta_{cpurp}$), and (\textit{cp}:$\theta$ $\notin$ $\pi_{use}$.\textit{cpurp}), then whenever \textbf{ConformanceCheck}(\textit{initgoal, Architecture, Rulesets, N}) == 1, the architecture does not DPR conform with the policy.  

\item If  \text{initgoal} = UPURPOSE($\theta$, \textit{up}) (i.e. \text{initgoal} $\in$ $\mathcal{G}^\theta_{upurp}$), and (\textit{up}:$\theta$ $\in$ $\pi_{use}$.\textit{upurp}), then whenever \textbf{ConformanceCheck}(\textit{initgoal, Architecture, Rulesets, N}) == 1, the architecture functionally conforms with the policy. 

\item If  \text{initgoal} = UPURPOSE($\theta$, \textit{up}) (i.e. \text{initgoal} $\in$ $\mathcal{G}^\theta_{upurp}$), and (\textit{up}:$\theta$ $\notin$ $\pi_{use}$.\textit{upurp}), then whenever \textbf{ConformanceCheck}(\textit{initgoal, Architecture, Rulesets, N}) == 1, the architecture does not DPR conform with the policy. 

\item If \text{initgoal} = FWPURPOSE($\theta$, \textit{fwp}) (i.e. \text{initgoal} $\in$ $\mathcal{G}^\theta_{fwpurp}$), and  (\textit{fwp}:$\theta$ $\in$ $\pi_{fw}$.\textit{fwpurp}),  then whenever \textbf{ConformanceCheck}(\textit{initgoal, Architecture, Rulesets, N}) == 1, the architecture functionally conforms with the policy.  

\item If \text{initgoal} = FWPURPOSE($\theta$, \textit{fwp}) (i.e. \text{initgoal} $\in$ $\mathcal{G}^\theta_{fwpurp}$), and  (\textit{fwp}:$\theta$ $\notin$ $\pi_{fw}$.\textit{fwpurp}),  then whenever \textbf{ConformanceCheck}(\textit{initgoal, Architecture, Rulesets, N}) == 1, the architecture does not DPR conform with the policy.   

\item If  \text{initgoal} = \{STORE(\textit{E}, $\theta$, \textit{EV}$_{\textit{from}}$),  STOREAT(\textit{E}, $\theta$, \textit{EV}$_{\textit{from}}$,  \textbf{Time}(\textit{TT}))\} (i.e. \text{initgoal} $\in$ $\mathcal{G}^\theta_{places}$), and  \textbf{(\textit{E} $\in$  $\pi_{str}$.\textit{where})}, then whenever \textbf{ConformanceCheck}(\textit{initgoal, Architecture, Rulesets, N}) == 1, the architecture functionally conforms with the policy.  

\item If  \text{initgoal} = \{STORE(\textit{E}, $\theta$, \textit{EV}$_{\textit{from}}$),  STOREAT(\textit{E}, $\theta$, \textit{EV}$_{\textit{from}}$,  \textbf{Time}(\textit{TT}))\} (i.e. \text{initgoal} $\in$ $\mathcal{G}^\theta_{places}$), and  \textbf{(\textit{E} $\notin$  $\pi_{str}$.\textit{where})}, then whenever \textbf{ConformanceCheck}(\textit{initgoal, Architecture, Rulesets, N}) == 1, the architecture does not DPR conform with the policy.  

\item If \text{initgoal} = DELETEWITHIN(\textit{E}, $\theta$, \textit{EV}$_{\textit{from}}$,  \textbf{Time}(\textit{dd})) (i.e. \text{initgoal} $\in$ $\mathcal{G}^\theta_{within}$) and \textbf{(\textit{dd} $\leq$ $\pi_{del}$.\textit{deld})} and \textbf{(\textit{E} $\in$  $\pi_{del}$.\textit{fromwhere})}, then whenever \textbf{ConformanceCheck}(\textit{initgoal, Architecture, Rulesets, N}) == 1, the architecture functionally conforms with the policy.  

\item If \text{initgoal} = DELETEWITHIN(\textit{E}, $\theta$, \textit{EV}$_{\textit{from}}$,  \textbf{Time}(\textit{dd})) (i.e. \text{initgoal} $\in$ $\mathcal{G}^\theta_{within}$) and \textbf{(\textit{dd} $\geq$ $\pi_{del}$.\textit{deld})} and \textbf{(\textit{E} $\in$  $\pi_{del}$.\textit{fromwhere})}, then whenever \textbf{ConformanceCheck}(\textit{initgoal, Architecture, Rulesets, N}) == 1, the architecture does not DPR conform with the policy.   

\item If \text{initgoal} = RECEIVE($E$, $\theta$, \textit{EV}$_{\textit{from}}$) (\text{initgoal} $\in$ $\mathcal{G}^\theta_{fwto}$), and $E$ $\in$ $\pi_{fw}$.\textit{fwto}, then whenever \textbf{ConformanceCheck}(\textit{initgoal, Architecture, Rulesets, N}) == 1, the architecture functionally conforms with the policy.  

\item If \text{initgoal} = RECEIVE($E$, $\theta$, \textit{EV}$_{\textit{from}}$) (\text{initgoal} $\in$ $\mathcal{G}^\theta_{fwto}$), and $E$ $\notin$ $\pi_{fw}$.\textit{fwto}, then whenever \textbf{ConformanceCheck}(\textit{initgoal, Architecture, Rulesets, N}) == 1, the architecture does not DPR conform with the policy.  

\end{enumerate}
\end{ttp}

\begin{proof}

\textbf{ConformanceCheck}(\textit{initgoal, Architecture, Rulesets, N}) == 1 means that a proof of \textit{initgoal} can be found with \textit{Architecture}. Whenever \textit{initgoal} can be proved with a rule \textit{rule} = $H$ $\vdash$ $T_1$,\dots,$T_n$ (in Algorithm~\ref{alg2}), there is at least one fact in \textit{Architecture}  that can be used to prove the sub-goals $T_1$,\dots, $T_n$.  Besides, since \textit{Data} includes the entity who originally sent it, i.e. \textit{Data} = ($\theta V$, \textit{EV}$_{\textit{from}}$)
, we can avoid that in the rules for consent collections (e.g. D1-D2), the consent contains a different data from the one can be received.   In addition, in rules \textit{D1}-\textit{D7}, \textit{P1}, \textit{P4}-\textit{P5}, \textit{P11}, and \textit{P15}-\textit{P16}, \textit{Data} is a pair of a data type and the entity who originally sent it, i.e. \textit{Data} = ($\theta V$, \textit{EV}$_{\textit{from}}$), which can be differentiated from the other data pairs.       

Therefore, in case of points 1 and 3 (of Property~\ref{ttp:1}),  the first two points of Definition~\ref{def:func} are satisfied, respectively. In case of points 2 and 4, the two points of Definition~\ref{def:priv} are unsatisfied, respectively. In case of point 5, \textbf{ConformanceCheck}(\textit{initgoal, Architecture, Rulesets, N}) == 1 means that the first point of   Definition~\ref{def:dpr} is satisfied. Points 6, 8, 10 of Property~\ref{ttp:1} correspond to the satisfaction of point 4 of Definition~\ref{def:func}, while points 7, 9, 11  mean that point 2 in Definition~\ref{def:dpr} is unsatisfied. Point 12 of Property~\ref{ttp:1} correspond to the satisfaction of point 5 of Definition~\ref{def:func}, while point 13 correspond to (the unsatisfied) point 3 in Definition~\ref{def:dpr}. Point 14 of Property~\ref{ttp:1} correspond to the satisfaction of point 6 of Definition~\ref{def:func}, while point 15 correspond to (the unsatisfied) points 4-5 in Definition~\ref{def:dpr}. Point 16 corresponds to the satisfactory of point 7 of Definition~\ref{def:func}. Finally, point 17 corresponds to (the unsatisfied) point  6 in Definition~\ref{def:dpr}.      
\end{proof}

\begin{ttp} (Termination up-to $N$)
\label{ttp:2}
Let $N$ be the maximum number of nested layers of cryptographic functions that the verification engine will examine. Assume that the nested layers of the defined data types are finite, beside a finite $N$, the proof process never gets into an infinite loop. 
\end{ttp}

\begin{proof} The verification engine performs resolution steps between the goals and the rules in \textit{Rulesets},  as well as the (action) facts in \textit{Architecture}. If there is an infinite loop in the proof process, then we would have an infinite number of resolution steps. We will show that the number of resolution steps is always finite during the proof of \textit{initgoal}.   

As a result of a resolution step \textit{goal} $\circ_{(goal,\ head\ of\ rule)}$ \textit{rule}, where \textit{rule} $\in$ \{\textit{P8}, \textit{P9}, \textit{P10}\},  we get the two new (sub-)goals in the tails of the rules (e.g. \textit{goal} $\circ$ \textit{P8} =  HAS(\textit{EV},\textbf{Senc}($\theta V$,\textit{K}))$\sigma$, HAS(\textit{EV},\textit{K})$\sigma$). Since the verification engine does not prove/examine any goal with more than $N$ layers of cryptograpghic functions (e.g. HAS(sp,\textbf{Senc}(\textbf{Senc},...(\textbf{Mac}(name,key))),\dots,key),key), there are maximum $N$ recursive calls of the resolution step \textit{goal} $\circ_{(goal,\ head\ of\ rule)}$ \textit{rule}, beside \textit{rule} $\in$ \{\textit{P8}, \textit{P9}, \textit{P10}\}. Each recursive call produces two (sub-)goals, hence, $N$ recursive calls result in at most $2^N$ (sub-)goals to be proved. In the worst case scenario, this would mean $2^N$*$|$\textit{Rulesets}$|$ resolution steps (between each goal and rule pair, where $|$\textit{Rulesets}$|$ is the number rules in \textit{Rulesets}). 

In case \textit{rule} is one of \textit{P3}-\textit{P7} or \textit{P15}-\textit{P18}, a resolution step \textit{goal} $\circ_{(goal,\ head\ of\ rule)}$ \textit{rule} would generate a single goal (e.g. \textit{goal} $\circ_{(goal,\ head\ of\ \textit{P4})}$ \textit{P4} = RECEIVEAT(\textit{EV}, \textit{Data}, \textbf{Time}(\textit{TT}))$\sigma$). Then, the resulted (sub-)goals will be checked against the facts in \textit{Architecture}, which yields $|$\textit{Architecture}$|$ + $1$ resolution steps for each rule (where $|$\textit{Architecture}$|$ is the number elements in \textit{Architecture}). 

In case \textit{rule} is one of \textit{D1}-\textit{D7} or \textit{rule} $\in$ \{\textit{P1}, \textit{P11}\}, 2*$|$\textit{Architecture}$|$ + $1$ resolution steps are carried out. For \textit{rule} $\in$ \{\textit{P2}, \textit{P12}, \textit{P13}, \textit{P14}\}, a step \textit{goal} $\circ_{(goal,\ head\ of\ rule)}$ \textit{rule} generates a single (sub-)goal. The (sub-)goals are then be checked against the rule set (\textit{Rulesets}), including  rules \textit{P3}-\textit{P7} (or \textit{P15}-\textit{P18}), which yields 2*$|$\textit{Architecture}$|$ + $1$ resolution steps in each case.  In addition, when these (sub-)goals are checked against \textit{P8}-\textit{P10}, it yields $2^N$*$|$\textit{Rulesets}$|$ resolution steps in each case. We note that in rules \textit{P2} and \textit{P12}-\textit{P14}, \textit{ds} is a value and \textit{P(ds)} is a function on \textit{ds} that represents the pseudonym. Therefore, we cannot have an infinite number of recursive resolution steps between these rules and the resulted sub-goals, because \textit{ds} cannot be unified with \textit{P(ds)}, and \textit{$\theta$V} cannot be unified with either \textit{ds} or \textit{P(ds)} being of different types.     

In case \textit{rule} is one of \textit{L1}-\textit{L8},  a resolution step \textit{goal} $\circ_{(goal,\ head\ of\ rule)}$ \textit{rule} generates two (sub-)goals. Each (sub-)goal will be examined against every rule (in \textit{Rulesets}), but a resolution step can only be successful in case of \textit{P3}-\textit{P10}. The resolution with each of these rules results in a finite number of further resolution steps (as we argued above). Similarly, the case of  \textit{U1}-\textit{U8} only yields a finite number of resolution steps.

\end{proof}

The completeness property can be stated as a consequence of the termination property (Property~\ref{ttp:2}), as follows:

\begin{ttp} (Completeness)\\ 
\label{ttp:3}
If all the data types specified in Architecture  contain at most $N$ layers of nested cryptographic functions, for some finite $N$, and all the defined data types contain a finite number of layers of other data types, then: 
\begin{enumerate}
\item If \textit{initgoal} $\in$ \{HAS(E,$\theta$), HASUPTO(E,$\theta$,Time(dd))\}, and $E$ $\in$ $\pi_{\theta}$.$\pi_{has}$ at the policy level, then whenever \textbf{ConformanceCheck}(\textit{initgoal, Architecture, Rulesets, N}) == 0, the architecture does not functionally conform with the policy. 

\item If \text{initgoal} $\notin$ $\mathcal{G}^\theta_{link}$ and $(E,\theta')$ $\in$ $\pi_{\theta}$.$\pi_{link}$, then whenever \textbf{ConformanceCheck}(\textit{initgoal, Architecture, Rulesets, N}) == 0, \textit{Architecture} does not functionally conform with the policy.  

\item If \text{initgoal} $\in$ $\mathcal{G}^\theta_{ccons}$ $\cup$ $\mathcal{G}^\theta_{ucons}$ $\cup$ $\mathcal{G}^\theta_{scons}$ $\cup$ $\mathcal{G}^\theta_{fwcons}$, and $\pi_{col}$.\textit{cons} = $Y$, $\pi_{use}$.\textit{cons} = $Y$, $\pi_{str}$.\textit{cons} = $Y$, $\pi_{fw}$.\textit{cons} = $Y$ in $\pi_{\theta}$, respectively, then the architecture does not DPR conform with the policy whenever \textbf{ConformanceCheck}(\textit{initgoal, Architecture, Rulesets, N}) == 0.   

\item If \text{initgoal} = CPURPOSE($\theta$, \textit{cp})  (i.e. \text{initgoal} $\in$ $\mathcal{G}^\theta_{cpurp}$), and (\textit{cp}:$\theta$ $\in$ $\pi_{use}$.\textit{cpurp}), then whenever \textbf{ConformanceCheck}(\textit{initgoal, Architecture, Rulesets, N}) == 0, the architecture \textbf{does not} functionally conform with the policy.  

\item If  \text{initgoal} = UPURPOSE($\theta$, \textit{up}) (i.e. \text{initgoal} $\in$ $\mathcal{G}^\theta_{upurp}$), and (\textit{up}:$\theta$ $\in$ $\pi_{use}$.\textit{upurp}), then whenever \textbf{ConformanceCheck}(\textit{initgoal, Architecture, Rulesets, N}) == 0, the architecture \textbf{does not} functionally conform with the policy. 

\item If \text{initgoal} = FWPURPOSE($\theta$, \textit{fwp}) (i.e. \text{initgoal} $\in$ $\mathcal{G}^\theta_{fwpurp}$), and  (\textit{fwp}:$\theta$ $\in$ $\pi_{fw}$.\textit{fwpurp}),  then whenever \textbf{ConformanceCheck}(\textit{initgoal, Architecture, Rulesets, N}) == 0, the architecture \textbf{does not} functionally conform with the policy.  

\item If  \text{initgoal} = \{STORE(\textit{E}, $\theta$, \textit{EV}$_{\textit{from}}$),  STOREAT(\textit{E}, $\theta$, \textit{EV}$_{\textit{from}}$,  \textbf{Time}(\textit{TT}))\} (i.e. \text{initgoal} $\in$ $\mathcal{G}^\theta_{places}$), and  \textbf{(\textit{E} $\in$  $\pi_{str}$.\textit{where})}, then whenever \textbf{ConformanceCheck}(\textit{initgoal, Architecture, Rulesets, N}) == 0, the architecture \textbf{does not} functionally conform with the policy.  

\item If \text{initgoal} = DELETEWITHIN(\textit{E}, $\theta$, \textit{EV}$_{\textit{from}}$,  \textbf{Time}(\textit{dd})) (i.e. \text{initgoal} $\in$ $\mathcal{G}^\theta_{within}$) and \textbf{(\textit{dd} $\leq$ $\pi_{del}$.\textit{deld})} and \textbf{(\textit{E} $\in$  $\pi_{del}$.\textit{fromwhere})}, then whenever \textbf{ConformanceCheck}(\textit{initgoal, Architecture, Rulesets, N}) == 0, the architecture \textbf{does not} functionally conforms with the policy.  

\item If \text{initgoal} = RECEIVE($E$, $\theta$, \textit{EV}$_{\textit{from}}$) (\text{initgoal} $\in$ $\mathcal{G}^\theta_{fwto}$), and $E$ $\in$ $\pi_{fw}$.\textit{fwto}, then whenever \textbf{ConformanceCheck}(\textit{initgoal, Architecture, Rulesets, N}) == 0, the architecture \textbf{does not} functionally conform with the policy.  

\end{enumerate}
\end{ttp}

Property~\ref{ttp:3} says that completeness can only be ``achieved" up to the maximum allowed nested layers of cryptographic functions, $N$. 

\begin{proof}

If \textbf{ConformanceCheck}(\textit{initgoal, Architecture, Rulesets, N}) == 0, then \textit{initgoal} cannot be proved by any fact in \textit{Architecture} provided that all facts in \textit{Architecture} contain at most $N$ nested layers of functions \textit{Senc}, \textit{Aenc}, and \textit{Mac}, and nested layers of other data types. The latter assumption is required for a resolution step to be successful, while the first is required to make the verification terminates. Otherwise, if there is a set of facts in \textit{Architecture}, which can be used to prove \textit{initgoal}, then there would be a derivation tree meaning that \textbf{ConformanceCheck}(\textit{initgoal, Architecture, Rulesets, N}) == 1.  

Therefore, point 1 of Property~\ref{ttp:3} does not satisfy the first point of Definition~\ref{def:func}. Similarly, point 2 of Property~\ref{ttp:3} does not satisfy the second point of Definition~\ref{def:func}. Point 3 of Property~\ref{ttp:3} does not satisfy the first point of Definition~\ref{def:dpr}. Points 4-6 of Property~\ref{ttp:3} correspond to point 4 of Definition~\ref{def:func}. Points 7, 8, and 9  of Property~\ref{ttp:3} correspond to points 5, 6, and 7 of Definition~\ref{def:func}, respectively. 

Moreover, we show that the inference rules cover all the possible data types (format) may be defined in the  architectural actions. 
\begin{itemize}
\item In case of consents, rules D1-D5 capture the case when an action RECEIVE(AT) is defined on a data type $\theta V$ and the corresponding consent is also defined on this data type. 

\item In case of forward and collection consents, the fact Anytypeinccrypto[$\theta V$] in rules D6 and D7 covers all data types that contains $\theta V$. 

\item Similarly, in the LINK and LINKUNIQUE cases, the rules in Figures 
the rules in Figure~\ref{fig:inf3b} are defined on data types of form Anytypeinccrypto[$\theta V$]. 

\item For the ``trivial" HAS, LINK and LINKUNIQUE facts, the set \textit{TrivialHASLINKFacts} contains all the possible generation of data types (as shown in Figures~\ref{fig:trivi}-\ref{fig:trivi2}).  
\end{itemize}

\end{proof}

\section{Discussion}
\label{sec:discussion}

As most of the laws and articles in the GDPR are complex, formally specifying them without simplification is either cumbersome or impossible. In this paper, we attempt to capture some basic requirements in an abstract way. There are several ways to improve or extend the proposed formal sepcifications. For instance, practically, depending on the context of a consent (e.g. health-care or education contexts), a consent may contain different pieces of information that need to be modelled. Furthermore, in our languages we do specify the deletion of a consent, but only when the data itself is deleted (see the last rule in Figure~\ref{fig:semevent}). A more detailed study of the consent revocation process can be addressed in the future, for example, when the collected data has not been deleted yet, but the consent for transfer has been revoked. This could be addressed by changing the last rule in Figure~\ref{fig:semevent} such that only the consent in question is deleted. 

There are areas to improve regarding the transfer sub-policy as well, for example, the GDPR covers the case when personal data is transferred to a third country or an international organisation, and appropriate agreement and arrangement must be done prior data transfer \cite{Gdpr46}. This agreement could be specified in the form of a sticky policy between a service provider and an international organisation. Sticky policies are used in PPL \cite{PPL} to match the expectation of a client and the obligation offered by a service provider. Regarding the deletion sub-policy, in the GDPR, the data subject also has the right to request a deletion for their collected data. This can be modelled with an event/action that captures the reception of a deletion request (e.g. \textit{recvdelreq}($\theta$, \textit{place}, $t$)) and a corresponding deletion event within a specified delay. Finally, transparency is also an important part of the GDPR as it captures the ``right to be informed", which can be defined by the event/action ``notify" that happens before the data collection, usage, storage and transfer. 

To capture the (collection, usage, or transfer) purposes, for simplicity, the  architecture language proposed in this paper relies on only the two basic actions \textit{create} and \textit{calculate}. In the same way, additional actions can be added to specify purposes such as ``send some type of data" (defined by \textit{send}:$\theta$ such as \textit{send}:\textit{bill}), or ``notify about some type of data" (e.g. \textit{notify}:$\theta$ such as \textit{notify}:\textit{energyconsumption}). 

Besides the simplified data protection requirements, the strength of our approach is the data possession and data connection policies, as well as the automated verification of these. Although at the policy and architecture levels the verification process seems to be simpler than in case of verifying a program code, it is relevant to detect any design flaws at these higher levels. Manual and informal reasoning can be error-prone,  especially when there are many complex data types and entities in the system.  

\section{Implementation}
\label{sec:imp} 

DataProVe is written in Python, and is available for download from GitHub\footnote{\url{https://github.com/vinhgithub83/DataProVe}} and its website\footnote{\url{https://sites.google.com/view/dataprove/}}.    

\subsection{The System Architecture Specification Page}
\label{sec:arch} 
\begin{figure}[htb!]
    \begin{center}
        \includegraphics[width=1\textwidth]{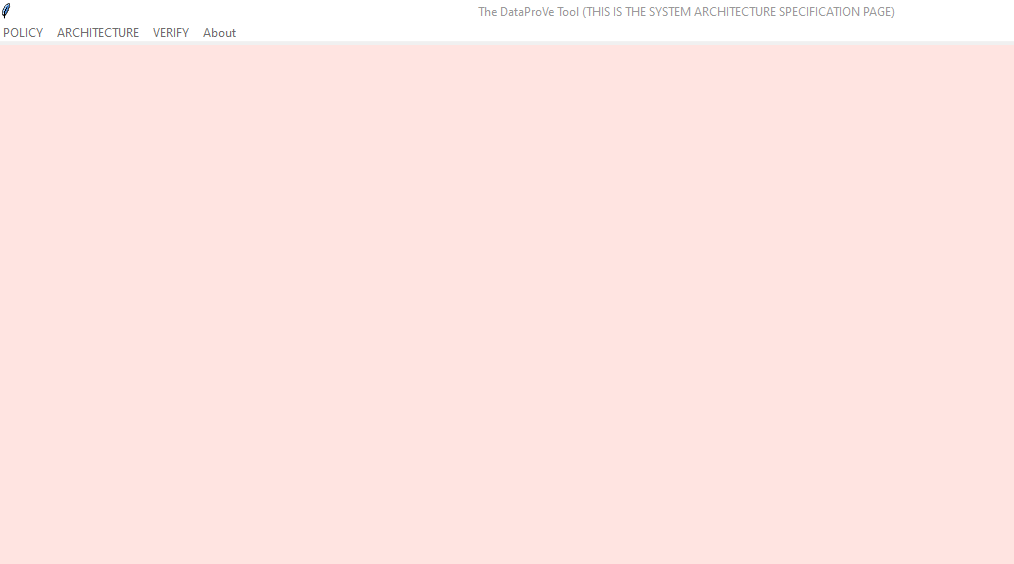}
    \end{center}
    \caption{After launching DataProVe, the system architecture specification page can be seen.}
    \label{fig:1}
\end{figure}

After launching the tool, as depicted in Figure~\ref{fig:1}, the default page can be seen, where the user can specify a system architecture. DataProVe supports two types of components, the so-called main components, and the sub-components. The main components can represent an entire organisation, system or entities that consists of several smaller components, such as a service provider, a customer, or authority (trusted third-party organisation). Sub-components are elements of a main component, for example, a service provider can have a server, a panel, or storage place. A main component usually has access to the data handled by its own sub-components, but this is not always the case, for instance, two main components can share a sub-component and only one main-component has access to its data.  This can happen, for example, when a service provider operates a device of a trusted third party, but it does not have free access to the content of the data stored inside the device.  

\begin{figure}[htb!]
    \begin{center}
        \includegraphics[width=1\textwidth]{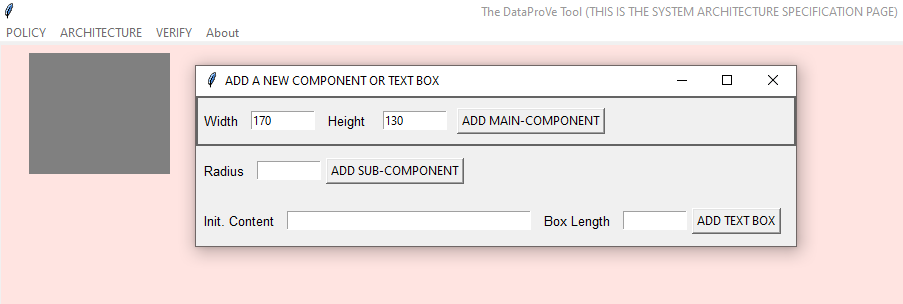}
    \end{center}
    \caption{Adding a main component of size 170x130.}
    \label{fig:2}
\end{figure}

\begin{figure}[htb!]
    \begin{center}
        \includegraphics[width=1\textwidth]{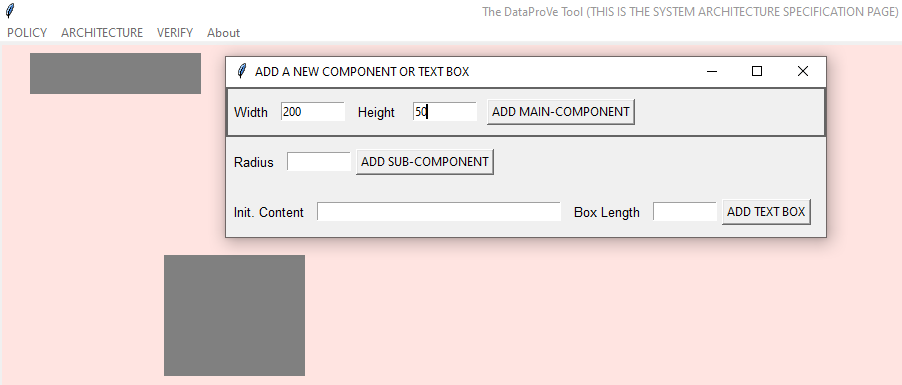}
    \end{center}
    \caption{Adding a new main component of size 200x50.}
    \label{fig:3}
\end{figure}

\begin{figure}[htb!]
    \begin{center}
        \includegraphics[width=1\textwidth]{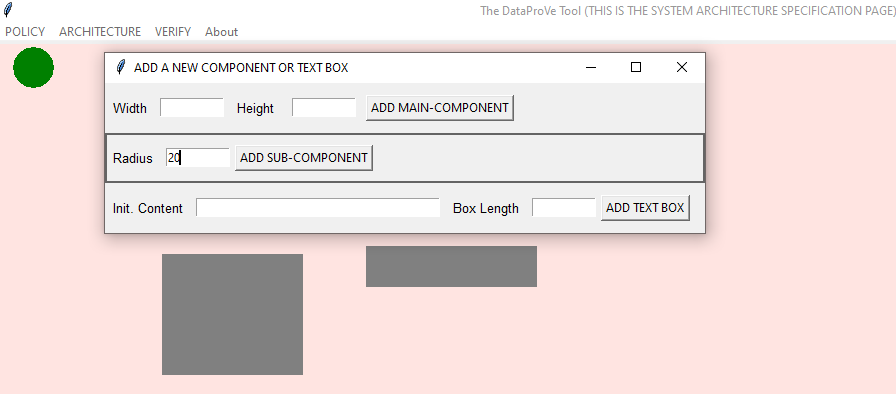}
    \end{center}
    \caption{Adding a new sub-component with a radius size of 20.}
    \label{fig:4}
\end{figure}

\begin{figure}[htb!]
    \begin{center}
        \includegraphics[width=0.6\textwidth]{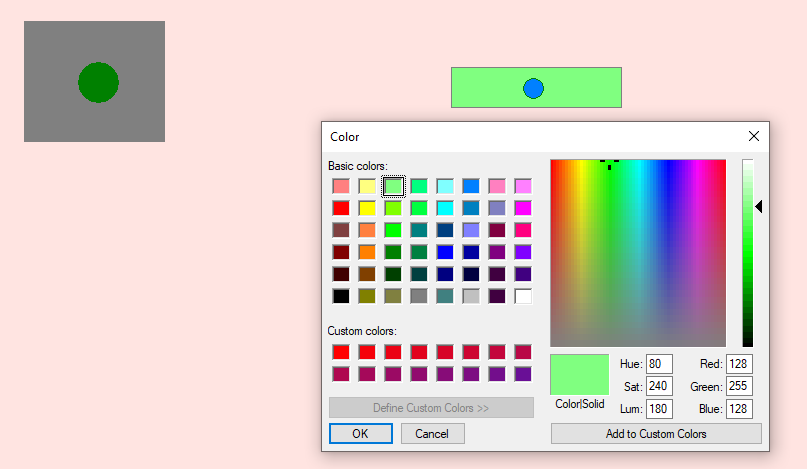}
    \end{center}
    \caption{Choosing the color for a component.}
    \label{fig:5}
\end{figure}

In the first version of DataProVe (v0.9), main components are represented by rectangular shapes, while sub-components are represented by circles. Examples can be seen in Figures~\ref{fig:2}-\ref{fig:5}.

In this report, we will interchange between the two terms entity and component, because the term entity has been used in our theoretical papers, while the tool uses the term component more. They refer to the same thing in our context.

\begin{figure}[htb!]
    \begin{center}
        \includegraphics[width=0.9\textwidth]{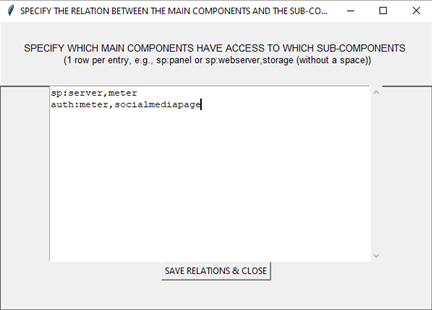}
    \end{center}
    \caption{Specify which main component has access to the data in which sub component (sp has access to server and meter, while auth has access to meter and socialmediapage).}
    \label{fig:8}
\end{figure}

In DataProVe one can specify which main-component can have access to which sub-component.  An example can be seen in Figure~\ref{fig:8}, where we specified the relation between sp and server, meter, as well as between the authority auth and meter, socialmediapage. 

\begin{figure}[htb!]
    \begin{center}
        \includegraphics[width=0.9\textwidth]{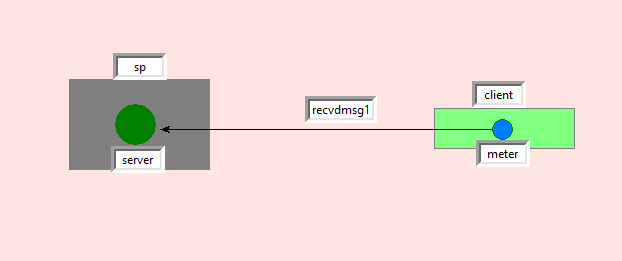}
    \end{center}
    \caption{Draw an arrow from the component meter to server.}
    \label{fig:11}
\end{figure}  

\begin{figure}[htb!]
    \begin{center}
        \includegraphics[width=0.9\textwidth]{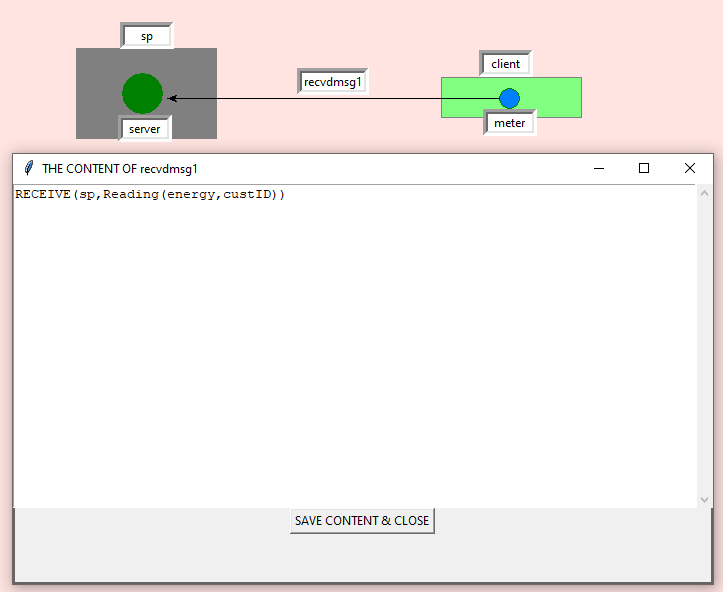}
    \end{center}
    \caption{Specify the message content of recvdmsg1 (through the action RECEIVE).}
    \label{fig:12}
\end{figure}

In Figure~\ref{fig:11}, a new text box is created with the name \textit{recvdmsg1}, which denotes that the server receives a message called \textit{msg1}. Its content (depicted in Figure~\ref{fig:12}) says that sp can receive a reading that contains the energy consumption (energy) and the customer ID (\textit{custID}).  

In the architecture level, we distinguish entity/component, actions and data, where actions specify what an entity/component can do on a piece of data (it may not perform this action  eventually during a low-level system run, but there are instances of the system run that where this action happens),  except for DELETEWITHIN, as we will see later.   

\subsubsection{ACTIONS} 
\label{sec:actions}

Based on the definition of actions and architectures in Figure~\ref{fig:sysarch}, we propose their corresponding formats that can be given in the text boxes/text editor in DataProVe.
  
Actions are words/string of all capital letters, and DataProVe supports the actions "OWN", "RECEIVE", "RECEIVEAT", "CREATE", "CREATEAT", "CALCULATE", "CALCULATEAT", "STORE", "STOREAT", "DELETE", "DELETEWITHIN". The syntax of each action in DataProVe is as follows.

\begin{center}
\begin{tabular}{|c|}
\hline\\ 
 \textit{Note: no space character is allowed when specifying the actions in the bullet points below.}  \\ \\ 
 \hline
\end{tabular}
\end{center}

The reserved/pre-defined keywords are highlighted in bold, while the non-bold text can be freely defined by the user: 
 
\begin{itemize}
\item \textbf{OWN}(component,Datatype) : 

This action defines that a component (e.g., \textbf{sp}, auth, server, meter etc.) can own a piece of data of type Datatype. For example, \textbf{OWN}(server,spkey) say they server can own the a piece of data of type service provider key (spkey). 

\item \textbf{RECEIVE}(component,Datatype): 

This action defines that a component can receive a piece of data of type Datatype, for example, \textbf{RECEIVE}(server,Sicknessrecord(name,insurancenumber)) says that server can receive a sickness record that contains a piece of data of type name and insurance number.

\item \textbf{RECEIVEAT}(component,Datatype,\textbf{Time(t)}): 

This action is similar to the previous one, except that here we also need to define the time when the data can be received. Since at the architecture level we do not intent to specify the concrete time value, the generic time construct, denoted by the keyword \textbf{Time(t)} specifies that component can receive a piece of data of type Datatype at some (not specific) time \textbf{t}. 
\textbf{RECEVEAT} is used to define when a consent (\textbf{Cconsent}(Datatype), \textbf{Uconsent}(Datatype), \textbf{Sconsent}(Datatype), \textbf{Fwconsent}(Datatype)) is received.

\item \textbf{CREATE}(component,Datatype): 

This action defines that a component can create a piece of data of type Datatype, for instance, \textbf{CREATE}(\textbf{sp},Account(name,address,phone)) defines that a service provider \textbf{sp} can create an account that contains three pieces of data of types name, address and phone number.   

\item \textbf{CREATEAT}(component,Datatype,\textbf{Time(t)}): 

This action defines that a component can create a piece of data of type Datatype at some (not specific) time \textbf{t}. For example, \textbf{CREATE}(\textbf{sp},Account(name,address,phone),\textbf{Time(t)}).    

\item \textbf{CALCULATE}(component,Datatype): 

This action defines that a component can calculate a piece of data of type Datatype, for instance, \textbf{CALCULATE}(\textbf{sp},Bill(energyconsumption)) defines that a service provider \textbf{sp} can calculate a bill using a piece of data of type energy consumption.   

\item \textbf{CALCULATEAT}(component,Datatype,\textbf{Time(t)}): 

This action defines that a component can calculate a piece of data of type Datatype at some (not specific) time \textbf{t}. For example, \textbf{CALCULATE}(\textbf{sp},Bill(energyconsumption),\textbf{Time(t)}).

\item \textbf{STORE}(storageplace,Datatype): 

This action defines that a service provider can store a piece of data of type Datatype in storageplace, where storageplace can be \textbf{mainstorage}, \textbf{backupstorage}. These reserved keywords define a collection of storage place(s) that can be seen as ``main" storage, or ``backup" storage, respectively.  

For example, \textbf{STORE}(\textbf{mainstorage},Account(name,address,phone)) defines that a service provider can store an account that contains name, address and phone number in its main storage place(s).   

\item \textbf{STOREAT}(storageplace,Datatype,\textbf{Time(t)}):
 
This action defines that a component can store a piece of data of type Datatype in the place(s) storageplace at some (not specific) time \textbf{t}. 

For example, \textbf{STORE}(\textbf{mainstorage},Account(name,address,phone),\textbf{Time(t)}) defines that an account with a name, address and phone number can be stored in the main storage of the service provider at some time \textbf{t}. 

\item \textbf{DELETE}(storageplace,Datatype):

The action delete is closely related to the action store, as it defines that a piece of data of type Datatype can be deleted from storageplace. 
 
For example, \textbf{DELETE}(\textbf{mainstorage},Account(name,address,phone)) captures that a service provider can.

\item \textbf{DELETEWITHIN}(storageplace,Datatype,\textbf{Time}(tvalue)): 

This action captures that once the data is stored, a component must delete a piece of data of type Datatype within the given time value tvalue (tvalue is a data type for time values). Unlike the non-specific \textbf{Time(t)}, which is a predefined construct, tvalue is defined by the user, and takes specific time values such as 3 years or 2 years 6 months.  

For example, \textbf{DELETE}(\textbf{mainstorage},Account(name,address,phone),\textbf{Time}(2y)) defines that the service provider must delete an account from its main storage within 2 years. 
\end{itemize}

\subsubsection{COMPONENTS/ENTITY} 
\label{sec:entities}

A component can be specified by a string of all lower case, for example, a service provider can be specified by \textbf{sp}, or a third-party authority by auth (obviously they can be specified with any other string).    

DataProVe supports some pre-defined or reserved components/entities, such as \textbf{sp}, \textbf{trusted}, \textbf{mainstorage}, \textbf{backupstorage}. 

\begin{itemize}
\item \textbf{sp}: this reserved keyword defines a service provider. DataProVe only allows a single service provider at a time (in the specification of a policy and architecture). 

\item \textbf{trusted}: this reserved keyword defines a trusted authority that is able to link a pseudonym to the corresponding real name. 

\item \textbf{mainstorage}: this reserved keyword defines the collection of main storage places of a service provider.
 
\item \textbf{backupstorage}: this reserved keyword defines the collection of backup storage places of a service provider.
\end{itemize}

\begin{center}
\begin{tabular}{|c|}
\hline\\ 
 \textit{Note:  An entity/component is always defined as the first argument of an action.}  \\ \\ 
 \hline
\end{tabular}
\end{center}

\subsubsection{DATA TYPES}
\label{sec:dttypes}

DataProVe supports two groups of data types, the so-called compound data types, and simple data types. 

\begin{itemize}
\item \textbf{Simple data types} do not have any arguments, and they are specified by strings of all lower cases, without any space or special character. Example simple data types include name, address, phonenumber, nhsnumber, etc.

\item \textbf{Compound data types} have arguments, and they are specified by strings that start with a capital letter followed by lower cases (again without any space or special character). For example, Account(name,address,phone) is a compound data type that contains three simple data types as arguments.  Another example compound data type can be Hospitalrecord(name,address,insurance). Any similar compound data types can be defined by the user. We note that the space character is not allowed in the compound data types. 

Nested compound data types are compound data types that contain another compound data types. For instance, Hospitalrec(Sicknessrec(name,disease),address,insurance) captures a hospital record that contains a sickness record of a name and disease, and an address, and finally, an insurance number.  

\begin{center}
\begin{tabular}{|c|}
\hline\\ 
 \textit{Note: The first version of DataProve (v0.9) supports three layers of nested data types.}  \\ \\ 
 \hline
\end{tabular}
\end{center}

\end{itemize}

DataProVe has pre-defined or reserved data types, such as 

\begin{itemize}
\item The types of consents: \textbf{Cconsent}(Datatype),  \textbf{Uconsent}(Datatype),  \textbf{Sconsent}(Datatype),  \textbf{Fwconsent}(Datatype).

We do not differentiate among the different consent format, it can be e.g. written consent, or online consent form, or some other formats. 

\begin{itemize}
\item \textbf{Cconsent}(Datatype): This is a type of collection consent on a piece of data of type Datatype.  For example, \textbf{Cconsent}(illness), \textbf{Cconsent}(Account(creditcard,address)) capture the collection consent on the illness information, and the account containing a credit card number and address. 

\item \textbf{Uconsent}(Datatype): A type of usage consent on a piece of data of type Datatype. For example, \textbf{Uconsent}(Energy(gas,water,electricity)), \textbf{Uconsent}(address).  

\item \textbf{Sconsent}(Datatype): A type of storage consent on a piece of data of type Datatype.  For example, \textbf{Sconsent}(personalinfo), \textbf{Sconsent}(Account(creditcard,address)) defines the types of storage consent on a type of personal information and account, respectively.     

\item \textbf{Fwconsent}(Datatype,component): A type of forward/transfer consent on a piece of data of type Datatype, and a component to whom the data is forwarded/transfered.  E.g. \textbf{Fwconsent}(personalinfo,auth), \textbf{Fwconsent}(Account(creditcard,address),auth) defines the type of forward consent on the type of personal information and account, respectively, as well as a third party authority (auth) to which the given data is forwarded.  
\end{itemize}

\item The types of time and time value: \textbf{Time(t)} or \textbf{Time}(tvalue), where \textbf{Time}() is a time data type, while the pre-defined special keyword t  denotes a type of non-specific time, and tvalue is a type of time value (such as 5 years, 2 hours, 1 minute, etc.). tvalue is a (recursive) type and takes the form of  

\begin{center}
\begin{tabular}{|c|}
\hline\\ 
 \textit{tvalue ::= y $|$ mo $|$ w $|$ d  $|$  h  $|$  m $|$ numtvalue $|$  tvalue $+$ tvalue}  \\ \\ 
 \hline
\end{tabular}
\end{center}

where  y specifies a year, mo a month, w a week, d a day, h an hour and m a minute. Further, numtvalue is the a number (num) before tvalue, for example if num = 3 and tvalue = y, then numtvalue  is 3y (i.e. 3 years).  Additional examples include tvalue = 5y + 2mo + 1d + 5m.  

It is important to note that \textbf{Time}(tvalue) can only be used in the action \textbf{DELETEWITHIN}, \textbf{RECEIVEAT}, \textbf{CREATEAT}, \textbf{CALCULATEAT}, \textbf{STOREAT} must contain the non-specific time \textbf{Time(t)}.  

For example, the actions  
\begin{itemize}
\item \textbf{DELETEWITHIN}(\textbf{sp},\textbf{mainstorage},Webpage(photo,job),\textbf{Time}(10y+6mo)) 
Any webpage must be deleted from the main storage of the service provider within 10 years and 6 months. 

\item \textbf{RECEIVEAT}(\textbf{sp},\textbf{Cconsent}(illness),\textbf{Time(t)})
The service provider can receive a collection consent on illness information at some non-specific time \textbf{t}.

\item \textbf{RECEIVEAT}(\textbf{sp},\textbf{Uconsent}(Webpage(photo,job)),\textbf{Time(t)}) 
The service provider can receive a usage consent on a webpage at some non-specific time \textbf{t}.

\item \textbf{STOREAT}(\textbf{sp},\textbf{backupstorage},Webpage(photo,job),\textbf{Time(t)})
The service provider can store a webpage in its back up storage places at some non-specific time \textbf{t}.

\item \textbf{CREATEAT}(server,Account(name,address),\textbf{Time(t)}):  
The service provider can create an account that contains a name and address in at some non-specific time \textbf{t}.

\item \textbf{CALCULATEAT}(\textbf{sp},Bill(tariff,Energy(gas,water,electricity)),\textbf{Time(t)}): 
The service provider can create an account that contains a name and address in at some non-specific time \textbf{t}. 
\end{itemize} 
\end{itemize} 

\begin{itemize}
\item The type of metadata and meta values:  \textbf{Meta}(Datatype).

This data type defines the type of metadata (information about other data), or information located in the header of the packets, the meta information often travels through a network without any encryption or protection, which may pose privacy concern. Careful policy and system design are necessary to avoid privacy breach caused by the analysis of metadata or header information. 

\begin{center}
\begin{tabular}{|c|}
\hline\\ 
 \textit{Note:  \textbf{Meta}(Datatype) is always defined as the last argument in a piece of data.}  \\ \\ 
 \hline
\end{tabular}
\end{center}

Example application of metadata includes: 
\begin{itemize}
\item \textbf{RECEIVE}(sp,Sicknessrec(name,disease,\textbf{Meta}(ip))):

This action defines that the service provider can receive a packet that containing a name and disease, but the packet also includes the metadata IP address of the sender computer. We note that this syntax is simplified in terms that it aims to eliminate the complexity of nested data type. Specifically, this syntax abstracts away from the definition of the so-called packet data type, an ``abbreviation" of the lengthy \textbf{RECEIVE}(\textbf{sp},Packet(Sicknessrec(name,disease),\textbf{Meta}(ip)))).

\item \textbf{RECEIVE}(\textbf{sp},Sicknessrec(name,disease,\textbf{Meta}(\textbf{Enc}(ip,k)))):
This action is similar to the previous one, but now the metadata IP address is encrypted with a key k.

\item \textbf{RECEIVEAT}(\textbf{sp},Sicknessrec(name,disease,\textbf{Meta}(ip)),\textbf{Time(t)}): 

This action is similar to the first one, but it includes the time data types at the end. It defines that the service provider receives the sickness record along with the IP address of the sender device, at some non-specific time \textbf{t}. 
\end{itemize}
Obviously, any metadata can be defined instead of IP address in the examples above. 

\item The type pseudonymous data:  \textbf{P}(Datatype $|$ component).

This data type defines the type of pseudonymous data, for example, a pseudonym. The argument can be either a data type or a component \footnote{This would be in the versions above v0.9. In the version 0.9, DataProVe preserves the keyword (all small letters) \textbf{ds} for data subject, and the user can define P(\textbf{ds}) to specify that the real data subject/identity has been pseudonymised.}. Pseudonym is a means for achieving a certain degree of privacy in practice as the real identity/name and the pseudonym can only be linked by a so-called trusted authority. DataProVe also captures this property, namely, only the component trusted can link the pseudonym to the real name/identity. 

For example, 
\begin{itemize}
\item \textbf{RECEIVE}(\textbf{sp},Sicknessrec(\textbf{P}(name),disease)):

This action defines that a service provider can receive a sickness record, but this time, the name in the record is not the real name but a pseudonym, hence, the service provider cannot link a real name to a disease. 

\item \textbf{RECEIVE}(\textbf{trusted},Sicknessrec(\textbf{P}(name),disease)):

This is similar to previous case, but the trusted authority can receive a sickness record instead of the service provider. 

\item \textbf{RECEIVE}(\textbf{sp},Sicknessrec(\textbf{P}(name),disease,\textbf{Meta}(ip))):
Again, this is similar to the first case, but with metadata.

\item \textbf{RECEIVEAT}(\textbf{sp},Sicknessrec(name,disease,\textbf{Meta}(ip)),\textbf{Time(t)}): 
This is similar to previous case, but also include the time data type. 

\end{itemize}

\item The types of cryptographic primitives and operations:  DataProVe supports the basic cryptographic primitives for the architecture. Again, we provide the reserved keywords in bold.  
\begin{itemize}
\item Private key: \textbf{Sk}(Pkeytype):
 
This data type defines the type of private key used in asymmetric encryption algorithms. Its argument has a type of public key (Pkeytype). We note that public key is not a reserved data type.    

\item Symmetric encryption: \textbf{Senc}(Datatype,Keytype):

This is the type of the cipher text resulted from a symmetric encryption, and has two arguments, a piece of data and a symmetric key (Keytype). 

For example,  
\begin{itemize}
\item \textbf{RECEIVE}(\textbf{sp},\textbf{Senc}(Account(name,address),key)): 

This specifies that a service provider can receive a symmetric key encryption of an account using a key of type key. 

\item \textbf{RECEIVE}(\textbf{sp},\textbf{Senc}(Account(\textbf{Senc}(name,key),address),key)):

This specifies that a service provider can receive a symmetric key encryption of an account that contains another encryption of a name, using a key of type key. 

\item \textbf{OWN}(\textbf{sp},key):

This specifies that a service provider can own a key of type key.  

\end{itemize}

\item Asymmetric encryption: \textbf{Aenc}(Datatype,Pkeytype): 

This is the type of the cipher text resulted from an asymmetric encryption, and has two arguments, a piece of data and a public key (Pkeytype). 

For example, 
\begin{itemize}
\item \textbf{RECEIVE}(\textbf{sp},\textbf{Aenc}(Account(name,address),pkey)):

This specifies that a service provider can receive an asymmetric key encryption of an account using a public key of type pkey.
 
\item \textbf{CALCULATE}(\textbf{sp},\textbf{Sk}(pkey)): 

This specifies that a service provider can calculate a private key corresponding to the public key (of type pkey).
 
\item \textbf{OWN}(\textbf{sp},pkey):

This specifies that a service provider can own a public key of type pkey.  
\end{itemize}

\item Message authentication code (MAC): \textbf{Mac}(Datatype,Keytype):

This is the type of the message authentication code that has two arguments, a piece of data and a symmetric key (Keytype). 

For example,
\begin{itemize}
\item \textbf{RECEIVE}(\textbf{sp},\textbf{Mac}(Account(name,address),key)):

This specifies that a service provider can receive a message authentication code of an account using a key of type key. 

\end{itemize}

\item Cryptographic hash: \textbf{Hash}(Datatype): 
 
This is the type of the cryptographic hash that has only one argument, a piece of data. 

For example,
\begin{itemize}
\item \textbf{RECEIVE}(server,\textbf{Hash}(password)):

This specifies that a server can receive a hash of a password. 

\item \textbf{STORE}(\textbf{sp},\textbf{mainstorage},\textbf{Hash}(password)):

This specifies that a service provider can store a hash of a password in its main storage place(s). 

\end{itemize}

\end{itemize}

\end{itemize}

\subsection{The Data Protection Policy Specification Page}
\label{sec:pol}
\begin{figure}[htb!]
    \begin{center}
        \includegraphics[width=1\textwidth]{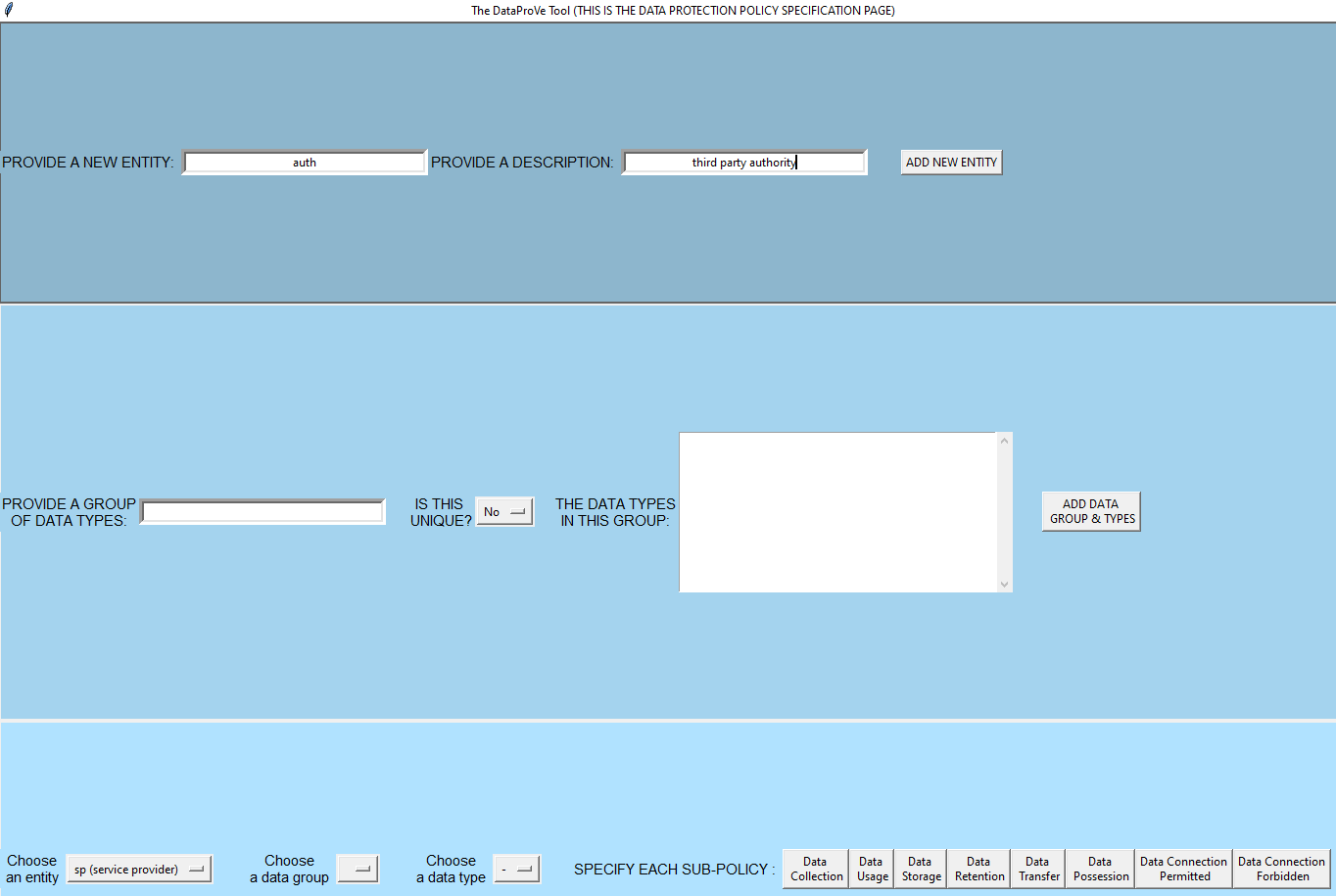}
    \end{center}
    \caption{The Policy Specification Page.}
    \label{fig:14}
\end{figure}

On the data protection policy specification page, we can define a high-level data protection policy (as shown in Figure~\ref{fig:14}). 

\subsubsection{Entities/Components (the top part)}
\label{sec:top}
The policy page has three parts, the top part is to specify the entities/components in the system, such as authority, client etc. On the left side, the user is expected to provide a short notation, and on the right side, the full name/description to help identifying the meaning of the notation.  For instance, in Figure~\ref{fig:14}, the notation is auth, and the description is third party authority. After adding a new entity, it will appear in the drop-down option menu in the bottom part. Note that \textbf{the entity sp (service provider) is a pre-defined entity} that is already added by default (hence, the user does not need to add). The user can specify any other entities. 

\subsubsection{Data groups/Data types (the middle part)}
\label{sec:middle}

\begin{figure}[htb!]
    \begin{center}
        \includegraphics[width=1\textwidth]{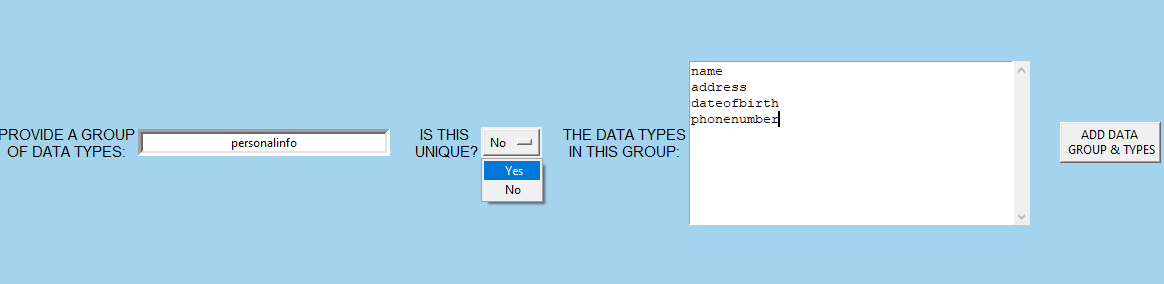}
    \end{center}
    \caption{Specifying data groups (personalinfo) and its data types.}
    \label{fig:15}
\end{figure}

The middle part in the policy specification page is for defining the data groups and data types. As shown in Figure~\ref{fig:15}, the user can define a group of data types, for instance, a data group denoted by \textit{personalinfo} is defined which includes four data types, name, address, \textit{dateofbirth}, and \textit{phonenumber}.

\begin{figure}[htb!]
    \begin{center}
        \includegraphics[width=1\textwidth]{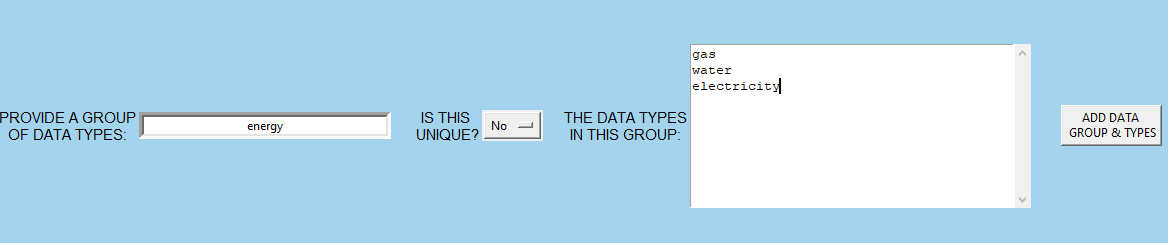}
    \end{center}
    \caption{Specifying data groups (energy) and its data types.}
    \label{fig:16}
\end{figure}

The option menu in the middle (called ``IS THIS UNIQUE") expects the user to provide if the data group together with its data types can be used to uniquely identify an individual. For instance, a name alone cannot be used to unique identify an individual, but a name together with an address, date of birth and phone number, can be, so the option ``Yes” was chosen. Another example is shown in Figure~\ref{fig:16}, with the data group called energy (refers to energy consumption) and its data types, gas, water, and electricity consumption. This type group together with its types cannot be used to uniquely identify an individual, hence, the option ``No” was chosen.

\subsubsection{Policy specification (the bottom part)}

Based on the syntax of the policy language given in Section~\ref{sec:syntaxpp}, we follow the seven sub-policies. However, here to avoid confusion we divide the last sub-policy, the data connection policy, into two categories,the data connection permit and data connection forbid policies. In the first one the user can specify which data link they allow, while in the second one for which they forbid.

\begin{figure}[htb!]
    \begin{center}
        \includegraphics[width=1\textwidth]{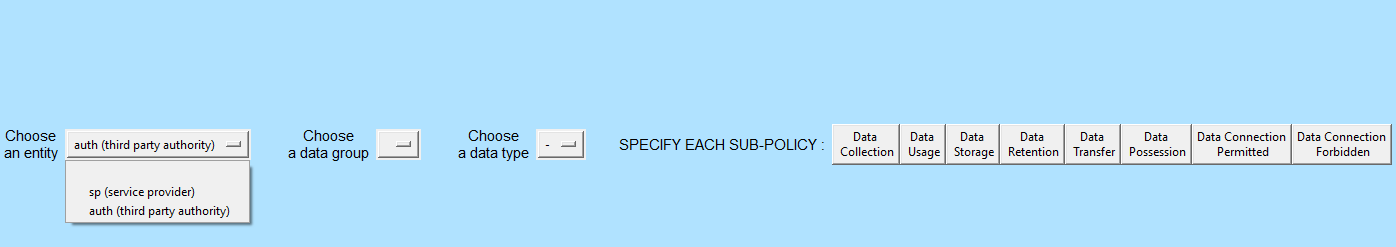}
    \end{center}
    \caption{The Policy Specification Page (entities).}
    \label{fig:19}
\end{figure}

A data protection policy is defined on a data group/type and an entity. In DataProVe, each policy consists of eight sub-policies, to achieve a fine-grained requirement specification (Figure~\ref{fig:19}). The users do not have to define all the eight sub-policies, but they can if it is necessary. Both the policies and architectures can be saved, and opened later to modify or extend.   

\begin{figure}[htb!]
    \begin{center}
        \includegraphics[width=1\textwidth]{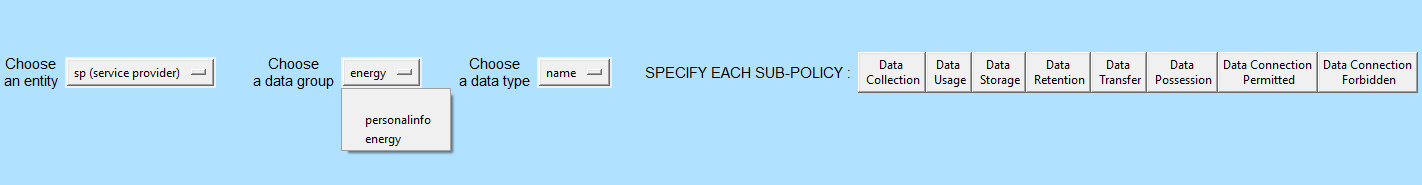}
    \end{center}
    \caption{The Policy Specification Page (data groups).}
    \label{fig:20}
\end{figure}

The first five sub-policies (collection, transfer) are defined \textit{only from the service provider's} perspective. For the rest three sub-policies (data possession and the two data connections policies), the user can specify from any entity's perspective.   

\begin{figure}[htb!]
    \begin{center}
        \includegraphics[width=1\textwidth]{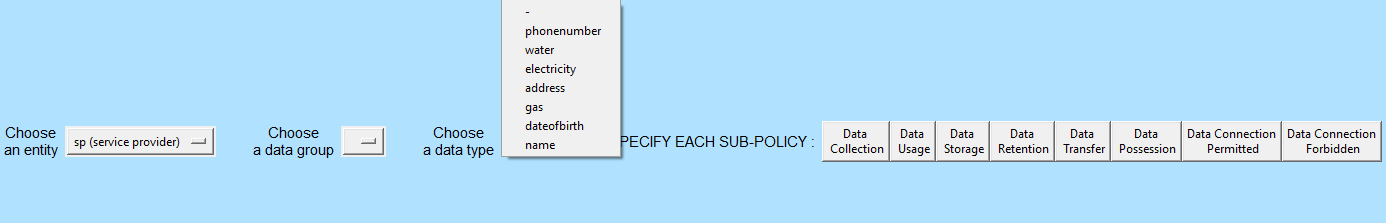}
    \end{center}
    \caption{The Policy Specification Page (choosing among data types).}
    \label{fig:21}
\end{figure}

The eight sub-policies are data collection, data usage, data storage, data retention, data transfer, data possession and the two data connection sub-policies. Below we only highlight four sub-policies, for the rest four the readers are referred to full manual in the GitHub repository\footnote{\url{https://github.com/vinhgithub83/DataProVe}}.

\textbf{The data collection sub-policy:} In the data collection sub-policy window, for a given entity and data group the user can specify whether consent is required to be collection when the selected entity collect a selected data group (Y for Yes/N for No), and then  specify the collection purposes.

\begin{figure}[htb!]
    \begin{center}
        \includegraphics[width=1\textwidth]{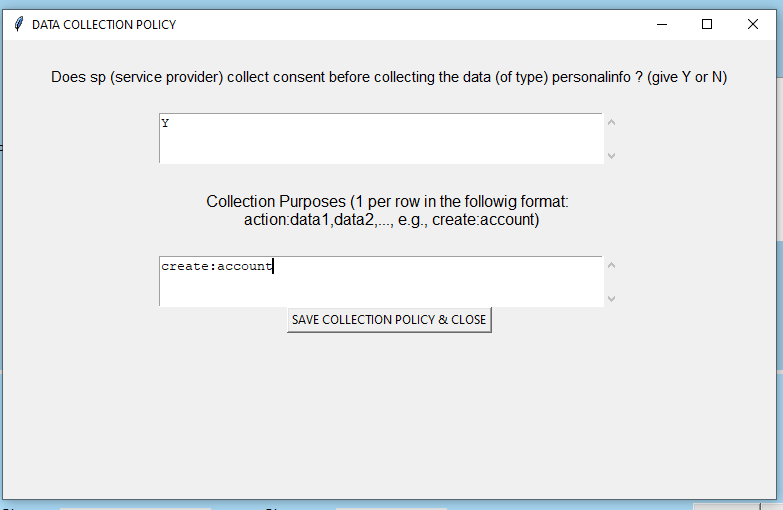}
    \end{center}
    \caption{The data collection sub-policy.}
    \label{fig:22}
\end{figure}

The collection purposes can be given row by row, each row with a different action in the format of: 

\begin{center}
\begin{tabular}{|c|}
\hline\\ 
 \textbf{\textit{action1:data1,data2,\dots,data\_n}}  \\ \\ 
 \hline
\end{tabular}
\end{center}

where action1 can be any action, while data1,\dots, data\_n are compound data types (note that these compound data types do not need to be specified/added in the policy).   For example, in Figure~\ref{fig:22}, the user sets that consent is required to be collected when the service provider collects the personal information. Then, the collection purpose for personal information is to create an account. The compound type account does not need to be defined in the policy.

\textbf{The data possession sub-policy:}

The data possession sub-policy defines who can have/possess a piece of data of a given group. The users only need to specify who are allowed to have or possess a given data group, DataProVe  automatically assumes that the rest entities/components are not allowed to have/possess the selected type of data.  

\begin{figure}[htb!]
    \begin{center}
        \includegraphics[width=1\textwidth]{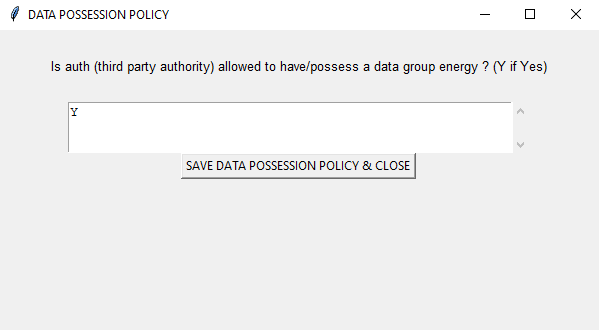}
    \end{center}
    \caption{The data possession sub-policy.}
    \label{fig:28}
\end{figure}

\textbf{The data connection permitted sub-policy:}
This sub-policy specifies which entity is permitted to connect or link two types/groups of data.

\begin{figure}[htb!]
    \begin{center}
        \includegraphics[width=1\textwidth]{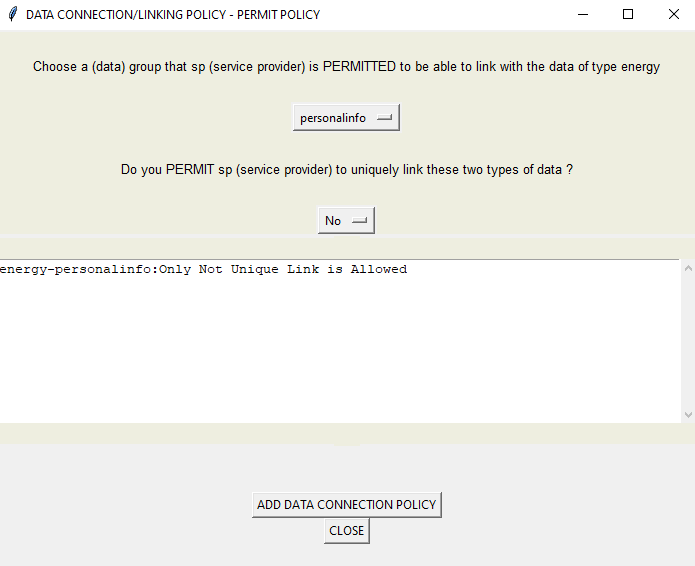}
    \end{center}
    \caption{The data connection permission sub-policy.}
    \label{fig:29}
\end{figure}

In the second drop-down option menu, the user can specify further if the selected entity is permitted to be able to link two pieces of data uniquely, meaning that it will be able to deduce that the two pieces of data belongs to the same individual. 
 
For example, in Figure~\ref{fig:29}, we specified that the service provider is permitted to be able to link the data group energy and the data group personalinfo. However, we do not allow the service provider to be able to uniquely link the two data groups.  Obviously, if personalinfo was defined as unique, then unique link would be possible, so there is chance that the architecture always violates this requirement of the policy.    

\textbf{The data connection forbidden sub-policy:}
This sub-policy is the counterpart of the permitted policy. While in case of the data possession policy, the user only needs to specify which entity is allowed to have or possess certain type of data, and DataProVe automatically assumes that the rest are not allowed, here the user needs to explicitly specify which pair of data types/groups are an entity is forbidden to be able to link together.   

\begin{figure}[htb!]
    \begin{center}
        \includegraphics[width=1\textwidth]{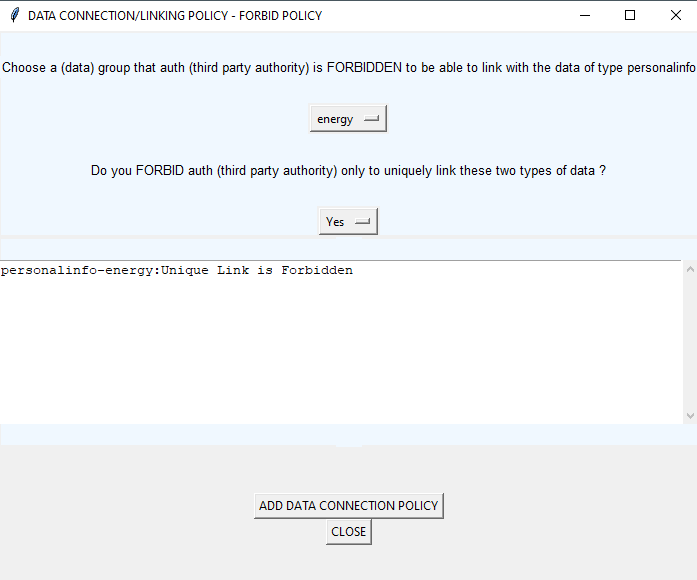}
    \end{center}
    \caption{The data connection permission sub-policy. The case when only unique link is forbidden.}
    \label{fig:30}
\end{figure}

For example, in Figure~\ref{fig:30}, we forbid for the third-party authority to be able to link the data group personalinfo with the data group energy. Here, we forbid the unique link-ability of these  two data groups for the third-party authority.  

\begin{figure}[htb!]
    \begin{center}
        \includegraphics[width=1\textwidth]{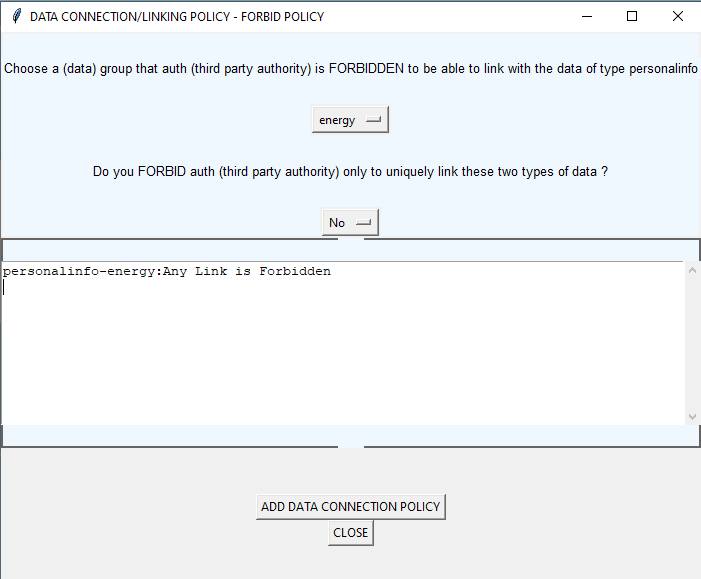}
    \end{center}
    \caption{The data connection permission sub-policy.}
    \label{fig:31}
\end{figure}

If we choose “No” (Figure~\ref{fig:31}), then it means that any ability to link any two pieces of data of the given data groups, is forbidden (not just unique link). Hence, this option is stricter than the previously one.

\subsection{Conformance verification}
\label{sec:dtconformance}
 
\begin{figure}[htb!]
    \begin{center}
        \includegraphics[width=1\textwidth]{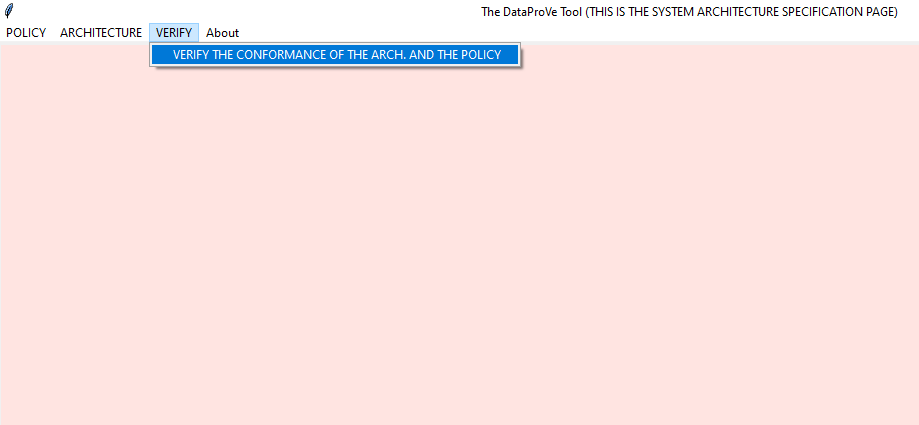}
    \end{center}
    \caption{To verify the conformance between the specified system architecture and policy.}
    \label{fig:32}
\end{figure}

We define three types of conformance, namely, functional conformance, privacy conformance and the so-called DPR conformance. 

\subsubsection{Functional conformance}

The functional conformance captures if an architecture is functionally conforming with the specified policy. Namely:

\begin{enumerate}
\item If in the policy, we allow for an entity to be able to have a piece of data of certain data type/group, then in the architecture the same entity can have a piece of data of the same type/group.
\item If in the policy, we allow for an entity to be able to link/uniquely link two pieces of data of certain types/groups, then in the architecture the same entity can link/uniquely link two pieces of data of the same types/groups. 
\item If in the policy, the (collection, usage, storage, transfer) consent collection is not required for a piece of data of given type/group, then in the architecture there is no consent collection. 
\item If in the policy, we define 

\begin{enumerate}
\item a storage option ``Main and Backup Storage" for a piece of data of certain type/group, then in the architecture there is a \textbf{STORE} or \textbf{STOREAT} action defined for both \textbf{mainstorage} and \textbf{backupstorage}, and for the same data type/group; 
\item a storage option ``Only Main Storage", then in the architecture there is a \textbf{STORE} or \textbf{STOREAT} action defined for only \textbf{mainstorage}, and for the same data type/group.
\item If in the policy, we allow a piece of data of certain type/group, data, to be transferred to an entity ent, then in the architecture there is \textbf{RECEIVEAT}(ent,data,\textbf{Time(t)}) or \textbf{RECEIVE}(ent,data).   
\end{enumerate}

\end{enumerate}

\subsubsection{Violation of the functional conformance}
\begin{enumerate}
\item In the policy, we allow for an entity to be able to have a piece of data of certain data type/group, but in the architecture the same entity cannot have a piece of data of the same type/group.
\item In the policy, we allow for an entity to be able to link/uniquely link two pieces of data of certain types/groups, but in the architecture the same entity cannot link/uniquely link two pieces of data of the same types/groups. 

\item In the policy, the (collection, usage, storage, transfer) consent collection is not required for a piece of data of given type/group, but in the architecture there is a consent collection, namely, an action 
\begin{itemize}
\item \textbf{RECEIVEAT}(\textbf{sp},\textbf{Cconsent}(data),\textbf{Time(t)}), or 
\item \textbf{RECEIVEAT}(\textbf{sp},\textbf{Sconsent}(data),\textbf{Time(t)}), or 
\item\textbf{RECEIVEAT}(\textbf{sp},\textbf{Uconsent}(data),\textbf{Time(t)}), or 
\item \textbf{RECEIVEAT}(third,\textbf{Fwconsent}(data,third),\textbf{Time(t)}). 
\end{itemize}

\item In the policy, we define 
\begin{enumerate}
\item a storage option ``Main and Backup Storage” for a piece of data of certain type/group, but in the architecture, there is \textbf{STORE} or \textbf{STOREAT} action defined for only either \textbf{mainstorage} or \textbf{backupstorage}, or no store action defined at all, for the same data type/group; 
\item	a storage option ``Only Main Storage”, but in the architecture there is no \textbf{STORE} or \textbf{STOREAT} action defined at all, for the same data type/group.
\end{enumerate}

\item In the policy, we allow a piece of data of certain type/group, data, to be transferred to an entity ent, but in the architecture there is no \textbf{RECEIVEAT}(ent,data,\textbf{Time(t)}) or \textbf{RECEIVE}(ent,data) defined (i.e., data is not transferred to the entity ent).   

\end{enumerate}

\subsubsection{Privacy conformance}

The privacy conformance captures if an architecture satisfies the privacy requirements defined in the policy. Namely: 

\begin{enumerate}
\item If in the policy, we forbid for an entity to be able to have or possess a piece of data of certain type/group, then in the architecture the same entity cannot have or possess a piece of data of the same type/group.

\item If in the policy, we forbid for an entity to be able to link/uniquely link two pieces of data of certain types/groups, then in the architecture the same entity cannot link/uniquely link two pieces of data of the same types/groups. 
\end{enumerate}

\subsubsection{Violation of the privacy conformance}
\begin{enumerate}
\item In the policy, we forbid for an entity to be able to have or possess a piece of data of certain type/group, but in the architecture the same entity can/is be able to have or possess a piece of data of the same type/group.

\item In the policy, we forbid for an entity to be able to link/uniquely link two pieces of data of certain types/groups, but in the architecture the same entity can/is be able to link/uniquely link two pieces of data of the same types/groups.  
\end{enumerate}

\subsubsection{DPR conformance}
The privacy conformance captures if an architecture satisfies the data protection requirements defined in the policy. Namely: 

\begin{enumerate}
\item If in the policy, the (collection, usage, storage, transfer) consent collection is required for a piece of data of given type/group, then in the architecture there is a collection for the corresponding consent. 

\item If in the policy, we define a (collection, usage, storage) purpose action:data for a piece of data of certain type/group, then in the architecture there is the action action defined on a compound data type data. 
\end{enumerate}

\subsubsection{Violation of the DPR conformance}
\begin{enumerate}
\item In the policy, the (collection, usage, storage, transfer) consent collection is required for a piece of data of given type/group, but in the architecture, there is no collection for the corresponding consent. 
\item In the policy, we define a (collection, usage, storage) purpose action:data for a piece of data of certain type/group, but in the architecture there is not any action action defined on a compound data type data, or besides action, there are also other actions defined in the architecture on data that are not allowed in the policy.  
\item In the policy, we define 
\begin{enumerate}
\item	a storage option ``Main and Backup Storage" for a piece of data of certain type/group, but in the architecture there is a STORE or STOREAT action defined for some storage place, different from \textbf{mainstorage} and \textbf{backupstorage}, for the same data type/group; 
\item	a storage option ``Only Main Storage", but in the architecture there is a \textbf{STORE} or \textbf{STOREAT} action defined for some storage place, different from \textbf{mainstorage}, for the same data type/group.
\end{enumerate}

\item In the policy, we define 

\begin{enumerate}
\item a deletion option ``From Main and Backup Storage" for a piece of data of a certain data type/group, data, but in the architecture there is not any of the action 
\begin{itemize}
\item \textbf{DELETE}(\textbf{mainstorage},data) or 
\item \textbf{DELETEWITHIN}(\textbf{mainstorage},data,\textbf{Time}(tvalue)), or 
\item \textbf{DELETE}(\textbf{backupstorage},data) or
\item  \textbf{DELETEWITHIN}(\textbf{backupstorage},data,\textbf{Time}(tvalue));
\end{itemize}

\item a deletion option ``Only From Main Storage" for a piece of data of a certain data type/group, data, but in the architecture there is no action  \textbf{DELETE}(\textbf{mainstorage},data) or \textbf{DELETEWITHIN}(\textbf{mainstorage},data,\textbf{Time}(tvalue)).
\end{enumerate}

\item	In the policy, we allow a piece of data of certain type/group, data, to be transferred to an entity ent, but in the architecture there is also an action \textbf{RECEIVEAT}(ent1,data,\textbf{Time(t)}) or \textbf{RECEIVE}(ent1,data) defined for some ent1 to whom we do not allow data transfer in the policy. 
\end{enumerate}

\subsection{Application Examples}
\label{sec:examples} 

In this section, we highlight the operation of DataProVe using two very simple examples.

\subsubsection{Example 1 (Data retention policy)}
\label{ex:del}

\begin{figure}[htb!]
    \begin{center}
        \includegraphics[width=0.7\textwidth]{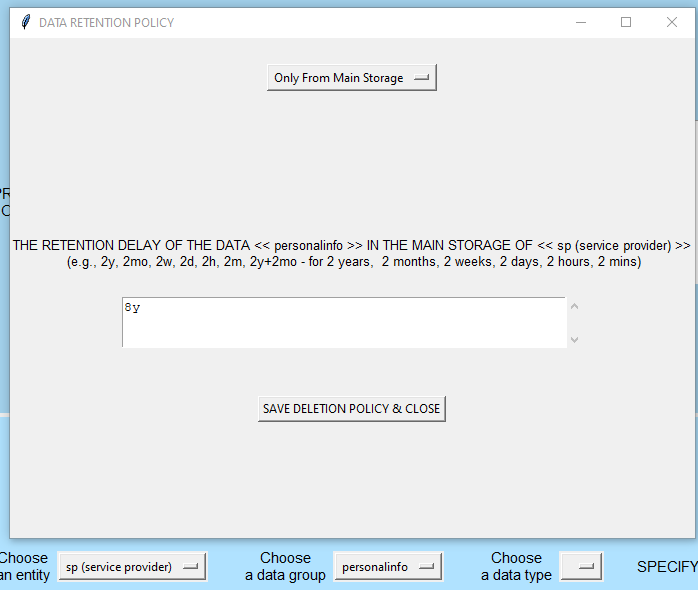}
    \end{center}
    \caption{We set that the data of type/group personal information must be deleted from the main storage places of the service provider within 8 year.}
    \label{fig:33}
\end{figure}

In this example, in the policy we specify a data group (a group of data types) called \textit{personalinfo}, which is stored centrally at the main storage places of the service provider. In the storage sub-policy, we also set that storage consent is required before the storage of \textit{personalinfo}. Finally, we do not give service provider (sp) the right to have the data of group/type \textit{personalinfo}. In the deletion policy, we set the retention delay in the main storage to 8 years (i.e. 8y in Figure~\ref{fig:33}).  

In the architecture level, we add an action that says a piece of data of type personalinfo must be deleted from the main storage within 10 years (action DELETEWITHIN, in the last line). 

\begin{center}
\begin{tabular}{|c|}
\hline\\ 
 Content of \textit{spmessages}: \textbf{RECEIVEAT}(\textbf{sp},\textbf{Sconsent}(personalinfo),\textbf{Time}(t))  \\ 
Content of \textit{storagemessages}: \textbf{RECEIVEAT}(\textbf{mainstorage},personalinfo,\textbf{Time}(t)) \\
Content of \textit{storemain}: \textbf{STOREAT}(\textbf{mainstorage},personalinfo,\textbf{Time}(t)) \\
Content of \textit{deletion}: \textbf{DELETEWITHIN}(\textbf{mainstorage},personalinfo,\textbf{Time}(10y)).  \\ \\ 
 \hline
\end{tabular}
\end{center}

\begin{figure}[htb!]
    \begin{center}
        \includegraphics[width=1\textwidth]{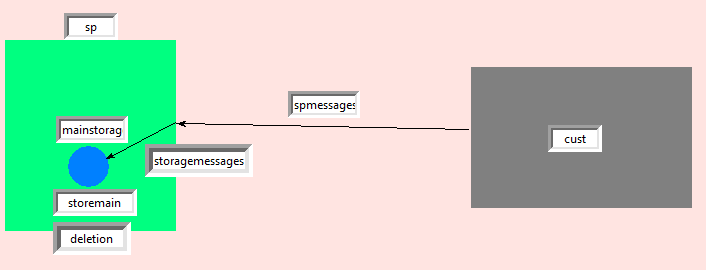}
    \end{center}
    \caption{The service provider (sp) stores the personal information in its main storage places.}
    \label{fig:34}
\end{figure}

In the architecture shown in Figure~\ref{fig:34}, the service provider (sp) can receive a storage consent for \textit{personalinfo} at some non-specific time \textit{t}. The main storage places of sp can receive the data at some non-specific time and store it. The data of this type/group is deleted within 10 years from the main storage places.   

\begin{figure}[htb!]
    \begin{center}
        \includegraphics[width=1\textwidth]{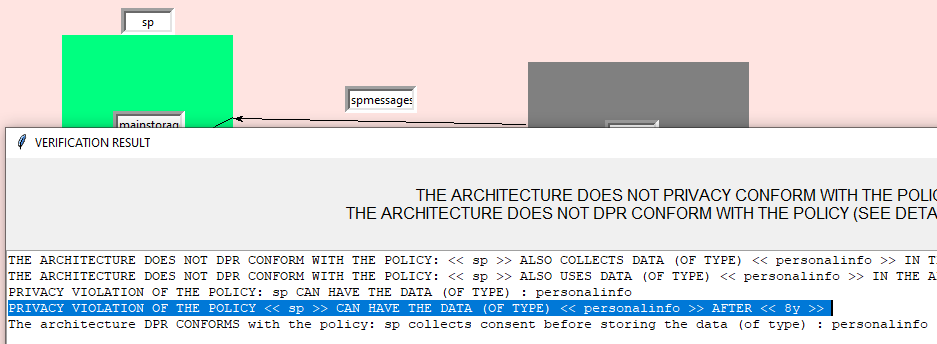}
    \end{center}
    \caption{The verification results show the violation of the privacy and DPR conformance properties. We also got the first two lines of DPR conformance because in this example, we did not specify the collection and usage sub-policies (we left them blank).}
    \label{fig:36}
\end{figure}

As a verification result (Figure~\ref{fig:36}), we got that the architecture violates the privacy conformance, as the architecture allows for sp to have the data of type personalinfo after 8 years, however, in the policy we set it to only 8 years.  In the last line of the verification result window, we can also see a DPR conformance property, namely, sp collects storage consent before the data is stored. 

\subsection{Example 2 (Data possession and connection policy)}
\label{ex:con}

In the second simple example, we focus on the data possession and data connection sub-policies. 
We present the receive action with the Meta construct (metadata or "packet" header data such as IP address, source, destination addresses, etc.). 

\begin{figure}[htb!]
    \begin{center}
        \includegraphics[width=1\textwidth]{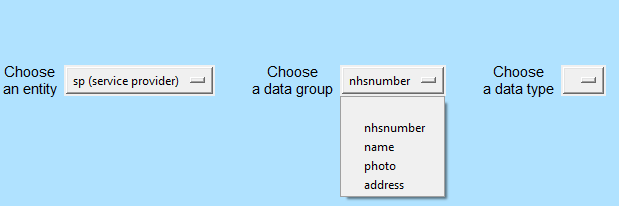}
    \end{center}
    \caption{The policy level with the four data types/groups.}
    \label{fig:37}
\end{figure}

In the policy, we define four data groups, nhsnumber (National Health Service number), name, photo, and address (see Figure~\ref{fig:37}). 

\begin{figure}[htb!]
    \begin{center}
        \includegraphics[width=0.8\textwidth]{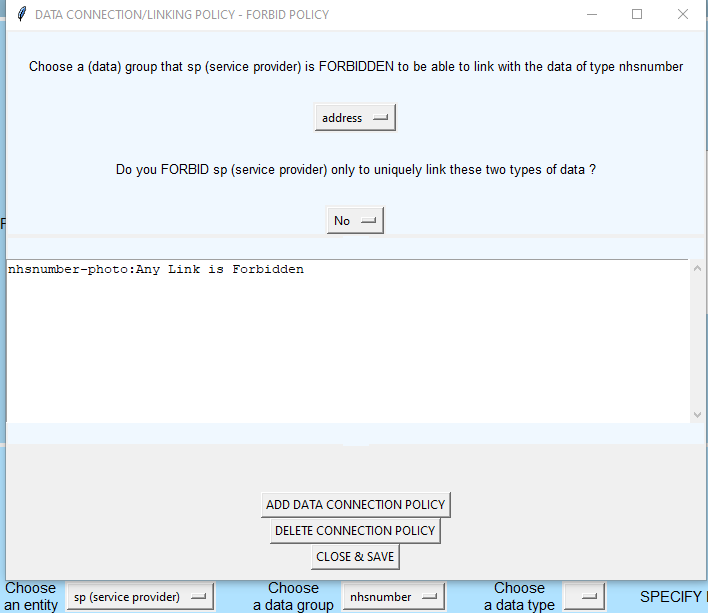}
    \end{center}
    \caption{The specified data connection sub-policy for example 2.}
    \label{fig:38}
\end{figure}

Then, we forbid (any kind of link-ability, not only unique link) for the service provider to be able to link two pieces of data of types nhsnumber, and photo (see Figure~\ref{fig:38}). Again, we also forbid for the service provider to be able to have all the four data types/groups. 

\begin{figure}[htb!]
    \begin{center}
        \includegraphics[width=1\textwidth]{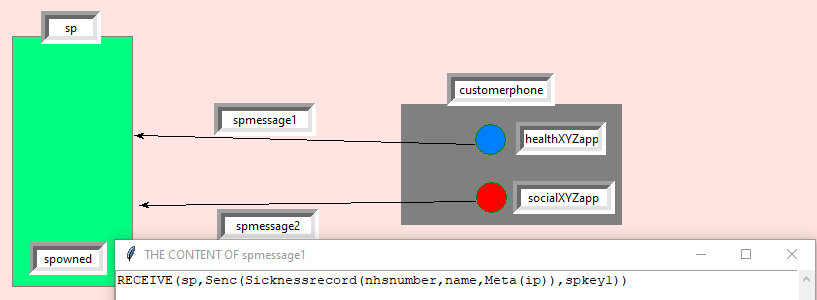}
    \end{center}
    \caption{The specified architecture for example 2.}
    \label{fig:39}
\end{figure}

In the architecture, a service provider collects data from two phone applications (Figure~\ref{fig:39}). The "HealthXYZ" app sends the service provider a sickness record with a public IP address (an unique IP of a phone) other app, called, "SocialXYZ" also sends the social profile with the same ip address (same phone). Both data types are encrypted (using symmetric key encryption) with the service provider keys (and sp owns the two keys). 

\begin{center}
\begin{tabular}{|c|}
\hline\\ 
 Content of \textit{spmessage1} in Figure~\ref{fig:39}:\\ 
 \textbf{RECEIVE}(\textbf{sp},\textbf{Senc}(Sicknessrecord(nhsnumber,name,\textbf{Meta}(ip)),spkey1))  \\\\
Content of \textit{spmessage2} in Figure~\ref{fig:39}:\\
 \textbf{RECEIVE}(\textbf{sp},\textbf{Senc}(Socprofile(photo,address,\textbf{Meta}(ip)),spkey2)) \\\\
Content of \textit{spowned} in Figure~\ref{fig:39}:\\
 \textbf{OWN}(\textbf{sp},spkey1)\\ 
 \textbf{OWN}(\textbf{sp},spkey2)\\ \\ 
 \hline
\end{tabular}
\end{center}

\begin{figure}[htb!]
    \begin{center}
        \includegraphics[width=1\textwidth]{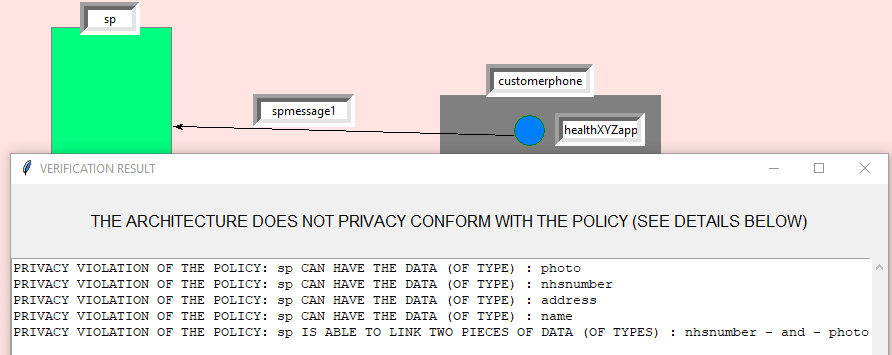}
    \end{center}
    \caption{The verification result for example 2.}
    \label{fig:40}
\end{figure}

As a result (Figure~\ref{fig:40}), we got that the service provider not only be able to link the data of types nhsnumber with the data of type photo, but it also has all the data of types nhsnumber, name, photo and address. The reason is that sp will be able to decrypt both messages and link, have the data inside them. Note that we only have linkability but not unique link, because the Apps can be used by different people in one family, so the set of possible individuals can be narrowed down, but \textit{sp} cannot be sure that \textit{nhsnummber} and \textit{photo} belong to the same individual.

\section{Conclusion and Future Work}
\label{sec:conc}
We addressed the problem of formal specification and automated verification of data protection requirements at the policy and architecture levels. Specifically, we proposed a variant of policy and architecture languages to specify a simple set of data protection requirements based on the GDPR. In addition, we proposed DataProVe, a tool based on the syntax of our languages and a logic based verification engine to check the conformance between a policy and an architecture. In this paper, our language variants and tool only cover a limited set of data protection requirements in an abstract way, hence, there are many possibilities to extend and improve their syntax and semantics to specify more complex laws. Regarding the conformance check of the privacy properties (the right to have and link data), a possible extension would be including the behaviour of the hostile attackers (e.g. steal personal data) in the verification.   Finally, we plan to improve the effectiveness of the conformance check algorithm for the data types with a large number of nested layers. 

\bibliographystyle{unsrt}
\bibliography{datatest}

\appendix 


\end{document}